\documentclass[12pt]{article}

\usepackage{amssymb,amsmath,amsmath,amsfonts,eurosym,geometry,ulem,graphicx,color,setspace,comment,footmisc,caption,pdflscape,array,hyperref}
\usepackage{booktabs}
\usepackage[table]{xcolor}
\usepackage{indentfirst}
\usepackage{natbib}
\usepackage[font=normalsize,labelfont=bf]{caption}
\usepackage{adjustbox}
\usepackage{multirow}
 \usepackage{ragged2e} 
\usepackage{mathtools}
\usepackage[flushleft]{threeparttable}
\usepackage{mwe}
\usepackage{lipsum}
\usepackage{longtable}
\usepackage{supertabular}
\usepackage{float} 
\usepackage{appendix}
\usepackage{xcolor}
\usepackage{multirow}
\usepackage{subcaption}
\usepackage{alphalph}
\usepackage{graphicx} 

\newcommand{\h}[1]{\boldsymbol{#1}}

\DeclareMathOperator{\Var}{Var}
\DeclareMathOperator{\E}{E}

\newcolumntype{L}[1]{>{\raggedright\let\newline\\arraybackslash\hspace{0pt}}m{#1}}
\newcolumntype{C}[1]{>{\centering\let\newline\\arraybackslash\hspace{0pt}}m{#1}}
\newcolumntype{R}[1]{>{\raggedleft\let\newline\\arraybackslash\hspace{0pt}}m{#1}}
\usepackage{authblk}
\renewcommand{\normalsize}{\fontsize{11}{11}\selectfont}

\usepackage{hyperref}
\hypersetup{
	colorlinks=true,
	linkcolor=blue,
	anchorcolor=blue,
	citecolor=blue} 

\newcommand{\hy}{\h{y}}

\newcommand{\hvarepsilon}{\h{\varepsilon}}

\newcommand{\hI}{\h{I}}

\begin{document}

\begin{titlepage}
\title{\bf\LARGE Monetary Policies on Green Financial Markets: Evidence from a Multi-Moment Connectedness Network}
\author[1,2,3,4]{\MakeUppercase{Tingguo Zheng}}
\author[3]{\MakeUppercase{Hongyin Zhang}}
\author[4,5*]{\MakeUppercase{Shiqi Ye}}
\affil[1]{Center for Macroeconomic Research, Xiamen University}
\affil[2]{Department of Statistics and Data Science, School of Economics, Xiamen University}
\affil[3]{Wang Yanan Institute for Studies in Economics, Xiamen University}
\affil[4]{Paula and Gregory Chow Institute for Studies in Economics, Xiamen University}
\affil[5]{Academy of Mathematics and Systems Science, Chinese Academy of Sciences}
\date{This version: December 2023}
\renewcommand*{\Affilfont}{\small\it} 
\renewcommand\Authands{ and } 

\maketitle

\begin{abstract}
\noindent 
This paper introduces a novel multi-moment connectedness network approach for analyzing the interconnectedness of green financial market. Focusing on the impact of monetary policy shocks, our study reveals that connectedness within the green bond and equity markets varies with different moments (returns, volatility, skewness, and kurtosis) and changes significantly around Federal Open Market Committee (FOMC) events. Static analysis shows a decrease in connectedness with higher moments, while dynamic analysis highlights increased sensitivity to event-driven shocks. We find that both tight and loose monetary policy shocks initially elevate connectedness within the first six months. However, the effects of tight shocks gradually fade, whereas loose shocks may reduce connectedness after one year. These results offer insight to policymakers in regulating sustainable economies and investment managers in strategizing asset allocation and risk management, especially in environmentally focused markets. Our study contributes to understanding the complex dynamics of the green financial market in response to monetary policies, helping in decision-making for sustainable economic development and financial stability.

\vspace{.5em}
\vspace{.5em}
\end{abstract}
\setcounter{page}{0}
\thispagestyle{empty}
\end{titlepage}
\pagebreak \newpage

\doublespacing

\section*{Highlights}

\begin{itemize}
\item We propose a novel multi-moment connectedness network approach that integrates the information conveyed by the connectedness
of different moments.
\item We quantify the short- and long-term impacts of monetary policy shock on the multi-moment connectedness within the green financial market.
\item Static analysis shows that connectedness decreases with higher moments.

\item Our dynamic analysis reveals significant variations in the projected total connectedness surrounding the FOMC event.

\item Higher moment connectedness might be more sensible under event-driven shocks.

\item Tight and loose monetary policy shocks all increase the overall connectedness during the first six months, the influence of tight monetary policy shock will disappear, and the loose monetary policy shock tends to have an adverse effect.
\end{itemize}

\clearpage

\section{Introduction} \label{sec:introduction}

The issue of climate warming has become a critical concern around the world \citep{lin2023tracking}. Nations are progressively adopting measures to reduce emissions, thereby facilitating the transition of their economies towards a low-carbon model. These initiatives aim to achieve innovation, job creation, increased international influence, and the promotion of long-term sustainable economic growth. Concurrently, environmentally-friendly financial instruments, including green bonds and green equities, are increasingly attracting the interest of investors \citep{sun2019fossil}. There is a significant interconnection between these green bonds and green assets, reflecting the uncertainties present in the green financial market during the low-carbon transition process. Such a connectedness provides valuable assistance in monitoring and mitigating the uncertainty or potential risks associated with low carbon transformation \citep{su2022green, su2022liquidity, lin2023uncertainties}.

Monetary policy shocks have long been a critical source of uncertainty in financial markets. In recent years, a growing body of literature has focused and confirmed the significant impact of monetary policy shocks on the fluctuation of the prices of financial assets \citep{lin2019effectively, li2022tracking, sun2022analysis}. In particular, existing research has identified that monetary policy shocks have a significant impact on connectedness within global or local financial markets \citep{zhou2022does, raza2023connectedness, chen2023determinants}. However, due to differences in investment types, investment focus and investment objectives between the green financial market and the global or local financial market, their responses to monetary policy shock may not necessarily be the same \citep{lin2023emerging}. 
Understanding the impact of monetary shocks on the green financial market is critical for advancing environmentally sustainable investments, enhancing the effectiveness and transparency of policies, increase investor confidence, and addressing global environmental and economic challenges \citep{dafermos2018climate}.
Consequently, it is important and necessary to examine the impact of monetary policy shocks on connectedness within the green financial market.

Existing research on connectedness within the green financial market typically estimates the return connectedness or volatility connectedness of green bonds or green equities \citep{reboredo2020network, reboredo2020price, mensi2022spillovers}. Recent studies have also found that connectedness of higher-order moments (skewness or kurtosis) can usually provide useful information about the uncertainty in the green financial market \citep{dogan2022investigating, zhang2023impact, hao2023dynamic}. However, the existing literature discusses the results of connectedness from different moments separately, and there is no effective way to integrate the information conveyed by the connectedness of different moments of green bonds or equities. To address this limitation, this paper proposes a multi-moment connectedness network approach that aims to examine the uncertainty in the green financial market from a broader perspective, utilizing a richer set of information.

Moreover, as mentioned above, although the existing literature has examined the impact of monetary policy shocks on global or local financial markets, there are few studies exploring the effects of these shocks on the green financial market. The limited existing literature mainly employs qualitative analysis methods and lacks adequate quantification of monetary policy shocks \citep{dafermos2018climate, desalegn2022effect}. This paper employs the methodology developed by \cite{jarocinski2020deconstructing, zheng2023global} to extract monetary policy shocks and examines their impact on connectedness within the green finance market.

Building on the aforementioned discussions, the contributions of this paper are notably reflected in three aspects: research methodology, research perspective, and research findings. Firstly, in terms of research methodology, this paper proposes a novel method for the construction and analysis of multi-moment connectedness networks. This method involves a general discussion of the networks of connectedness in green bonds and green asset prices based on returns (first moment), volatility (second moment), skewness (third moment) and kurtosis (fourth moment). Specifically, we construct weights based on the information density of each layer of the network and then build a projection network based on the weights and the connectedness networks of different moments. This approach allows for the integration of information extracted from multi-moment networks, enabling a comprehensive portrayal of the uncertainties or connectedness in the green finance market.

The second contribution of this paper lies in its unique research perspective. Specifically, drawing on the work of \cite{jarocinski2020deconstructing} and \cite{zhou2022does}, we utilize the variations around the Federal Reserve's Federal Open Market Committee (FOMC) meetings in 3-month Fed Funds Futures and the S\&P 500 index to extract monetary policy shocks based on a Bayesian Vector Autoregression (VAR) model. Further, we use local projection to explore the impact of these monetary policy shocks on the multi-moment connectedness in the green finance market. This approach provides both a quantitative tool and a novel research perspective to examine the effects of monetary policy on the green finance market.

The third key contribution of this paper is the discovery of several important empirical insights. In our static analysis of the multi-connectedness network, we observe a trend where the level of connectedness decreases at higher moments. In terms of dynamic analysis, we note a substantial shift in the projected total connectedness around the time of FOMC events. While higher moments demonstrate lower connectedness, they appear to be more sensitive to event-driven shocks, offering additional insights for risk identification. In our local projection analysis, we find that tight monetary policy shocks lead to a significant increase in the green financial market's interconnectedness during the initial six months, followed by a gradual reduction in their impact. On the other hand, loose monetary policy shocks also increase total connectedness over a similar six-month period, but they may result in a decrease in connectedness after one year.

Understanding the impact of monetary policy shocks on connectedness within the green finance market is crucial for both policymakers and investment managers. For policymakers, recognizing the short-, medium-, and long-term effects of monetary policy shocks on the connectedness of the green finance market is vital for sustainable economic regulation. Especially in economies where green finance is a significant component of the financial system, it is important to closely monitor how monetary policy impacts the connectedness between green assets, as these can have wider implications for economic stability and the transition towards a more sustainable economy. For countries with a robust green finance market, policymakers should be particularly attentive to how changes in monetary policy might influence these connectedness, in order to manage systemic risks effectively. For investment managers, understanding the influence of monetary policy shocks on green finance connectedness can aid in predicting market movements, allowing for more informed asset allocation and risk management strategies. This knowledge can be especially valuable in navigating the uncertainties of a market increasingly influenced by sustainability and environmental considerations.






We organize the remaining content as follows: Section 2 provides a brief review of related literature, Section 3 describes the data and provides some summary statistics, Section 4 introduces the primary methodologies in this paper, Section 5 documents some interesting findings and Section 6 concludes the paper.

\section{Literature review}

The study of connectedness between financial assets has become a hot topic in the field of financial market stability and risk measurement. Since the seminal paper of \cite{diebold2014network}, which introduced the concept of measuring inter-market connectedness within a vector autoregression model framework, a series of studies have made beneficial attempts around the topic of financial market connectedness \citep{diebold2015trans, maghyereh2016directional, yang2017quantitative, barigozzi2017network, hale2019monitoring, barunik2020asymmetric, ando2022quantile,zheng2024cholesky}. Among them, \cite{diebold2015trans} investigate the volatility connectedness within and between U.S. and European financial institutions; \cite{hale2019monitoring} estimate the connectivity between banks around the world to monitor financial systemic risk; \cite{barunik2020asymmetric} measure the asymmetric connectedness of fears obtained from the implied variance of assets; \cite{ando2022quantile} analyze the tail behavior of the connectedness between sovereign CDS spreads.  It is evident that the connectedness measurements serves as an effective tool for measuring uncertainty in financial markets, as well as for assessing the spillover relationships between financial assets.


Moreover, existing research has made a series of beneficial attempts to measure and analyze connectedness within the green finance market, particularly between green bonds or green equities. Among these, some studies have focused on the discussion and analysis of the connectedness between returns and volatility of green bonds or equities \citep{lundgren2018connectedness, reboredo2020network,su2022green, wang2023energy,zheng2023global}, while others have examined higher-moment connectedness \citep{zhang2023impact} within the energy finance market. The findings indicate that connectedness in the green finance market can effectively reflect the overall uncertainty of the market and also provide insights into climate transition risks \citep{dai2021multiscale, caporin2023measuring, bouri2023connectedness}. However, the existing literature has not yet examined the characteristics of connectedness in the green finance market from a comprehensive perspective, incorporating information from multiple moments. In light of this,  this paper employs multi-moment connectedness indices and the construction of multi-moment connectedness networks to measure and analyze connectedness in the green finance market from a more comprehensive, holistic, and generalized perspective.






The existing literature has examined the impact of monetary policy, especially that of the United States, on connectedness in financial markets from various perspectives. For example, \cite{yang2017quantitative} investigate the impact of quantitative easing on the volatility connectedness of financial assets across countries; \cite{brunetti2019interconnectedness} studied the response of bank interconnectedness to ECB announcements and interventions; \cite{zhou2022does} examine how U.S. monetary policy affects global financial markets’ connectedness; \cite{chan2023optimal} study the optimal monetary policy respond to green bubbles. However, the aforementioned studies have not examined the impact of monetary policy shocks on connectedness within the green finance market.  Investigating this issue holds important value for policy makers in devising robust low-carbon transition strategies, as well as for investors in making risk assessments amid macroeconomic fluctuations. This study supplements the existing literature with such an important empirical perspective.




\section{Data and motivating evidence}
\label{sec: data and summary}

\subsection{Green financial market}
Our main dataset consists of several indices that represent the financial performance of global green finance markets in terms of green bonds and green equity markets.
For the green bond market, we use three sub-indices of Bloomberg MSCI Global Green Bond Index, namely, Gov-Related (GBIG), Corporate (GBIC) and Financial (GBIF). For the green equity market, we use S\&P Kensho Cleantech Index (KCTI), S\&P Kensho Clean Energy Index (KCEI) and S\&P Global 1200 ESG Index (ESGI). Our analysis also includes the Bloomberg Barclays Global Treasury Bond Index (TBI) and Bloomberg Barclays Global Corporate Bond Index (CBI) to capture global bond market performance, along with the MSCI Energy Sector Index (MEI) to monitor global energy equity market conditions \citep{pham2021frequency}. We use the daily indicators from January 1, 2015, to November 1, 2023, obtained from Bloomberg Terminal. 

For each series, we first calculate the daily log returns by taking the log difference of its daily closing prices $r_{i,t} = \log(P_{i,t}-P_{i,t-1})\times 100$, then use $r_{i,t}$ to extract the time-varying volatility, skewness, and kurtosis, as will be shown later. Table \ref{tab: summary_rn} reports the summary statistics for the return series\footnote{Appendix Table \ref{tab: summary_vol} - \ref{tab: summary_kurt} provide summary statistics of the extracted daily variance, skewness, and kurtosis series.}. 
We observe that most series are stationary under ADF tests and exhibit characteristics of left skewness and leptokurtosis. In particular, bond indices generally show lower kurtosis, while government bond-related indices (GBIG and TBI) exhibit a weaker left skewness, with TBI showing a slight right skewness (0.05).

\begin{table}[!t] \tiny
\caption{Summary statistics}
\label{tab: summary_rn}
\centering
\renewcommand\arraystretch{1.4}
\resizebox{\linewidth}{!}
{
\begin{tabular}{lcccccccc}

\hline
     & Mean  & Std. & Max   & Min    & Skew. & Kurt. & ADF test  & JB test     \\ \hline
GBIG & -0.01 & 0.42 & 2.35  & -2.68  & -0.18    & 3.64     & -43.79*** & 1282.2***   \\
GBIC & 0.00  & 0.38 & 2.30  & -3.56  & -0.54    & 6.78     & -30.45*** & 4511.21***  \\
GBIF & 0.00  & 0.38 & 2.25  & -3.34  & -0.39    & 5.99     & -30.31*** & 3495.9***   \\
TBI  & -0.01 & 0.38 & 2.31  & -2.07  & 0.05     & 3.22     & -43.53*** & 991.49***   \\
CBI  & 0.00  & 0.34 & 2.25  & -3.66  & -0.98    & 11.19    & -23.6***  & 12362.17*** \\
KCTI & 0.04  & 2.36 & 13.43 & -15.61 & -0.21    & 4.60     & -11.08*** & 2039.32***  \\
KCEI & 0.00  & 1.40 & 9.88  & -12.71 & -0.81    & 11.93    & -12.01*** & 13862.03*** \\
ESGI & 0.02  & 0.96 & 8.21  & -10.22 & -1.05    & 16.69    & -12.58*** & 27080.43*** \\
MEI  & 0.00  & 1.69 & 15.67 & -21.23 & -1.18    & 21.79    & -14.25*** & 45953.38*** \\ \hline
\end{tabular}}
\end{table}

\begin{figure}[!h]
    \centering
    \begin{subfigure}{1\textwidth}
        \includegraphics[width=\linewidth]{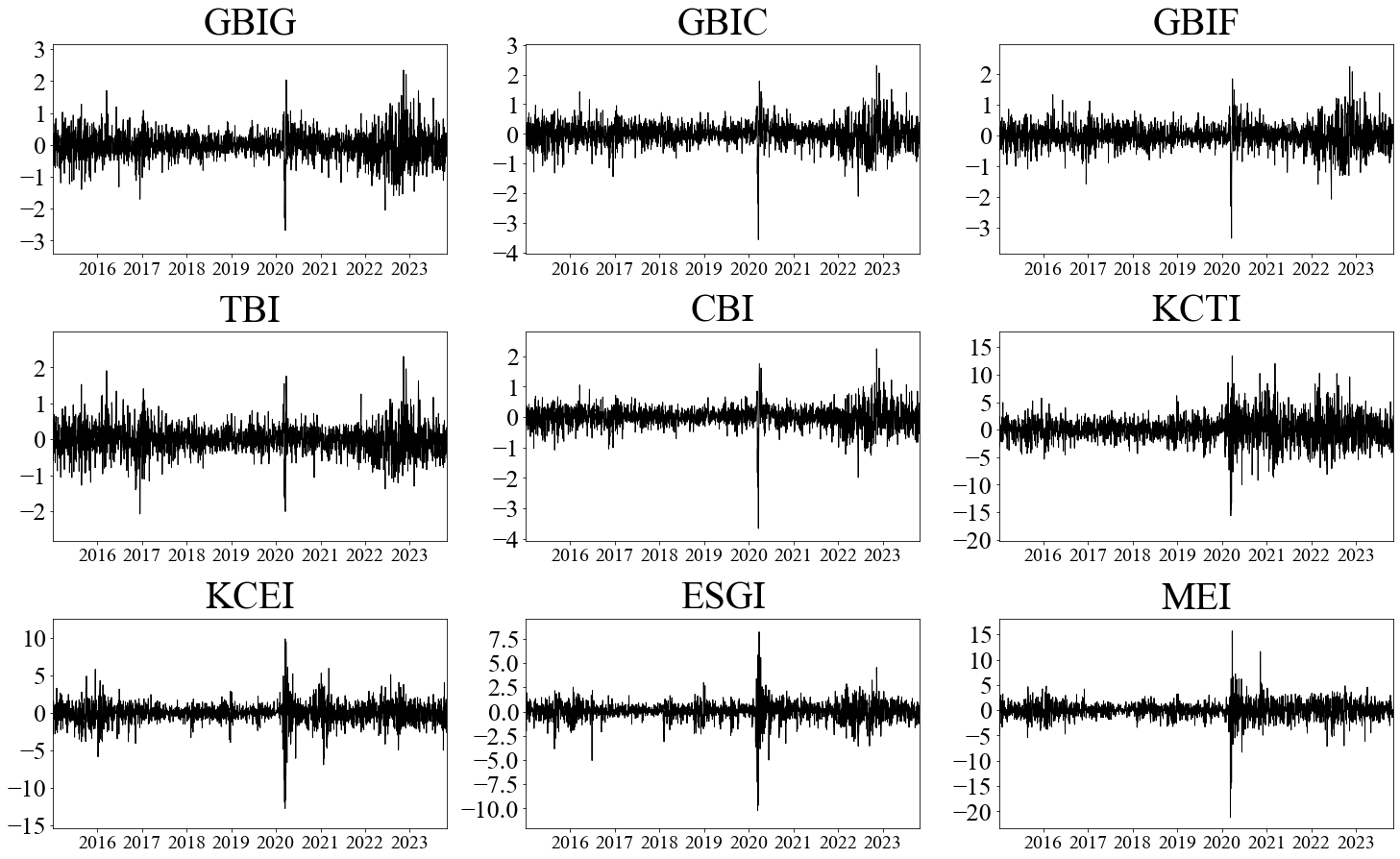}
    \end{subfigure}
    \caption{Return series}
    \label{figure: return}
\end{figure}

Figure \ref{figure: return} plots the return series. 
As we can see, the return of all indices experienced a significant bottom and drastic fluctuations in March 2020.
Since then, the green equity returns (KCTI, KCEI, ESGI) have been consistently highly volatile, while the green bond returns (GBIG, GBIC, GBIF) show enhanced volatility since 2022. 
This may be related to the restart of quantitative easing (QE) by the Fed in 2023 under the outbreak of COVID-19, as well as the start of interest rate hikes in 2022. 
Thus, it provides basic intuition for us to explore the impact of U.S. monetary policy on the green financial markets.

\subsection{Monetary Policy}

To capture U.S. monetary policy, following \cite{jarocinski2020deconstructing}, we extract monetary policy shocks using high-frequency co-movements of interest rates and stock prices around FOMC announcements. This involves two high-frequency surprise variables, including three-month Fed Funds Futures and S\&P 500 Index, both sourced from Barchart.com. The procedure for data processing is outlined as follows: Initially, compile the schedule of Federal Open Market Committee (FOMC) announcements. Subsequently, compute the price fluctuation within a 30-minute interval, spanning from ten minutes prior to twenty minutes following the announcement, and categorize this as the measure of surprise. In instances where a month does not encompass an FOMC meeting, the surprise variable should be designated as zero.

In addition, five low-frequency macroeconomic variables need to be controlled, including the 1-year treasury yield, S\&P 500 index, U.S. CPI, U.S. industrial production, and the excess bond premium \citep{gilchrist2012credit,jarocinski2020deconstructing,zhou2022does}. The ablve data come from the U.S. Department of the Treasury website, the Federal Reserve website, and Bloomberg Terminal. Since the available S\&P 500 Index intra-day data starts from April 2009, the sample period for measuring U.S. monetary policy spans from April 2009 to October 2023. This ensures the effectiveness of the estimation of the SVAR model and the accuracy of the identified monetary policy shocks. 

\section{Methodology}
\label{sec: method}

\subsection{Estimating time-varying volatility, skewness and kurtosis}

Our study initially requires obtaining estimates of the conditional volatility, skewness, and kurtosis of the daily return time series. To do this, we adopt the Gaussian dynamic adaptive mixture models (Gaussian-DAMMs, G-DAMMs) approach proposed by \cite{catania2021dynamic}. Let $\{r_t\}$ denote the return time series of interest, $\mathcal{F}_{t-1}$ denote a collection of past information, and $\h \theta_t$ denotes some potentially time-varying parameters. Let $p(y_t\vert \mathcal{F}_{t-1}, \h \theta_t)$ be the conditional distribution of $r_t$, the G-DAMM method assumes $p(\cdot)$ to be a finite mixture of $J$ real-valued conditional normal distributions
\begin{align}\label{equ: definition of DAMMs}
    p(\h y_t\vert \mathcal{F}_{t-1}, \h \theta_t) = \sum_{j=1}^J w_{j,t}\phi_j\left(y_t\vert \mu_{j,t}, \sigma_{j,t}^2\right), \quad j = 1,\ldots, J,
\end{align}
where $w_{j,t} \in (0,1)$, $\sum_{j=1}^J w_{j,t} = 1, \quad t=1,\ldots, T$ are time-varying weights, $\phi_j(\cdot)$ is univariate normal distribution with mean $u_{j,t}$ and variance $\sigma_{j,t}^2$, and is called the $j$th mixture component of $p(\cdot)$, and $\h \theta_t = \left(\h \theta_{j,t}', w_{j,t}\right)'$ where $\h \theta_{j,t} = (u_{j,t}, \sigma_{j,t})$ for $j = 1,\ldots, J$. 

\cite{catania2021dynamic} uses the score-driven (SD) method to specify the time evolution path of the weight $w_{j,t}$s and model parameters $\h \theta_{j,t}$s. To do so, consider a $(J-1)$-dimensional vector $\widetilde{\h{w}}_t$ such that $\h{\Lambda}^{w}(\widetilde{\h{w}}_t) = \h{w}_t$. Here, $\h{w}_t = (w_{1,t},\ldots, w_{J,t})'$, and $\h{\Lambda}^w : \mathbb{R}^{J-1} \rightarrow \mathcal{S}^J$, where $\mathcal{S}^J$ is the standard $J$-dimensional unit simplex. Similarly, let $\widetilde{\h{\theta}}_{j,t} \in \mathbb{R}^{d_j}$ be a $d_j$-dimensional vector such that for each time point $t$, $\h{\Lambda}^{j}(\widetilde{\h{\theta}}_{j,t})  = \h{\theta}_{j,t}$. Here, $\h{\Lambda}^{j} : \mathbb{R}^{d_j}\rightarrow \h{\Omega}^{j}, \ j = 1,\ldots, J$. Combining these, we can define:
\begin{align*}
    \widetilde{\h{\theta}}_t &= \left(\widetilde{\h{\omega}}', \widetilde{\h{\theta}}_{j,t}', j=1,\ldots, J\right)', \qquad \h{\Lambda}(\widetilde{\h{\theta}}_t) = \h{\theta}_t, \\
    \h{\Lambda} &: \mathbb{R}^{J-1}\times \mathbb{R}^{d_1}\times \cdots \times \mathbb{R}^{d_J} \rightarrow \mathcal{S}^J \times \h{\Omega}^{1}\times\cdots \times \h{\Omega}^J
\end{align*}

Based on $\widetilde{\h{\theta}}_t$, the SD update process is given as follows:
\begin{align}
    \widetilde{\h{\theta}}_{t+1} & = \h{\kappa} + \h{A} \h{\Xi}_t \widetilde{\nabla}(\h{y}_t\vert \widetilde{\h{\theta}}_t) + \h{B} \widetilde{\h{\theta}}_t \label{equ: SD for theta}\\
    \widetilde{\nabla}(\h{y}_t\vert \mathcal{F}_{t-1}, \widetilde{\h{\theta}}_t) & = \left.\frac{\partial \ln p(\h{y}_t\vert \widetilde{\h{\theta}})}{\partial \widetilde{\h{\theta}}}\right\vert_{\widetilde{\h{\theta}} = \widetilde{\h{\theta}}_t} = \mathcal{J}(\widetilde{\h{\theta}}_t)' \nabla(\h{y}_t\vert {\h{\theta}}_t) \notag\\
    \h{\Xi}_t &= \operatorname{E}_{t-1}\left[\widetilde{\nabla}(\h{y}_t\vert \widetilde{\h{\theta}}_t)\widetilde{\nabla}(\h{y}_t\vert \widetilde{\h{\theta}}_t)'\right]^{-\delta}, \quad 
    \delta \in \left\{0, \frac{1}{2}, 1\right\} \notag
\end{align}

Here, $\mathcal{J}(\widetilde{\h{\theta}}_t)$ is the Jacobian matrix of the mapping function $\h{\Lambda}$, $\h{\kappa}$ is a $L$-dimensional vector of parameters to be estimated, with $L = (J-1 + \sum_{j=1}^J d_j)$, and $\h{A}$ and $\h{B}$ are $L\times L$ dimensional diagonal matrices of dimensions $L\times L$ of parameters to estimate. To ensure the stationarity conditions of the SD update process, all elements of $\h{A}$ must be greater than 0, and the absolute values of all elements in $\h{B}$ must be less than 1. Since the computation of $\h{\Xi}$ requires the Fisher information matrix of the model's conditional log-likelihood with respect to $\widetilde{\h{\theta}}_t$, which often does not have a closed-form solution. Following the approach of \cite{bernardi2019switching}, we consider a block-diagonal structure for $\h{\Xi}$, and decompose equation \ref{equ: SD for theta} in $J+1$ different SD update forms:
\begin{equation}
    \begin{split}
        \widetilde{\h{w}}_{t+1} & = \h{\kappa}^w + \h{A}^w \h{\Xi}_t^w \mathcal{J}(\widetilde{\h{w}}_t)' \nabla(\h{y}_t\vert {\h{w}}_t) + \h{B}^w \widetilde{\h{w}}_t,  \\
        \widetilde{\h{\theta}}_{j,t+1} & = \h{\kappa}^j + \h{A}^j \h{\Xi}_t^j \mathcal{J}(\widetilde{\h{\theta}}_{j,t})' \nabla(\h{y}_t\vert {\h{\theta}}_{j,t}) + \h{B}^j \widetilde{\h{\theta}}_{j,t}, \quad j = 1,\ldots, J,
    \end{split}
\end{equation}
where $\h{\kappa}^w \in \mathbb{R}^{J-1}, \ \h{A}^w, \h{B}^w \in \mathbb{R}^{(J-1)\times (J-1)}$, and $\h{\kappa}^j \in \mathbb{R}^{d_j}, \ \h{A}^j, \h{B}^j \in \mathbb{R}^{d_j \times d_j}, \ j = 1,\ldots, J.$ By adopting this block-diagonal structure, the estimation of the parameters in the model becomes simpler, and the computational burden is reduced. 

The DAMM method has three advantages: First, the mixed distribution can fit distributions of various shapes, allowing us not to assume any specific distribution for the return series, making the model robust to some extent against misspecifications in distribution. Second, the time-varying weights $\h w_{j,t}$ and parameters $\h \theta_t$ of the mixture components allow us to estimate the time-varying distribution at each point in time, leading to estimates of time-varying volatility, skewness, and kurtosis. Finally, the DAMM method utilizes a score-driven process setting to update time-varying parameters, providing the model’s likelihood with a closed-form solution. This allows us to use maximum likelihood estimation to estimate model parameters, resulting in efficient and stable estimation results.
 
According to \cite{catania2021dynamic}, we let $\h \Xi_t^w = \h I$ and
\begin{align*}
    \h \Xi_t^j = \left[\mathcal{J}^j\left(\tilde{\h \theta}_{j,t}\right)'\mathcal{I}_{p_j}\left(\h \theta_{j,t}\right)\mathcal{J}^j\left(\tilde{\h \theta}_{j,t}\right)\right]^{-1/2},
\end{align*}
then the model exhibits the most stable fitting performance.

Next, we discuss the setting of $\h \Lambda^w$ and $\h \Lambda^j$. For the mixture of normal specification in this paper, we simply let $\widetilde{\h \theta}_{j,t} = \h \theta_{j,t}$, and let
\begin{equation*}
    \h \Lambda^w \coloneqq \left\{
    \begin{array}{l}
         w_{j,t} =  \lambda_{[0, b_{j,t}]}(\tilde{w}_{j,t}),\quad j = 1,\ldots, J-1 \\
         w_{J,t} = 1 - \sum_{b=1}^{J-1}w_{b,t}
    \end{array}
    \right.
\end{equation*}
where $b_{j,t} = b_{j-1,t} - w_{j-1,t}$ and $b_{1,t} = 1$. $\lambda_{[U,L]}$ denotes the modified logistic function, satisfying
$
    \lambda_{[L,U]}(x) = L + \frac{(U-L)}{1+\exp(-x)}
$. Based on the settings of $ \h \Lambda^w(\cdot)$ and $\lambda_{[L,U]}(\cdot)$, the $(j,b)$-th element of the $J\times(J-1)$ Jacobian matrix $ \mathcal{\h J}^w(\cdot)$ takes the following form:
\begin{align*}
    \mathcal{\h J}^w(\tilde{\h w}_t)_{(j,b)} = 
    \left\{
    \begin{array}{ll}
         \frac{b_{j,t}\exp(-\tilde{w}_{j,t})}{\left(1+\exp(-\tilde{w}_{j,t})\right)^2}, & \mbox{if\ }b=j\\
         \frac{-\sum_{k=1}^{J-1}\mathcal{\h J}^w(\tilde{\h w}_t)_{(k,b)}}{1+\exp(-\tilde{w}_{j,t})}, & \mbox{if\ }b<j \ \mbox{and}\ j\neq J \\
         -\sum_{k=1}^{J-1}\mathcal{\h J}^w(\tilde{\h w}_t)_{(k,b)}, & \mbox{if\ } j = J \\
         0, & \mbox{if\ } b > j
    \end{array}
    \right.
\end{align*}

After implementing the maximum likelihood estimation (MLE) approach to obtain $\hat{w}_{j,t}$, $\hat{u}_{j,t}$ and $\hat{\sigma}_{j,t}$, we can calculate the time-varying conditional variance (Vol), skewness (Skew), and kurtosis (Kurt) by
\begin{align}
           \widehat{\sigma}_t  = \widehat{\operatorname{Vol}}_t =  &\sum_{j=1}^J w_{j,t}\left(\hat{\sigma}_{j,t}^2 + \hat{\mu}_{j,t}^2\right), \\
\widehat{\operatorname{Skew}}_t =& \frac{1}{\widehat{\sigma}_t^2} \sum_{j=1}^J w_{j,t}\hat{\mu}_j\left(3\hat{\sigma}_{j,t}^2 + \hat{\mu}_j^2\right), \\
\widehat{\operatorname{Kurt}}_t = &\frac{1}{\hat{\sigma}_t^4}\sum_{j=1}^J w_{j,t}\left(\hat{\mu}_j^4 +6\hat{\mu}_j \hat{\sigma}_{j,t}^2 +3\hat{\sigma}_{j,t}^4\right) - 3.
\end{align}

\subsection{TVP-VAR approach and time-varying connectedness measurements}

Let $\h y_t = (y_{1,t},\ldots, y_{N,t})'$, where $\h y_t$ is a $N\times 1$ vector of observations that represents the time-varying return, volatility, skewness, or kurtosis. We first construct a time-varying parameter vector autoregressive (TVP-VAR) model:
\begin{align}\label{equ: TVP-VAR(p)}
       \h y_t = \h c_t+\h B_{1t}\h y_{t-1}+\ldots+\h B_{pt}\h y_{t-p}+\h \varepsilon_t,\qquad
	\h \varepsilon_t\sim \h{\mathcal{N}}(\h 0,\h \Sigma_t),
\end{align}
where $\h c_t$ denotes a $N\times 1$ vector of intercept, $\h B_{k,t}, k =1,\ldots, p$ denotes $N\times N$ matrix of time-varying autoregressive coefficients, $\h \varepsilon_t$ are $N\times 1$ vector of random disturbances that follows multivariate Gaussian distribution with mean $\h 0$ and time-varying covariance matrix $\h \Sigma_t$.  Given the model in \eqref{equ: TVP-VAR(p)}, by some linear transformations and rearrangements, we can obtain the following contemporaneous-form state space representation:
\begin{align*}
	\h y_t &= \h Z_t\h \beta_t + \h\varepsilon_t,\qquad \h\varepsilon_t\sim N(0,\h \Sigma_t),\\
	\h \beta_t &= \h \beta_{t-1} + \h v_t,\qquad \h v_t\sim N(0,\h Q_t),
\end{align*}
where $\h \beta = \mathrm{vec}([\h c_t, \h B_{1t},\ldots, \h B_{pt}]')$ and $\h Z_t = \hI_N\otimes [\h 1', \hy_{t-1}',\ldots, \hy_{t-p}']'$. Following the specification of \cite{koop2013large}, we assume that $\h \beta$ follows a random walk process, which can not only capture smooth structural changes, but also identify sudden jumps. Random errors $\h v_t$, also known as ``state equation errors'', are assumed to follow a multivariate Gaussian distribution with mean $\h 0$ and time-varying covariance matrix $\h Q_t$. We follow \cite{koop2013large} to estimate the TVP-VAR model by introducing a forgetting factor into the Kalman filter. Detailed estimation procedures are discussed in the appendix \ref{apdx: TVP-VAR formula}. We follow the recommendation of \cite{akyildirim2022connectedness} and \cite{zheng2023fast} to set $\lambda = \kappa = 0.99$, and use BIC to determine the lag order of the TVP-VAR model. 

Given the estimates of $\h c_t$ and $\h B_{lt} $ for $l = 1,\ldots, p$, we follow \cite{koop1996impulse} and \cite{diebold2014network}, and calculate $j$th variable's attribute to $i$th's $H$ step ahead generalized forecast error variance by:
\begin{align*}
	\theta_{ij,t}(H) = \frac{\sigma_{jj,t}^{-1}\sum_{h=0}^{H-1}(e_{i}^{\prime}\h\Psi_{h, t}\h \Sigma_t e_{j})^2}{\sum_{h=0}^{H-1}(e_{i}^{\prime}\h\Psi_{h,t}\h \Sigma_t\h\Psi_{h,t}^{\prime}e_{i})},
\end{align*}
where $\sigma_{jj,t}$ is the $j$th diagonal element of $\h \Sigma_t$. We set $H=12$, which provides a sufficiently long forecast horizon. Notice that the sum of $\theta_{ij,t}^{g}(H)$ given $i$ may not be equal to one; we normalize it and obtain $d_{ij,t}(H) = {\theta_{ij,t}(H)}/{\sum_{k=1}^{N}\theta_{ik,t}(H)}$. The $d_{ij,t}(H)$ represent the time-varying \textit{pairwise directional connectedness} from variable $j$ to $i$. Based on $d_{ij,t}(H)$, one can calculate the time-varying \textit{net} pairwise directional connectedness ($d_{ij,t}^{Net}(H)$) and the time-varying \textit{total} pairwise directional connectedness ($d_{ij,t}^{Total}$) by:
\begin{align} \label{equ: total pairwise connectedness}
    d_{ij,t}^{Net}(H) = d_{ij,t}(H) - d_{ji,t}(H), \quad d_{ij,t}^{Total}(H) = d_{ij,t}(H) + d_{ij,t}(H).
\end{align}

Moreover, we can sum up individual connectedness and obtain some aggregated connectedness measurements. The time-varying \textit{total connectedness index} ($\mathcal{C}_t(H)$), which measure the degree of connectedness of the whole system, can be calculated by: 
\begin{equation}
	\mathcal{C}_{t}(H) = \frac{1}{N}\sum_{\substack{i,j=1, i\neq j}}^{N}d_{ij,t}(H)\times 100.
\end{equation}

The time-varying \textit{To others} connectedness index ($\mathcal{C}_{\bullet j,t}(H)$) that captures the directional connectedness from variable $j$ to other variables, and the time-varying \textit{From others} connectedness index ($\mathcal{C}_{i\bullet,t}(H)$) that captures the directional connectedness from other variables to variable $i$, can be calculated by
\begin{align}
    \mathcal{C}_{\bullet j,t}(H) = \frac{1}{N}\sum_{\substack{i=1}}^{N}d_{ij,t}(H)\times 100, \qquad  \mathcal{C}_{i\bullet,t}(H) =  \frac{1}{N}\sum_{\substack{j=1}}^{N}d_{ij,t}(H)\times 100
\end{align}

Similarly, the aggregated time-varying \textit{Net} connectedness index ($\mathcal{C}_{i,t}^{Net}$) and the time-varying \textit{Total} connectedness index ($\mathcal{C}_{i,t}^{Total}$) of the variable $i$ can be calculated by:
\begin{align} \label{equ: aggregate connectedness}
    \mathcal{C}_{i,t}^{Net}(H) = \mathcal{C}_{\bullet i,t}(H) - \mathcal{C}_{i \bullet,t}(H), \qquad \mathcal{C}_{i,t}^{Total}(H) = \mathcal{C}_{\bullet i,t}(H) + \mathcal{C}_{i \bullet,t}(H)
\end{align}

\subsection{Constructing multi-moment connectedness network}

As we have mentioned before. Different to the existing literature, which analyzes individual connectedness networks of return (first moment), volatility (second moment), skewness (third moment) and kurtosis (forth moment), respectively. This section proposes a multi-moment connectedness network modeling approach, which simultaneously considers the connectedness networks of returns, volatility, skewness, and kurtosis. This allows us to examine the connectedness of the green financial market from a global to local perspective.

In this paper, we consider four different layers of the connectedness network, including the return layer, volatility layer, skewness layer, and kurtosis layer. Specifically, the multi-moment connectedness network encompasses both within-layer and cross-layer relationships. The within-layer relationships are captured by the nodes and edges of each network layer, according to the seminar work of \cite{diebold2014network}, we specify them as follows:

(1) \textit{ Node name and node color}.

We name the nodes in each layer based on the individual variable.  For example, the node ``EGCI'' in different layers denotes different moments of the S\&P Global 1200 ESG Index (ESGI). The color of the node denotes the category to which the variable belongs. 
The green nodes represent the variables belonging to the green equity market, while the red nodes indicate the variables associated with the green bond market.

(2) \textit{ Edge and edge thickness}.

We consider the total pairwise connectedness index ($d_{ij,l}^{Total}$) as the edges between the variables in each layer $l$, $l =1,\ldots, L$. Note that here we ignore the subscript $t$ because in the empirical study we examine not only the networks at each time point, but also the average network across all time points. As we have shown in \eqref{equ: total pairwise connectedness}, $d_{ij,l}^{Total}$ measures the total level of uncertainty transmitted between variables $i$ and $j$ in each layer. The thickness of the edge represents the degree of $d_{ij, l}^{Total}$, where a thicker edge indicates a greater $d_{ij, l}^{Total}$ between the variables $i$ and $j$.

For cross-layer relationships, we project the four layers into one artificially synthesized layer, called the projection layer. The projection layer network effectively assimilates the information present in the return, volatility, skewness, and kurtosis connectedness networks. This enables a comprehensive measurement of interconnectedness within the green finance market, taking into account the relationships between various moments of green bonds and green equities.

(3) \textit{ Cross-layer projection}

To extract information from each layer of the network and project it onto a new layer, it is necessary to determine the weights of each layer. For each layer $L_l, l =1,\ldots, L$, we first calculate its network density $d_l = \sum_{i}d_{ij, l}^{Total}$, which reflects the degree of connectivity between nodes in the network and serves as a measure of the information density. Based on $d_l$, we can calculate the weights of each layer by $w_l = d_l/\sum_l d_l$. The edges in the projection layer can then be calculated by $d_{ij, P}^{Total} = \sum_l w_ld_{ij, l}^{Total} $.

Note that based on the weight $d_l$ of each layer, we can also calculate a projected connectedness measurement according to 
\eqref{equ: total pairwise connectedness}-\eqref{equ: aggregate connectedness}. This approach allows us to derive indices for measuring the interconnectedness of financial markets at the multi-moment level.

(3) \textit{Visualization and layout.}

We use the networkx package in Python 3.8.8 to visualize the multi-moment network. For the layout of the network, we have adopted a spherical network layout and positioned variables of different categories on two sides of the network. Such an arrangement allows for a clear observation of the strength of connectedness both within the same variable category and between different variable categories.

 
\subsection{Evaluation of the monetary policy shock} \label{subsec: monetary shock}

\begin{figure}[!t]
    \centering
    \begin{subfigure}{1\textwidth}
        \includegraphics[width=\linewidth]{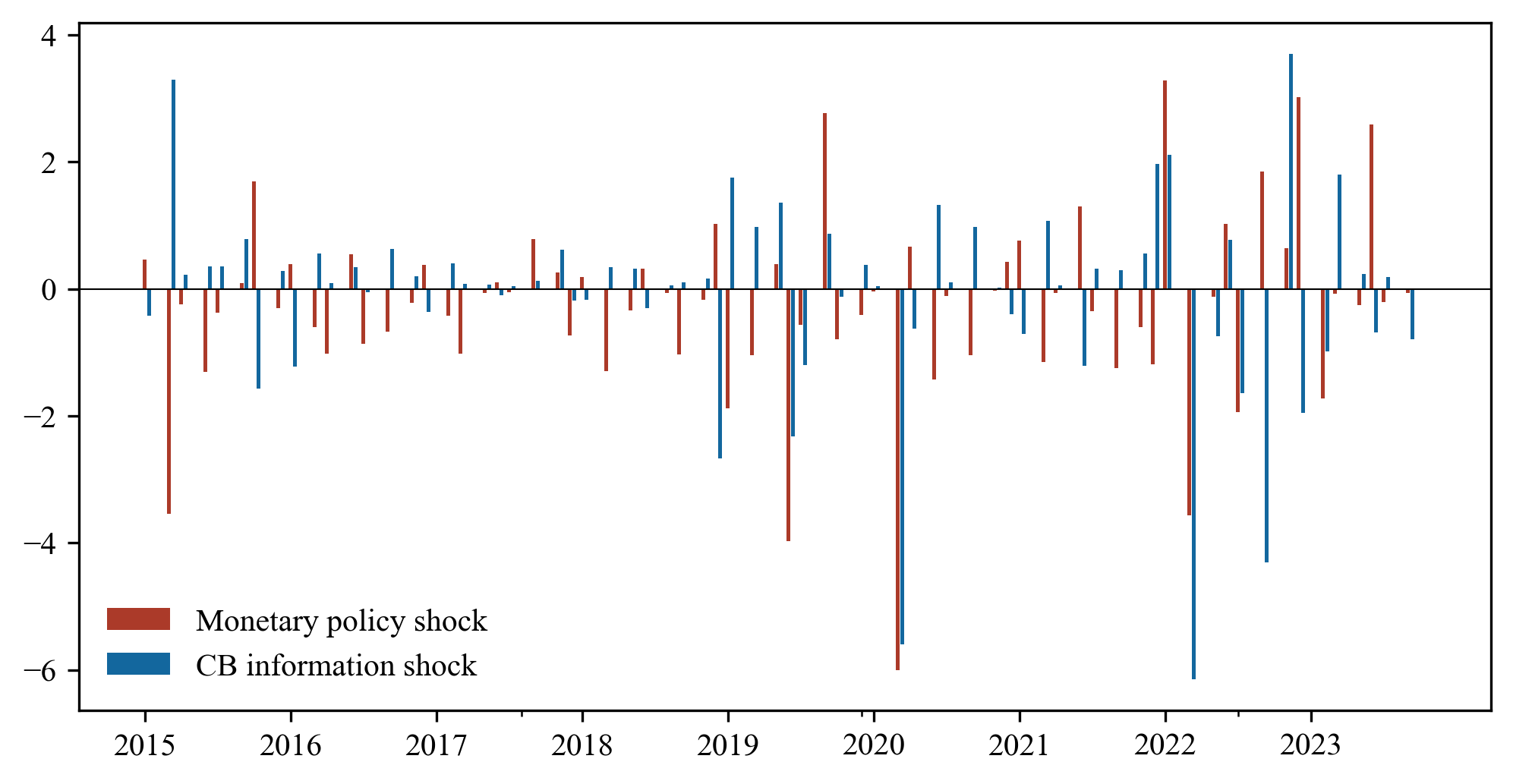}
    \end{subfigure}
    \caption{U.S. monetary policy and central bank information shocks}
    \label{figure: mpshock}
\end{figure}

We use the aforementioned two series, the 30 minute change in three-month fed fund future prices and the S\&P 500 index around the FOMC meeting to evaluate the Fed's monetary policy. The basic idea is to extract monetary shocks from the monetary surprise based on a Bayesian vector autoregressive model. Following \cite{jarocinski2020deconstructing} and \cite{zhou2022does}, the model takes the form:
\begin{equation}
\left(\begin{array}{c}
\h m_t \\
\h y_t
\end{array}\right)=\sum_{i=1}^p\left(\begin{array}{cc}
0 & 0 \\
\h B_1^i & \h B_2^i
\end{array}\right)\left(\begin{array}{c}
\h m_{t-i} \\
\h y_{t-i}
\end{array}\right)+\left(\begin{array}{l}
0 \\
C
\end{array}\right)+\left(\begin{array}{c}
u_t^{\h m} \\
u_t^{\h y}
\end{array}\right),\quad \left(\begin{array}{c}
u_t^{\h m} \\
u_t^{\h y}
\end{array}\right) \sim \h{\mathcal{N}}(0, \h \Sigma),
\end{equation}
where $\h m_t$ contains the monthly aggregated changes on the fed fund future rate and S\&P 500 index over the 30 minutes before and after each FOMC meeting. $\h y_t$ includes 1-year government bond yield, S\&P 500 index, CPI, excess bond premium, and industrial production index. 

The monetary shock and the information shock are identified by sign restrictions. The basic belief is that the first shock (monetary policy shock) and the second shock (information shock) generate a different direction of comovement. The sign restriction is achieved by rotating the first $2\times 2$ block matrix $\h C$, where $\h C$ denotes the Cholesky decomposition of $\h \Sigma$. Then, given the scaled structural shocks, the changes on the fed fund future rate $S_t$ can be decomposed by
\begin{align*}
    S_t = S_t^{MP} + S_t^{IF},
\end{align*}
where $S_t^{MP}$ denotes the monetary policy shock. When $S_t^{MP}$ is positive,it signifies a tightening of monetary policy, while a negative $S_t^{MP}$ indicates an easing of monetary policy. Figure \ref{figure: mpshock} presents the estimated results. It can be seen that during March 2020, when the Fed restarted quantitative easing, there was a significant negative monetary policy shock, which corresponds to notable changes in connectedness within the green financial market (as shown in Figure \ref{figure: tspl_proj}(a)). These findings potentially verify our
assumption that U.S. monetary policy can increase global green financial market connectedness.



\section{Empirical results}
\label{sec: empirical results}
In this section we provide a static and dynamic analysis of the multi-moment connectedness network in the green finance market, and then examine the impact of U.S. monetary policy shocks on the multi-moment connectedness network.

\subsection{Static sample analysis}



\begin{figure}[!p]
    \centering
    \begin{subfigure}{0.85\textwidth}
        \includegraphics[width=\linewidth]{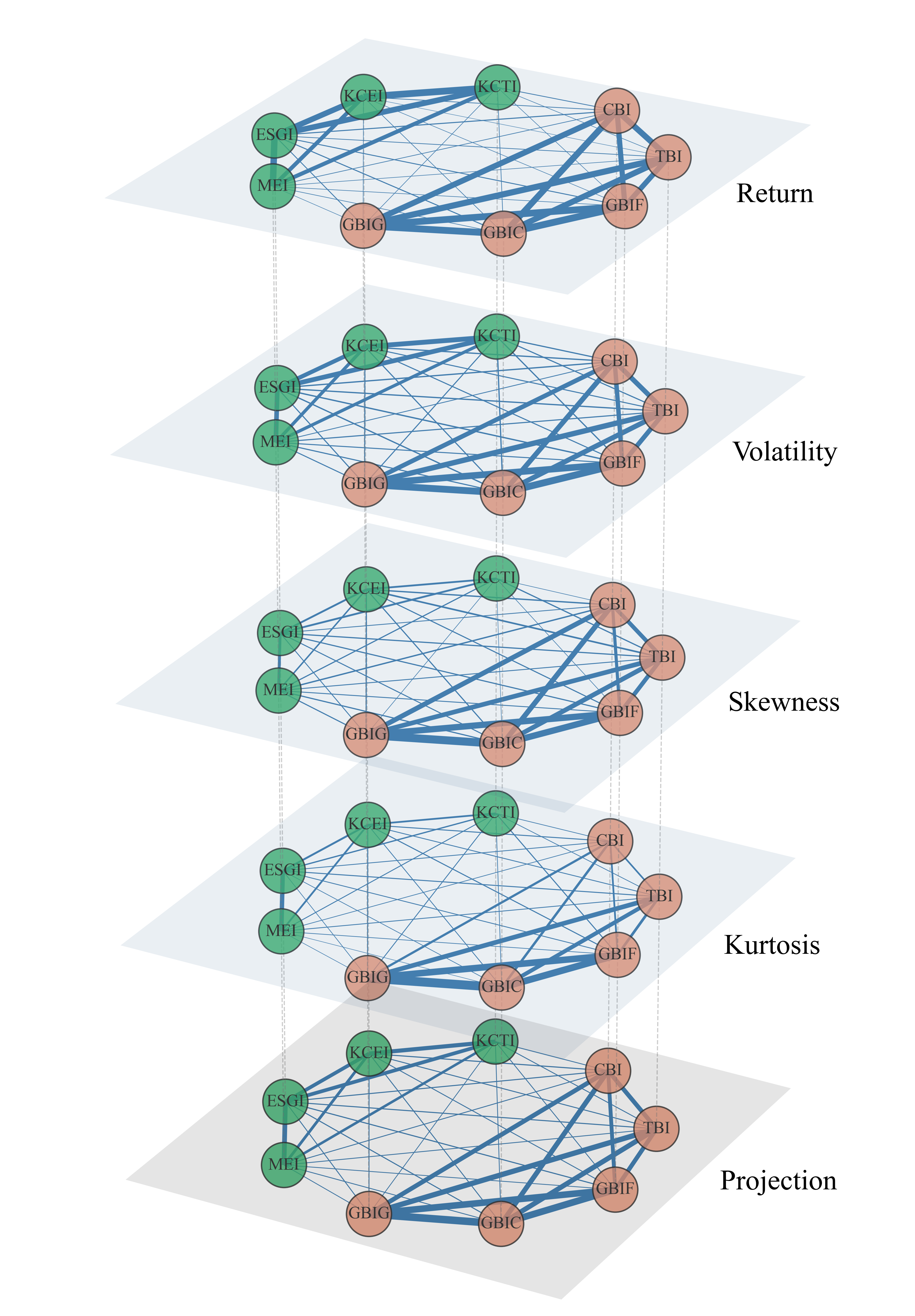}
    \end{subfigure}
    \caption{Static multi-moment connectedness network}
    \raggedright{\justifying{\footnotesize{\textit{Notes}:  The top four layers are connectedness networks of the first four moments. The bottom layer is the projection connectedness network. In each layer, red nodes represent bond market indices and blue nodes represent stock market indices. Edges denote the pairwise total connectedness between the indices, with thicker edges indicating stronger connectedness.}}}
    \label{figure: mul_nw}
\end{figure}

To provide a comprehensive evaluation of multi-moment connectedness within the green financial market, Figure \ref{figure: mul_nw} illustrates the aggregated static multi-layer connectedness network, averaged across the entire sample. This analysis specifically encompasses the first through the fourth moments, aligning respectively with the layers of return, volatility, skewness, and kurtosis. Additionally, the density of each network layer is quantitatively assessed and assigned as weights. This methodology enables the systematic projection of each moment's information onto the foundational projection layer.

Overall, across various layers of the connectedness network, including the projection layer, we observe a stronger connectedness within the market, covering both the green bond and green equity sectors. Figure \ref{figure: mul_nw} illustrates this, with nodes of similar colors being linked by thicker lines, indicating stronger connectedness. Upon examining the heterogeneity between the layers, it is clear that the connectedness within each layer becomes less dense as we move from the return layer to the kurtosis layer. In the green equity market (green nodes), the level of connectedness decreases with higher moments, showing a significant connection only between ESGI and MEI in the skewness and kurtosis layers. Similarly, the green bond market (red nodes) displays a decreasing trend in connectedness as the moment increases. However, it still maintains a relatively strong level of connectedness overall, particularly among GBIG, GBIC, and GBIF.

\begin{figure}[!t]
    \centering
    \begin{subfigure}{0.49\textwidth}
        \includegraphics[width=\linewidth]{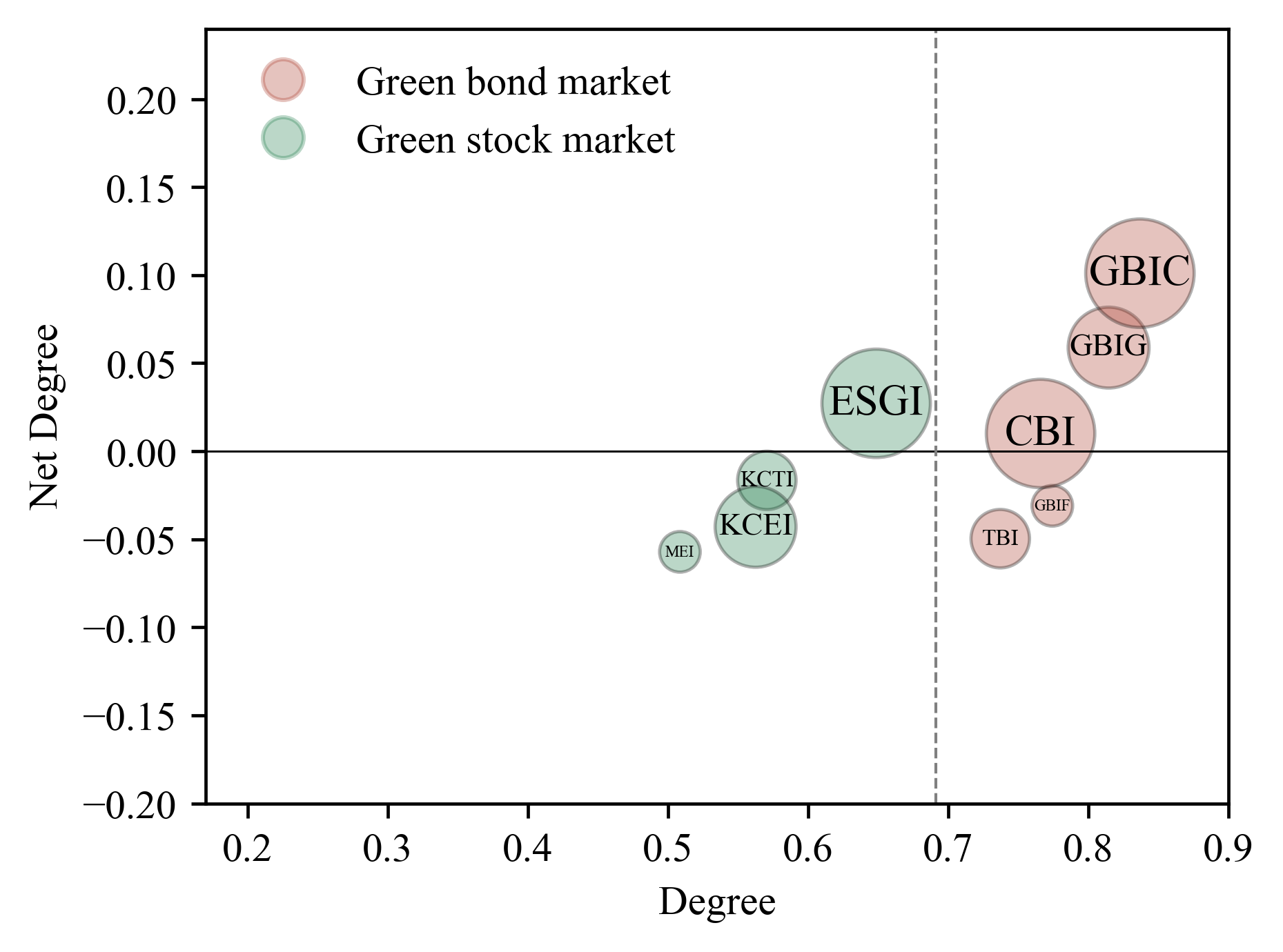}
        \caption{\footnotesize{Return layer}}
    \end{subfigure}
    \vspace{0.25cm}
    \begin{subfigure}{0.49\textwidth}
        \includegraphics[width=\linewidth]{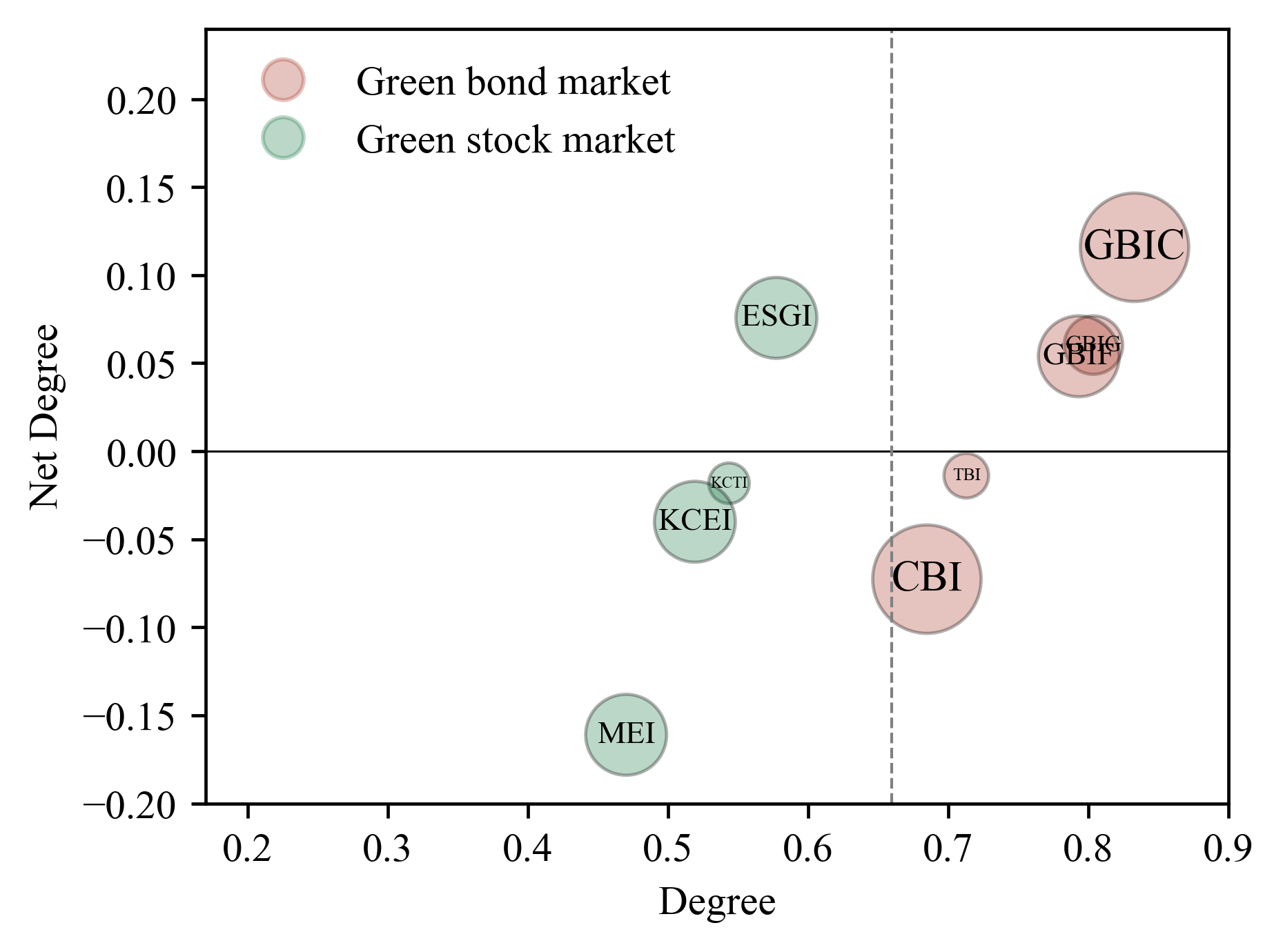}
        \caption{\footnotesize{Volatility layer}}
    \end{subfigure}
    
    \begin{subfigure}{0.49\textwidth}
        \includegraphics[width=\linewidth]{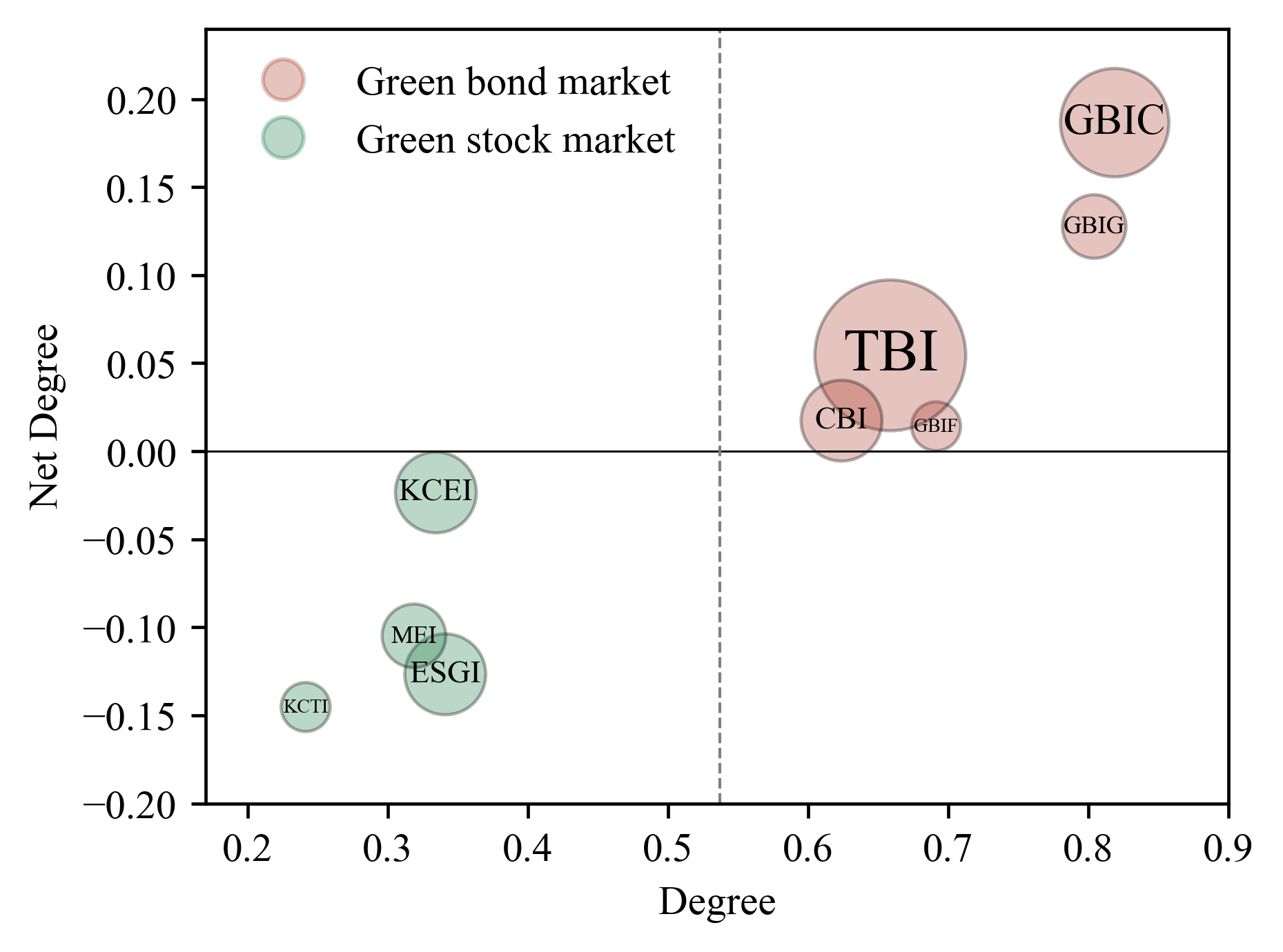}
        \caption{\footnotesize{Skewness layer}}
    \end{subfigure}
    \begin{subfigure}{0.49\textwidth}
        \includegraphics[width=\linewidth]{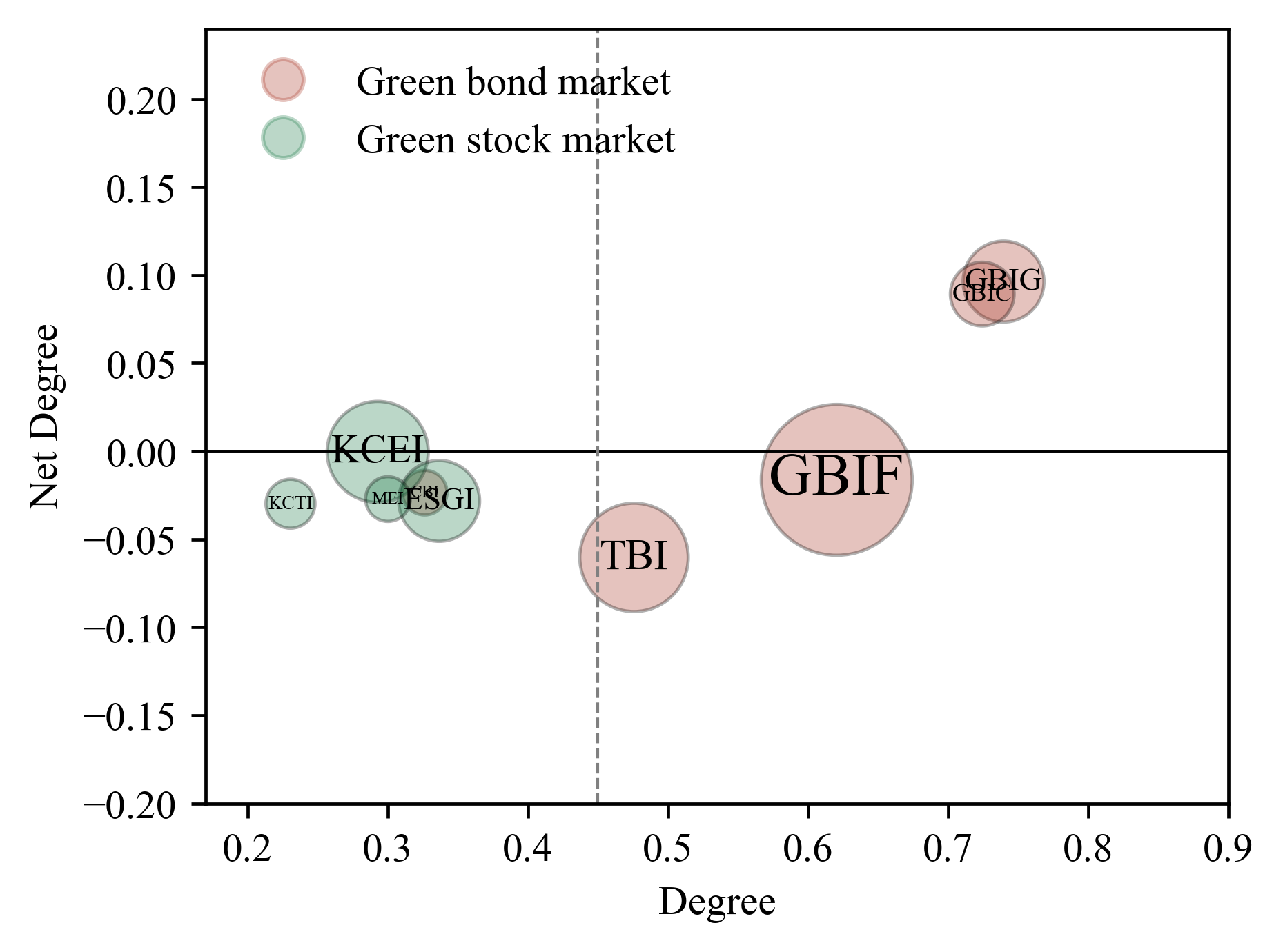}
        \caption{\footnotesize{Kurtosis layer}}
    \end{subfigure}
    \caption{Static multi-layer degree and centrality}
    \raggedright{\justifying{\footnotesize{\textit{Notes}:  In each layer, the vertical dashed line represents the average weighted degree of nodes. Node size represents bridge centrality. Larger nodes indicate higher systemic importance within the layer.}}}
    \label{figure: mul_cent}
\end{figure}

Figure \ref{figure: mul_cent} illustrates the relative importance of the nodes in each layer. We examine two measures of node importance within the multi-moment network: the weighted degree (Degree)  corresponds to the aggregated total connectedness ($\mathcal{C}_{i,t}^{Total}(H)$) in \eqref{equ: aggregate connectedness}, and the net weighted degree (Net Degree) corresponds to the aggregated net connectedness ($\mathcal{C}_{i,t}^{Net}(H)$) in \eqref{equ: aggregate connectedness}. The horizontal axis in Figure \ref{figure: mul_cent} represents the weighted degree, the vertical axis represents net weighted degree, and the vertical dashed line represents the average weighted degree of the layer. In addition, the node size signifies bridge centrality \citep{valente2010bridging, wu2022complex}, which is used to calculate the change in network cohesion caused by deleting edges to a given node. 

From the within-layer perspective, as shown in Figure \ref{figure: mul_cent}, the nodes related to green bonds (nodes in red) exhibit relatively high ``Degree'', mostly positioned to the right of the average (dashed line). Moreover, their overall ``Net Degree'' is positive, placing them in the role of risk contributors in each layer. The green equity nodes (nodes in green), in contrast, display lower ``Degree'' and negative  ``Net Degree'', acting as risk receivers. Additionally, despite the lower ``Degree'' of the green equity nodes, the relatively large node sizes of ESGI and KCEI indicate their high bridging centrality and a relatively significant systemic importance in the network, potentially serving as the linkages between the green equity and green bond markets. This association can be further validated in the subsequent pairwise connectedness table (Table \ref{table: static pro}).

From the cross-layer perspective, we find that from the return layer in Figure \ref{figure: mul_cent}(a) to the kurtosis layer in Figure \ref{figure: mul_cent}(d), as the moment order increases, the dashed line (the average ``Degree'') gradually shifts to the left. This suggests that the overall connectedness within layer gradually decreases, which is consistent with the findings in Figure \ref{figure: mul_nw}. Moreover, the distribution of nodes in ``Degree'' (horizontal axis) and bridge centrality (node size) also tends to be dispersed, indicating that risk tends to cluster towards a few nodes in high-order moment networks. For example, the degree range of the return layer in Figure \ref{figure: mul_cent}(a) (0.5-0.9) is narrower than that of the kurtosis layer in Figure \ref{figure: mul_cent}(d) (0.2-0.8), and the GBIC with the largest node size in Figure \ref{figure: mul_cent}(a) is smaller than the GBIF in Figure \ref{figure: mul_cent}(d).

Table \ref{table: static pro} shows the pairwise-directional connectedness table of the projection layer\footnote{The detailed connectedness table of the return, volatility, skewness, and kurtosis layers can be found in the Appendix Table \ref{table: rn} - \ref{table: kurt}.}. 
In the pairwise connectedness table, the sum of columns represents the connectedness from the specific node to others ($\mathcal{C}_{\bullet j}(H)$). The sum of rows indicates the received connectedness from other nodes ($\mathcal{C}_{i \bullet}(H)$). Additionally, the difference between ``From others" and ``To others" is reflected as ``Net" ($\mathcal{C}_{i}^{Net}(H)$) in the last row, corresponding to the Net Degree in Figure \ref{figure: mul_cent}. 

\begin{table}[!t]
\caption{Static pairwise connectedness table of the projection layer}
\label{table: static pro}
\centering
\footnotesize
\renewcommand\arraystretch{1.4}
\resizebox{\linewidth}{!}{
\begin{tabular}{l|ccccc|cccc|c}
\hline
                   & GBIG           & GBIC           & GBIF           & TBI            & CBI            & KCTI           & KCEI           & ESGI           & MEI            & \textbf{From others} \\ \hline
GBIG               & 24.66          & 22.17          & 18.46          & 14.29          & 12.42          & 1.97           & \textbf{2.32}  & 1.93           & 1.79           & \textbf{75.34}    \\
GBIC               & 21.69          & 25.14          & 18.79          & 13.07          & 13.20          & 1.90           & \textbf{2.30}  & 2.17           & 1.73           & \textbf{74.86}    \\
GBIF               & 19.80          & 20.46          & 27.22          & 12.37          & 10.98          & 2.23           & 2.52           & \textbf{2.61}  & 1.81           & \textbf{72.78}    \\
TBI                & 16.60          & 16.07          & 12.94          & 32.92          & 12.42          & 2.30           & 2.44           & \textbf{2.58}  & 1.73           & \textbf{67.08}    \\
CBI                & 14.03          & 15.67          & 11.37          & 13.01          & 36.52          & 2.08           & \textbf{2.64}  & 2.50           & 2.17           & \textbf{63.48}    \\ \hline
KCTI               & 2.47           & 2.74           & 2.65           & 2.90           & 2.75           & 55.36          & 12.17          & 11.73          & 7.24           & \textbf{44.64}    \\
KCEI               & 2.88           & 3.03           & 2.70           & 3.11           & \textbf{3.60}  & 12.02          & 53.93          & 11.19          & 7.56           & \textbf{46.07}    \\
ESGI               & \textbf{3.30}  & \textbf{3.78}  & \textbf{3.70}  & \textbf{3.40}  & 3.37           & 9.72           & 9.99           & 49.97          & 12.78          & \textbf{50.03}    \\
MEI                & 2.80           & 3.21           & 2.80           & 3.16           & 2.94           & 7.52           & 8.75           & 14.81          & 54.02          & \textbf{45.98}    \\ \hline
\textbf{To others}  & \textbf{83.57} & \textbf{87.12} & \textbf{73.40} & \textbf{65.31} & \textbf{61.67} & \textbf{39.73} & \textbf{43.12} & \textbf{49.52} & \textbf{36.81} & \textbf{60.03}    \\ \hline
\textbf{Net} & \textbf{8.23}  & \textbf{12.26} & \textbf{0.62}  & \textbf{-1.77} & \textbf{-1.81} & \textbf{-4.91} & \textbf{-2.94} & \textbf{-0.51} & \textbf{-9.17} &       \textbf{--}           \\ \hline
\end{tabular}}
\end{table}

We summarize several important findings as follows: First, the total connectedness of the projection layer reaches 60.03\%, indicating the presence of interconnectedness within the multi-moment network of the green financial market. 
Second, the green bonds (GBIC, GBIG and GBIF) exhibit stronger outflows (``To others'') and act as net contributors (``Net''), contrasting with the green equities (KCTI, KCEI and ESGI), which act as receivers (``From others''). This aligns with the findings of Figure \ref{figure: mul_nw} and Figure \ref{figure: mul_cent}. 
Third, focusing on the bond market, as shown in the upper left block of Table \ref{table: static pro}, the green bonds (GBIC, GBIG, and GBIF) remain contributors, while the benchmark bonds (TBI and CBI) exhibit relatively weaker spillover effects. Among them, TBI demonstrates a stronger spillover effect on GBIG, both of which consist of government bonds; while CBI exhibits a stronger spillover effect on GBIC, both belonging to corporate bonds.
Similarly, for the within-market connectedness, as shown in the lower right block of Table \ref{table: static pro}, the benchmark index MEI acts as a receiver, with ESGI showing the strongest spillover effect (from ESGI to MEI, 14.81\%). Meanwhile, there is a strong connectedness between KCTI and KCEI.
Finally, for the cross-market connectedness, ESGI and KCEI show the strongest connection with the bond market, as indicated by the bold elements in the lower left and upper right blocks of Table \ref{table: static pro}. Within the bond market, the benchmark indices TBI and CBI demonstrate relatively stronger connections with the equity market\footnote{The detailed local connectedness table of the projection layer can be found in the Appendix Table \ref{table: local1} - \ref{table: local4}.}.

\subsection{Dynamic sample analysis}
\label{sec: connectedness}
\begin{figure}[!p]
    \centering
    \begin{subfigure}{1.0\textwidth}
        \includegraphics[width=\linewidth]{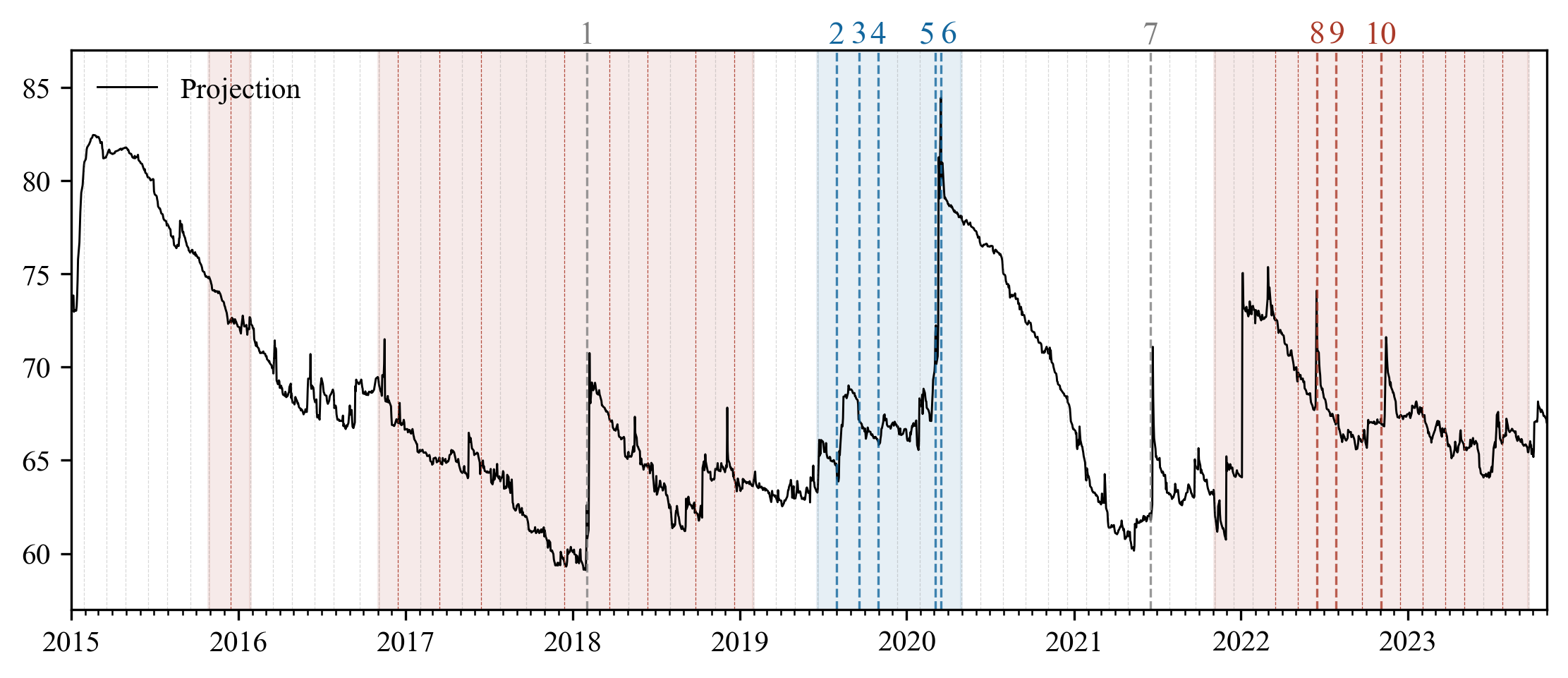}
        \caption{\footnotesize{Projection layer}}
    \end{subfigure}
    
    \begin{subfigure}{1\textwidth}
        \includegraphics[width=\linewidth]{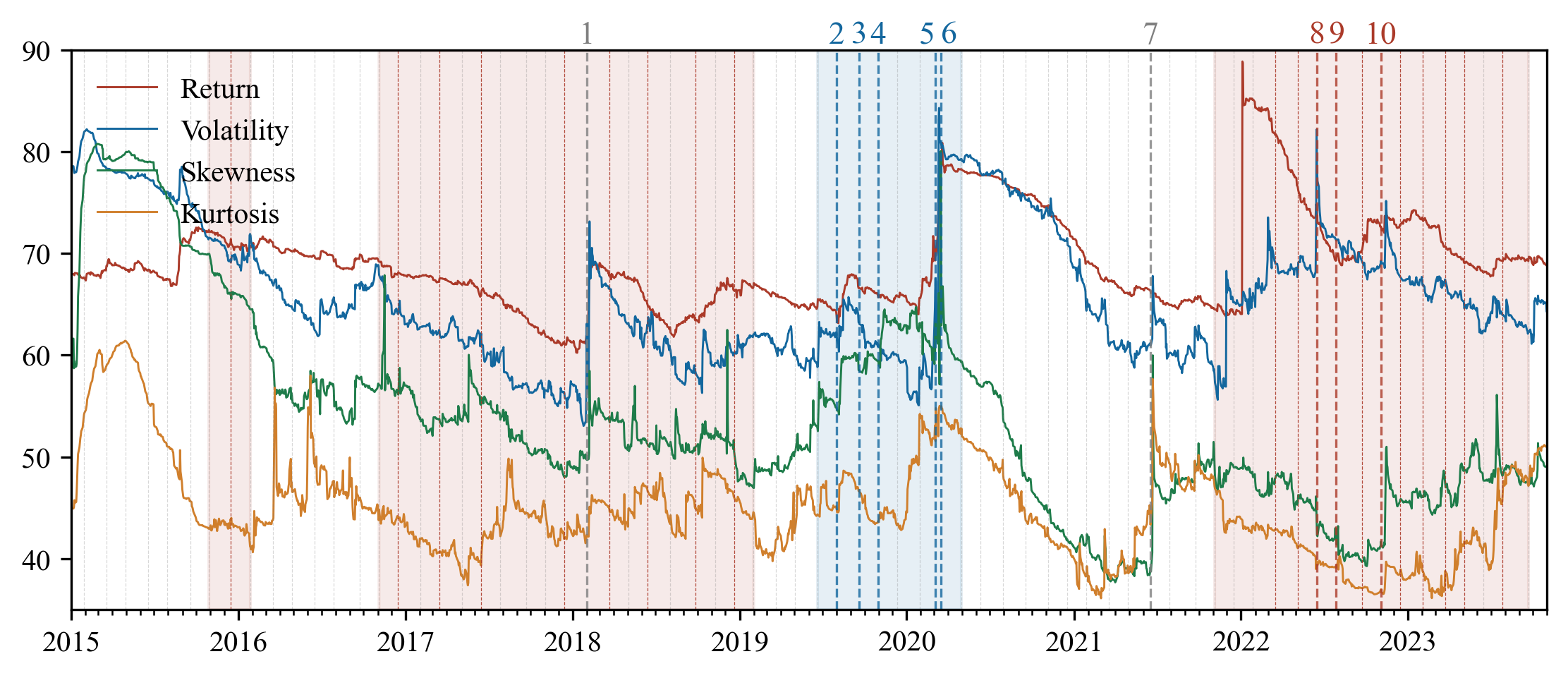}
        \caption{\footnotesize{Multi-moment layers}}
    \end{subfigure}
    \caption{Time-varying total connectedness of the projection layer}
    \raggedright{\justifying{\footnotesize{\textit{Notes}: (1) The vertical dashed lines denote the dates of each Federal Open Market Committee’s (FOMC) announcements of interest rate decisions, with red indicating rate hikes, blue indicating rate cuts, and grey indicating that the federal funds rate remains unchanged. (2) Among them, we highlight with bold dashed lines the dates when the total spillover index shows significant changes, namely Event 1: 2018.02.01; Event 2: 2019.08.01; Event 3: 2019.09.19; Event 4: 2019.10.31; Event 5: 2020.03.03; Event 6: 2020.03.15; Event 7: 2021.06.17; Event 8: 2022.06.16; Event 9: 2022.07.28; Event 10: 2022.11.03.}}}
    \label{figure: tspl_proj}
\end{figure}

Now we move to the dynamic feature of the multi-moment green finance market connectedness, with a specific focus on its relationship with the U.S. monetary policy. Figure \ref{figure: tspl_proj}(a) illustrates the time-varying total connectedness index ($\mathcal{C}_{t}(H)$) of the projection layer, annotated with all dates of FOMC interest rate announcements during the sample period from January 1st, 2015 to November 1st, 2023. The red part corresponds to an increase in the federal funds rate, the blue part corresponds to a cut, and the gray part represents the unchanged federal funds rate.

It can be observed from Figure \ref{figure: tspl_proj}(a) that during some of the FOMC announcements, such as Event 1 to Event 10 as labeled in the figure, the total connectedness index changes significantly. 
For example, following the outbreak of the COVID-19 pandemic, the FOMC held two unscheduled meetings on March 3rd and March 15th, 2020 (Event 5 and Event 6), and lowered the target range for the federal funds rate to near-zero levels. At the same time, the total connectedness index surged, reaching its peak within the sample period at 84.44\% on March 16th. 
Since March 2022, the Fed commenced a new round of interest rate hikes. Notably, between June 16th and November 3rd of the same year, the FOMC consecutively rtaised rates four imes, each time by 75 basis points. Consequently, the total connectedness index exhibited a considerable increase following the announcements of these two rate decisions (Event 8 and Event 10).
Furthermore, some FOMC meetings have decided not to alter interest rate targets, but there is still a considerable increase in the total connectedness index, as observed in Event 1 and Event 7. This phenomenon might be associated with market expectations regarding the Fed's monetary policy. Unchanged interest rates could also act as a shock, impacting the interaction in green finance markets.

Moreover, Figure \ref{figure: tspl_proj}(b) shows the evolution in the total connectedness index in the returns, volatility, skewness, and kurtosis layers, respectively. 
On the one hand, there is a similarity in the trends among the total connectedness across different layers. For instance, during Event 1, Event 2, Event 5, and Event 6, there is a noticeable increase in connectedness across all layers. On the other hand, there are also some heterogeneous features. Overall, from the return layer to the kurtosis layer, the total connectedness gradually decreases, consistent with the conclusions drawn from static analysis. However, there are reversals observed after certain FOMC announcements. For example, during Event 1, Event 6, Event 7, Event 9, and Event 10, the volatility layer surpasses the return layer; during Event 4, the skewness layer surpasses the volatility layer; and during Event 7, the kurtosis layer surpasses the skewness layer. This suggests that although higher moments have lower connectedness, their sensitivity might be higher under event-driven shocks, providing additional information for risk identification.

\begin{figure}[!t]
    \centering
    \begin{subfigure}{1\textwidth}
        \includegraphics[width=\linewidth]{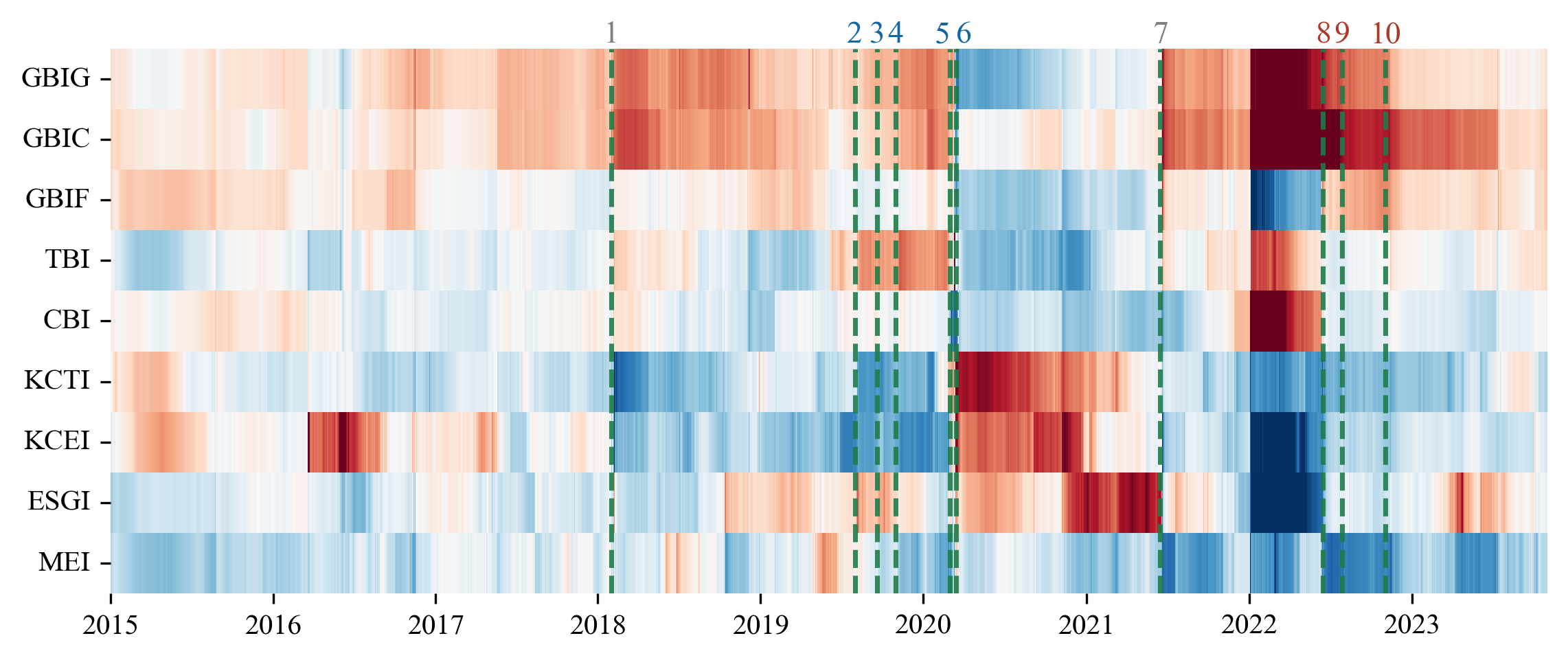}
    \end{subfigure}
    \caption{Time-varying net connectedness of the projection layer}
    \raggedright{\justifying{\footnotesize{\textit{Notes}: (1) Red represents net connectedness outflow, while blue represents net connectedness inflow, the darker the color the stronger the net connectedness. (2) The green vertical dashed lines mark the FOMC announcements date corresponding to significant changes in Figure \ref{figure: tspl_proj}.
    }}}
    \label{figure: netspl_proj}
\end{figure}

Given the importance of event-driven shocks, we turn to nodes within the projection layer, and identify the transmission roles played by each node in response to event shocks. Figure \ref{figure: netspl_proj} illustrates the time-varying net connectedness index for each node ($\mathcal{C}^{Total}_{i,t}(H)$) in the projection layer, where the red color indicates a positive net connectedness, while blue represents a negative net connectedness, with darker colors signifying larger absolute values of net connectedness. Overall, the green bond market is the contributor or transmitter of connectedness (GBIG, GBIC, GBIF appear in red), while the green equity is the receiver (KCTI, KCEI, ESGI appear in blue). Especially in early 2018 (Event 1) and since mid-2021 (Event 7, Event 8, Event 9, and Event 10), the net connectedness of the green bond has increased significantly, as shown in Figure \ref{figure: netspl_proj} by the deepening red color of GBIG, GBIC, and GBIF. 
However, during the quantitative easing policy (Event 5 and Event 6) under the COVID-19 pandemic, there is a shift in the transmission roles. KCTI, KCEI, and ESGI shift from blue to red, implying that the green equity market becomes a net contributor of connectedness.

\subsection{Effects of the monetary policies}
The preceding analysis explored the time-varying characteristics of multi-moment connectedness in the green finance market across different FOMC meetings. In this section, leveraging the extracted exogenous U.S. monetary policy shocks by the method described in Section \ref{subsec: monetary shock}, we apply local projection techniques proposed by \cite{jorda2005estimation} to further examine the significance and duration of the impact of monetary policy shocks on multi-moment connectedness of the green financial market. At horizon $h = 0,1,2,\cdots,18$, the impulse response of the connectedness index $\mathcal{C}_{t}$\footnote{Here we ignore the forecast horizon $H$ of the generalized variance decomposition for notation simplicity and to avoid confusion with local projection's prediction periods.} to the monetary policy shock $S_t^{MP}$ is reflected in coefficients $\beta^h$ in the following regression:
\begin{align} \label{equ: regression}
    \mathcal{C}_{t+h} = \alpha_0^{h} + \alpha_1^{h}\mathcal{C}_{t-1} + \beta_1^{h} S_t^{MP} d^{Hike} + \beta_2^{h} S_t^{MP} d^{Unch} + \beta_3^{h} S_t^{MP} d^{Cut} + \gamma_1^{h} Z_t^{CPI} + \gamma_2^{h} Z_t^{IP} + \epsilon_t^{h}
\end{align}
where $d^{Hike}$, $d^{Unch}$, and $d^{Cut}$ are three dummy variables representing Fed rate hike, Fed rate unchanged, and Fed rate cut periods\footnote{We identify periods of Fed rate hike, Fed rate unchanged, and Fed rate cut based on FOMC target rate announcements. In addition, if an unchanged target rate is announced at an adjacent FOMC meeting, but we capture a monetary policy shock in the same direction as the current rate announcement, the month of the adjacent FOMC is set to 1 in the corresponding dummy variable. This setup allows for a better incorporation of market expectations regarding monetary policy.}. Therefore, the coefficient for the interaction term between the dummy variable and the monetary policy shock $\beta_1^h$, $\beta_2^h$, and $\beta_3^h$ can respectively reflect the impulse responses under different periods of monetary policy shocks in the U.S.. 
For example, if the coefficient $\beta_1^h$ is significantly greater than 0, it means that during periods of Fed rate hike, the tight monetary policy shock increased the connectedness on the horizon $h$. Conversely, if $\beta_3^h$ is significantly less than 0, it indicates that during periods of Fed rate cut, the loose monetary policy shock heightened connectedness. 

Regarding $\mathcal{C}_{t}$, we consider the total connectedness index for the projection layer and multi-layer (return, volatility, skewness, kurtosis), as well as the directional net connectedness index of the projection layer, in line with Section \ref{sec: connectedness}. In addition, $Z_t^{CPI}$ and $Z_t^{IP}$ are control variables, which include  U.S. CPI and industrial production (IP) index\footnote{We take first-order difference of the CPI and IP to transform them into stationary time series.}.



\begin{figure}[!h]
    \centering
    \begin{subfigure}{0.325\textwidth}
        \includegraphics[width=\linewidth]{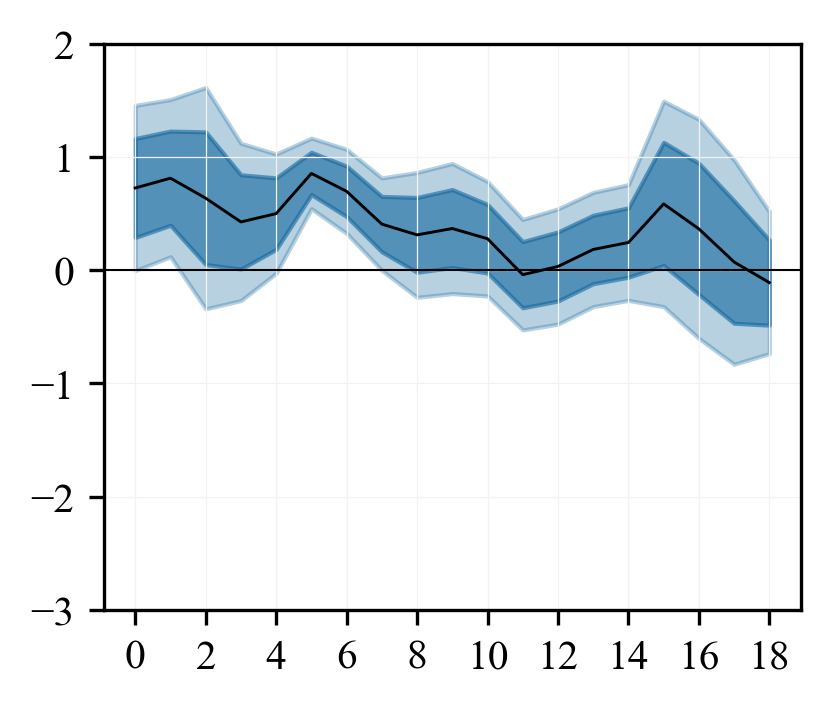}
        \caption{\footnotesize{Fed rate hike period}}
        \label{figure: ir_tspl_proj_a}
    \end{subfigure}
    \begin{subfigure}{0.325\textwidth}
        \includegraphics[width=\linewidth]{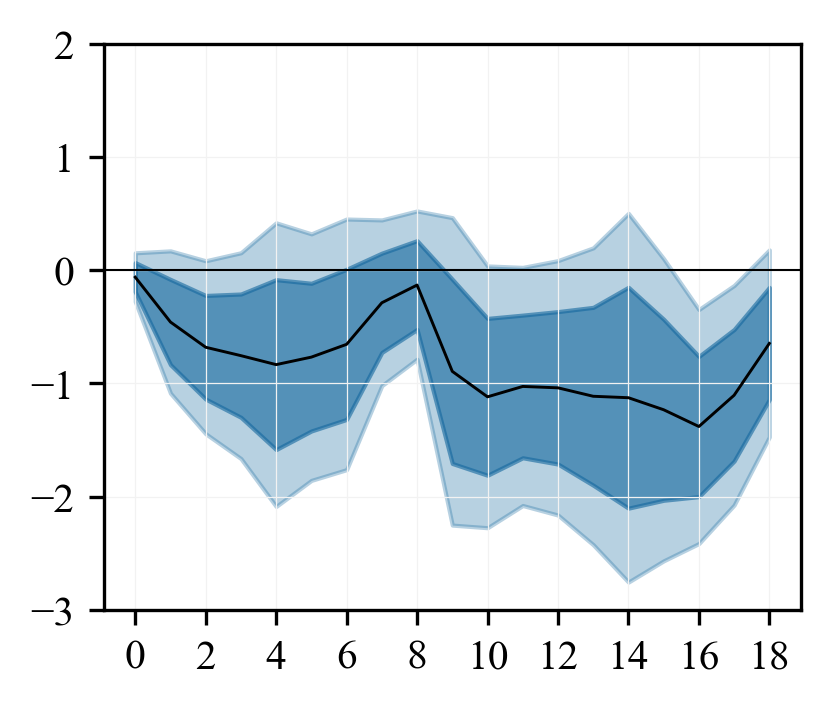}
        \caption{\footnotesize{Fed rate unchange period}}
        \label{figure: ir_tspl_proj_b}
    \end{subfigure}
    \begin{subfigure}{0.325\textwidth}
        \includegraphics[width=\linewidth]{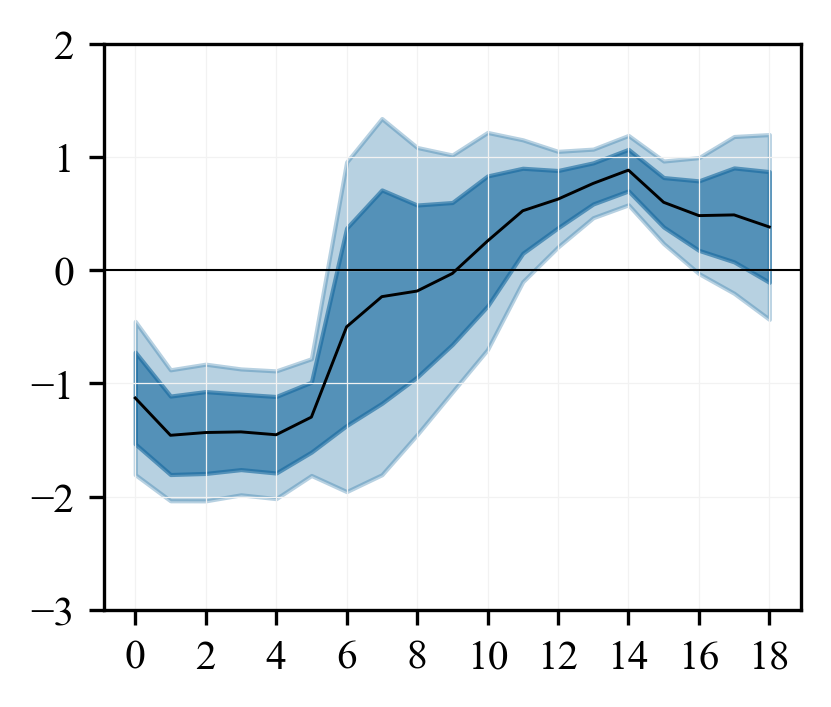}
        \caption{\footnotesize{Fed rate cut period}}
        \label{figure: ir_tspl_proj_c}
    \end{subfigure}
    \caption{Impulse response of the projection layer total connectedness}
    \label{figure: ir_tspl_proj}
    \raggedright{\justifying{\footnotesize{\textit{Notes}: The black line is the impulse response of the the projection layer total connectedness index to monetary policy shocks in a given period. The dark blue band represents the 68\% confidence interval, while the light blue band represents the 90\% confidence interval.}}}
\end{figure}

Figure \ref{figure: ir_tspl_proj} shows the response of the total connectedness of the projection layer to different types of monetary policy shocks across different periods, where the black solid line represents the estimated coefficients, and the dark blue area and the light blue area represent the 68\% and 90\% confidence interval , respectively. Overall, there exists significant heterogeneity in the effects of different types of monetary policy shocks (Fed rate hike, Fed rate unchanged, and Fed rate cut) on the total connectedness index of the projection layer. During the periods of Fed rate hikes and rate cuts, represented in Figures \ref{figure: ir_tspl_proj}(a) and (c), policy shocks have a significant effect on total connectedness. However, in Figure \ref{figure: ir_tspl_proj}(c), the effect is insignificant during periods of unchanged Fed rates. 

Specifically, during the period of Fed rate hikes (Figure \ref{figure: ir_tspl_proj}(a)), the impact of the monetary policy shock is positive and significant for the fifth and sixth months, then decreasing to zero. This indicates that tight monetary policy shocks significantly elevate the overall interconnectedness of the green financial market within the first six months, with their impact gradually fading. On the contrary, during Fed rate cuts (Figure \ref{figure: ir_tspl_proj}(c)), the impact of the monetary policy shock is notably negative and lasts for 6 months, gradually turning positive and becoming significant between 12 and 16 months. Since the interest rate decline represents a negative shock, the negative coefficient implies an increase in connectedness after the shock. This suggests that loose monetary policy shocks also notably enhance total connectedness within approximately six months but might lead to a decline in connectedness after a year. Such a phenomenon  may be due to the fact that negative shocks from interest rate cuts can heighten financial market uncertainty in the short run, but accommodating policies help improve financial conditions and alleviate market uncertainty in the long run. Furthermore, comparing the three subfigures in Figure \ref{figure: ir_tspl_proj}, it is evident that the coefficient in Figure \ref{figure: ir_tspl_proj}(c) is larger in absolute value and more significant, implying a relatively stronger impact of loose monetary policy during the sample period.

\begin{figure}[!p]
    \centering
    \begin{subfigure}{0.325\textwidth}
        \includegraphics[width=\linewidth]{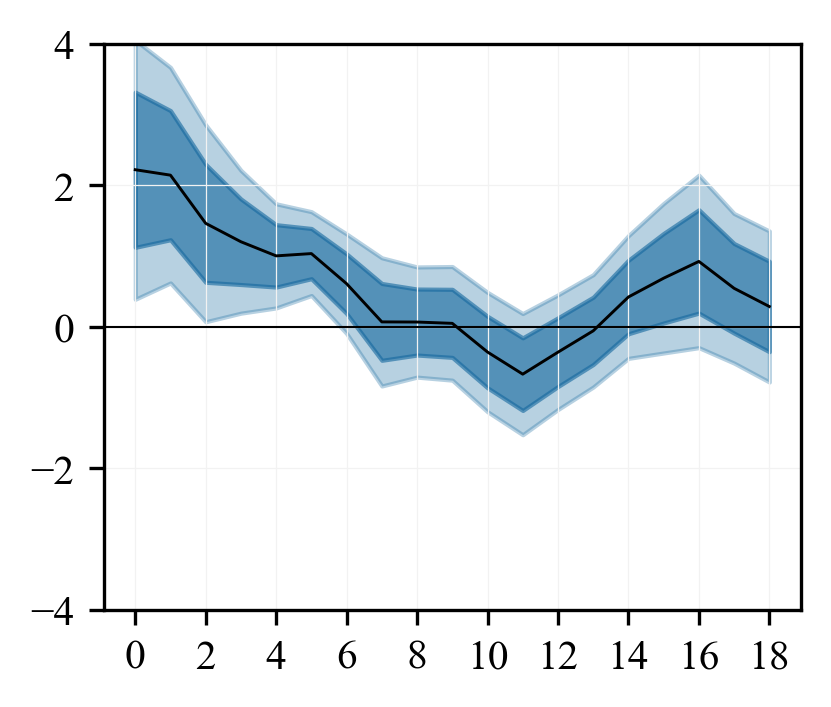}
        \caption{\footnotesize{Return: hike}}
        \label{figure: ir_tspl_multi_a}
    \end{subfigure}
    \vspace{0.1cm}
    \begin{subfigure}{0.325\textwidth}
        \includegraphics[width=\linewidth]{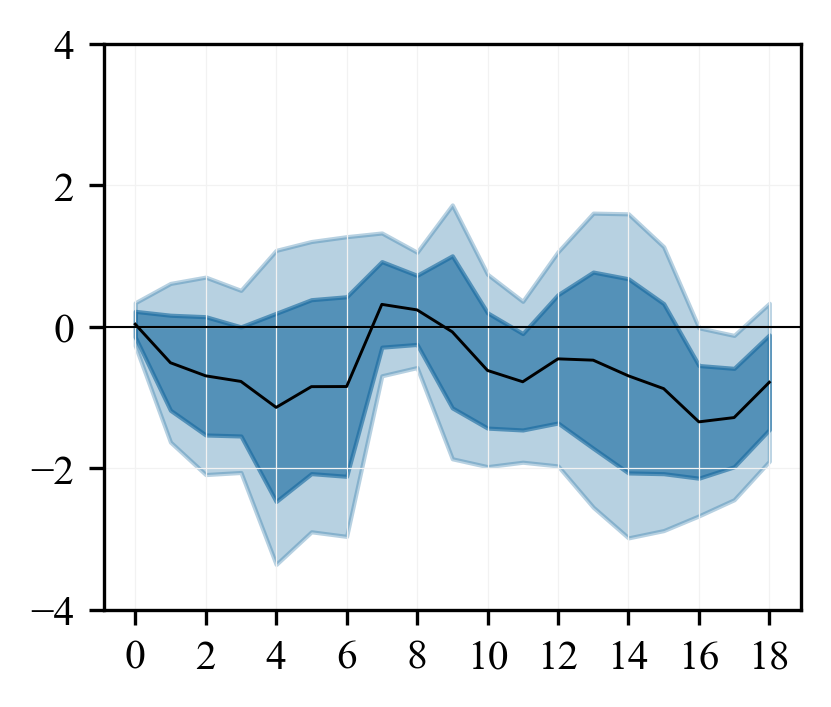}
        \caption{\footnotesize{Return: unchange}}
        \label{figure: ir_tspl_multi_b}
    \end{subfigure}
    \begin{subfigure}{0.325\textwidth}
        \includegraphics[width=\linewidth]{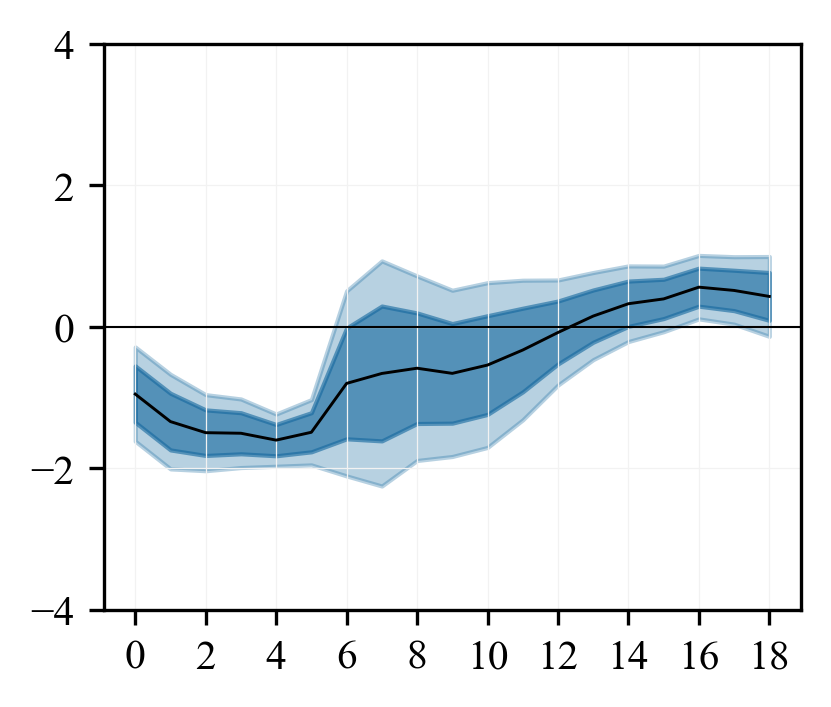}
        \caption{\footnotesize{Return: cut}}
        \label{figure: ir_tspl_multi_c}
    \end{subfigure}
    
    \begin{subfigure}{0.325\textwidth}
        \includegraphics[width=\linewidth]{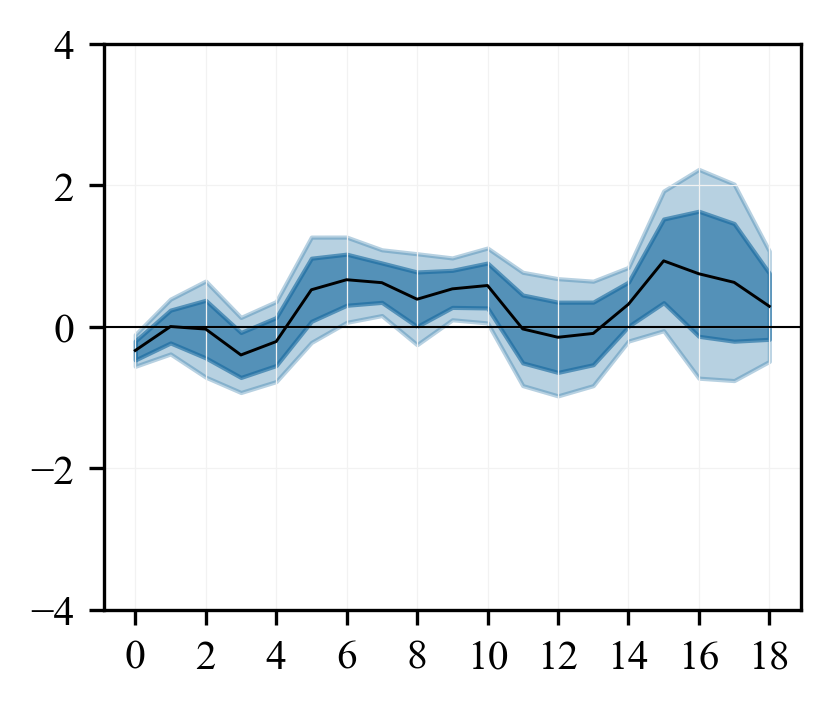}
        \caption{\footnotesize{Volatility: hike}}
        \label{figure: ir_tspl_multi_d}
    \end{subfigure}
    \vspace{0.1cm}
    \begin{subfigure}{0.325\textwidth}
        \includegraphics[width=\linewidth]{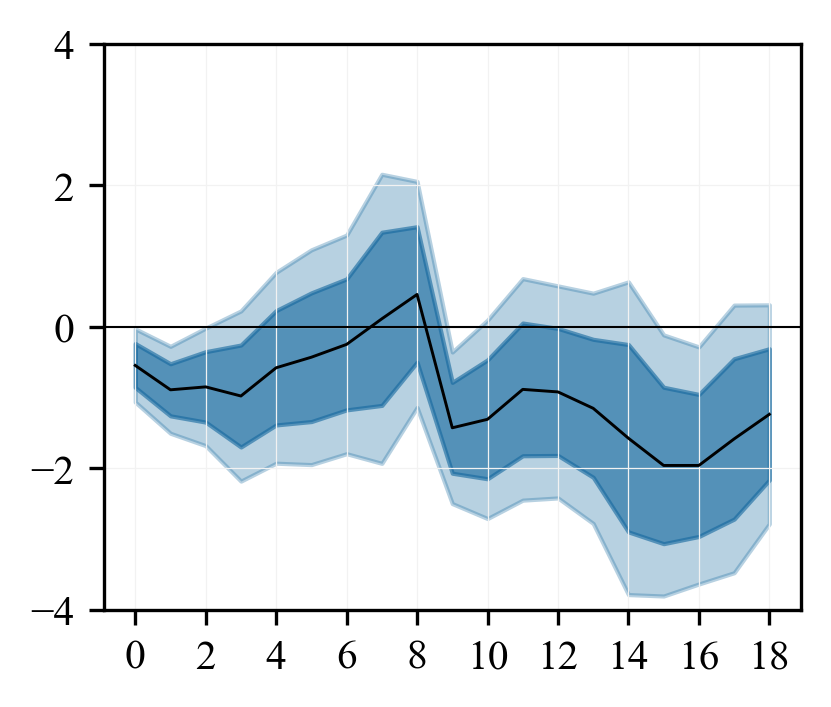}
        \caption{\footnotesize{Volatility: unchange}}
        \label{figure: ir_tspl_multi_e}
    \end{subfigure}
    \begin{subfigure}{0.325\textwidth}
        \includegraphics[width=\linewidth]{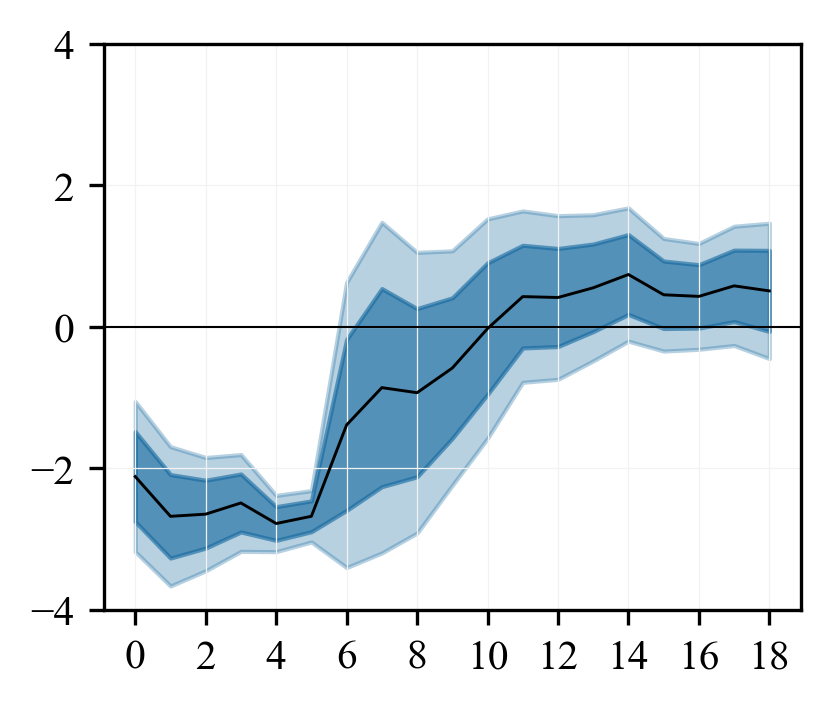}
        \caption{\footnotesize{Volatility: cut}}
        \label{figure: ir_tspl_multi_f}
    \end{subfigure}

    \begin{subfigure}{0.325\textwidth}
        \includegraphics[width=\linewidth]{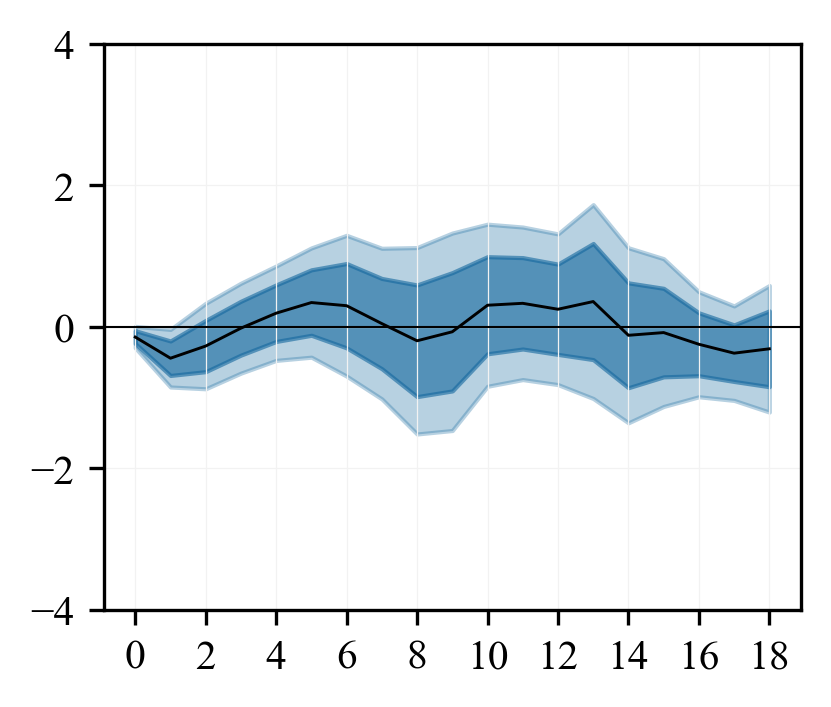}
        \caption{\footnotesize{Skewness: hike}}
        \label{figure: ir_tspl_multi_g}
    \end{subfigure}
    \vspace{0.1cm}
    \begin{subfigure}{0.325\textwidth}
        \includegraphics[width=\linewidth]{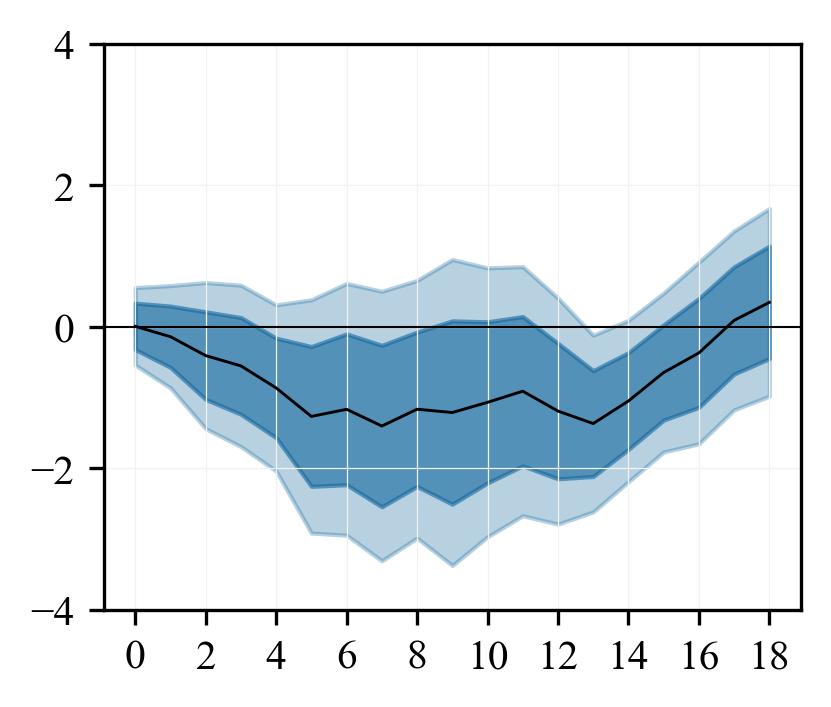}
        \caption{\footnotesize{Skewness: unchange}}
        \label{figure: ir_tspl_multi_h}
    \end{subfigure}
    \begin{subfigure}{0.325\textwidth}
        \includegraphics[width=\linewidth]{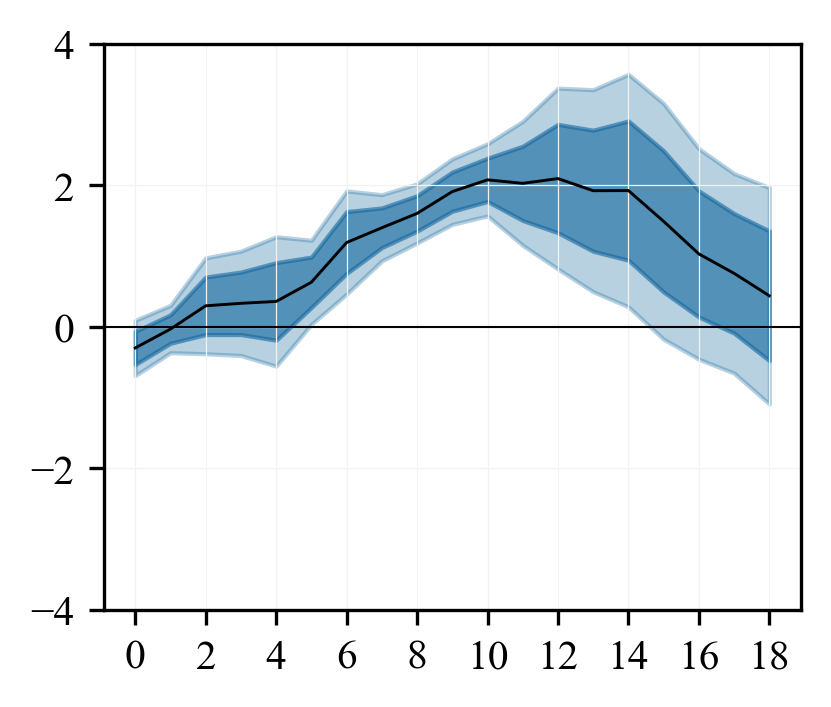}
        \caption{\footnotesize{Skewness: cut}}
        \label{figure: ir_tspl_multi_i}
    \end{subfigure}

    \begin{subfigure}{0.325\textwidth}
        \includegraphics[width=\linewidth]{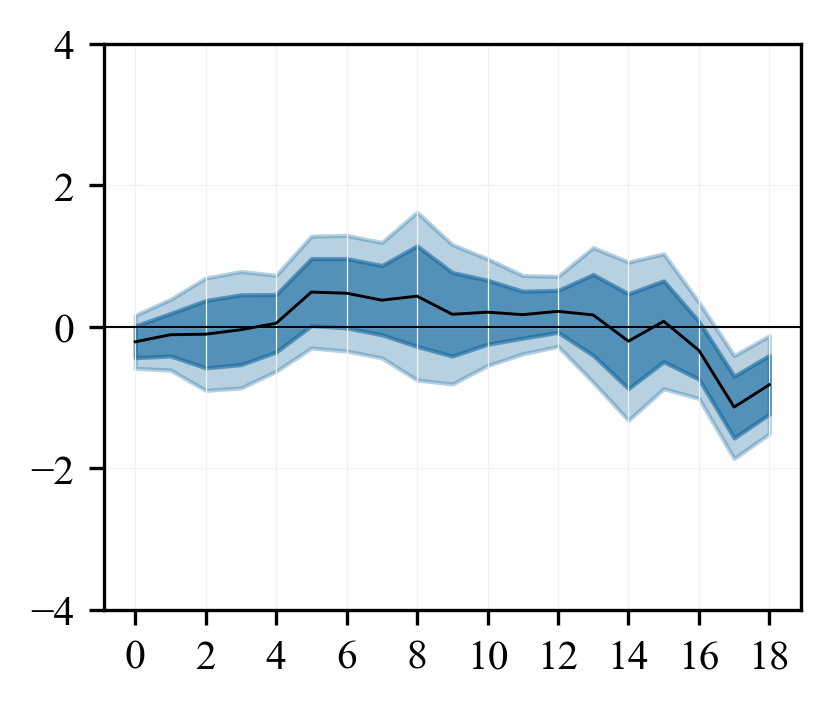}
        \caption{\footnotesize{Kurtosis: hike}}
        \label{figure: ir_tspl_multi_j}
    \end{subfigure}
    \begin{subfigure}{0.325\textwidth}
        \includegraphics[width=\linewidth]{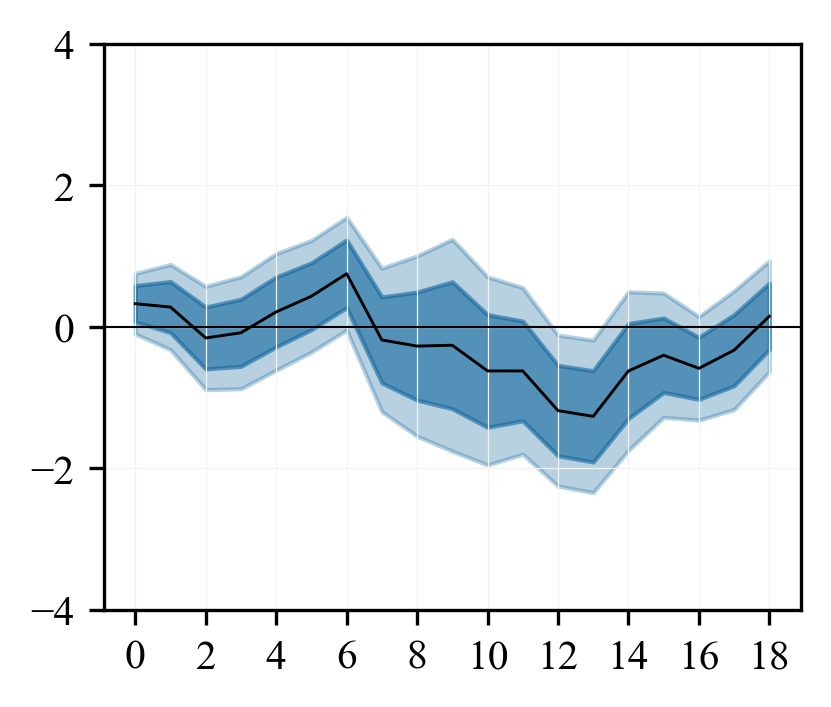}
        \caption{\footnotesize{Kurtosis: unchange}}
        \label{figure: ir_tspl_multi_k}
    \end{subfigure}
    \begin{subfigure}{0.325\textwidth}
        \includegraphics[width=\linewidth]{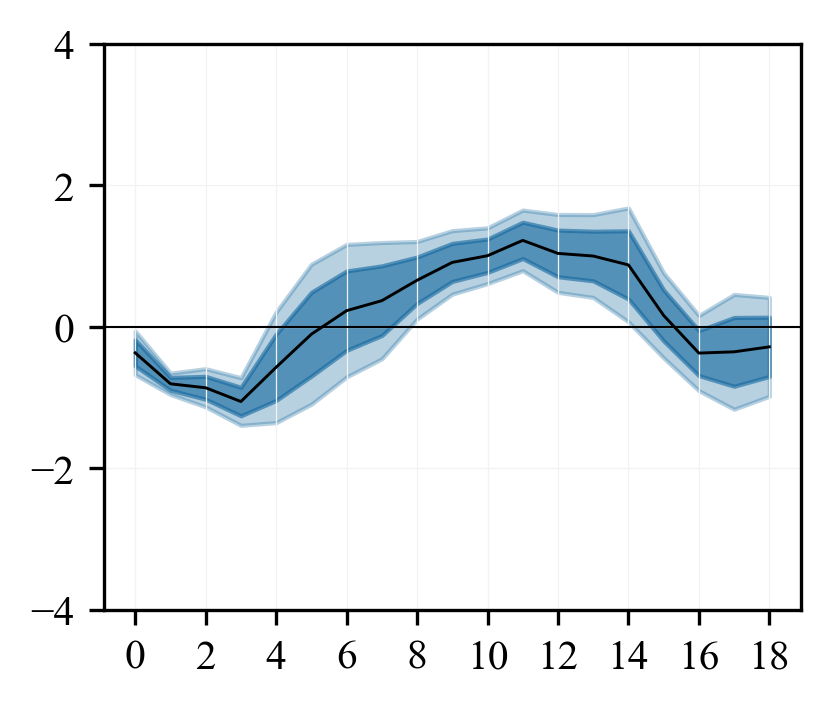}
        \caption{\footnotesize{Kurtosis: cut}}
        \label{figure: ir_tspl_multi_l}
    \end{subfigure}
    \caption{Impluse response of multi-moment layer total connectedness}
    \label{figure: ir_tspl_multi}
    \raggedright{\justifying{\footnotesize{\textit{Notes}: The black line is the impulse response of total connectedness index in a certain moment layer to monetary policy shocks in a given period. The dark blue band represents the 68\% confidence interval, while the light blue band represents the 90\% confidence interval.}}}
\end{figure}

Figure \ref{figure: ir_tspl_multi} presents the corresponding results for the multi-moment total connectedness indices. We highlight some important findings. First, in general, the impact of monetary policy on the total connectedness of different moments varies. 
During periods of Fed rate hikes, as depicted in the first column of Figure \ref{figure: ir_tspl_multi}, across the multi-moment connectedness network, monetary policy shocks exhibit a significant positive effect only within the return layer represented in Figure \ref{figure: ir_tspl_multi}(a) for a forecast horizon of up to 6 months. For the volatility layer (Figure \ref{figure: ir_tspl_multi}(d)), the overall effect is positive but not significant. Similarly, for the skewness layer (Figure \ref{figure: ir_tspl_multi}(g)) and kurtosis layer (Figure \ref{figure: ir_tspl_multi}(j)), the effects of the shocks fluctuate around zero, suggesting a limited impact.

\begin{figure}[!t]
    \centering
    \begin{subfigure}{1\textwidth}
        \includegraphics[width=\linewidth]{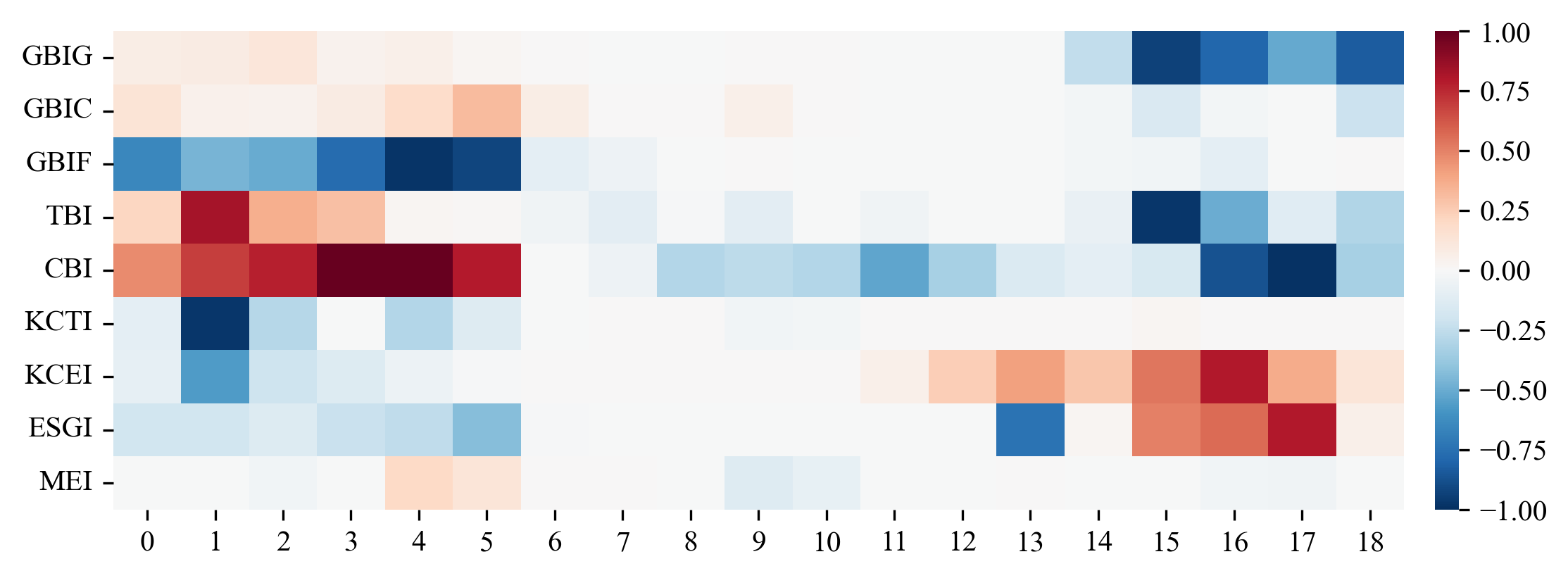}
        \caption{\footnotesize{Fed rate hike}}
    \end{subfigure}
    \begin{subfigure}{1\textwidth}
        \includegraphics[width=\linewidth]{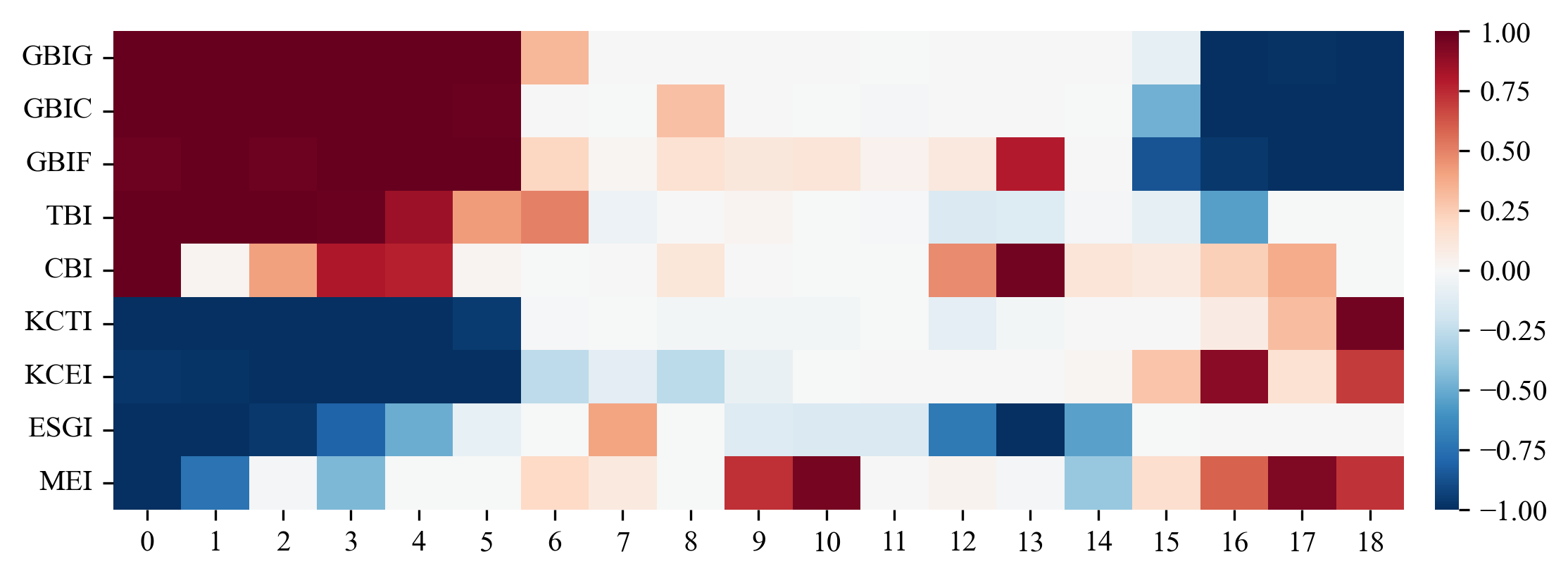}
        \caption{\footnotesize{Fed rate cut}}
    \end{subfigure}
    \caption{Impluse response of net connectedness}
    \label{figure: heat_up}
    \raggedright{\justifying{\footnotesize{\textit{Notes}: To illustrate the direction, significance, and heterogeneity of the impulse response succinctly and clearly, we calculate an indicator $s=d\times(1-p)^7$ for each impulse response coefficient, where $p$ represents the p-value, and $d$ denotes the direction of the impulse response. Thus, darker colors in the plot represent stronger significance, where red marks the positive response and blue marks the negative response. When the absolute value of $s$ is greater than 0.48, it indicates that the impulse response coefficient is significant at the 10\% significance level. }}}
\end{figure}

Second, during periods of Fed rate cuts, as shown in the third column of Figure \ref{figure: ir_tspl_multi}, the impacts of monetary policy shocks on the return and volatility layers (Figures \ref{figure: ir_tspl_multi}(c)(f)) are significantly negative within 6 months. The maximum absolute value appears in the fourth month for the volatility layer. Then, the impact turns to positive after 12 periods but remain insignificant. For the skewness layer (Figure \ref{figure: ir_tspl_multi} (i)), the response to total connectedness quickly turns positive at the second month and remains significant between 6 and 14 months. For the kurtosis layer (Figure \ref{figure: ir_tspl_multi}(l)), the impact of monetary policy shocks is a significant negative in 4 months, then turns positive and significant between the 8 and 14 months.
These results indicate that loose monetary policy shocks contribute significantly to the increasing total connectedness of the return, volatility, and kurtosis layers within 4 months, while having a decreasing effect on the total connectedness of the skewness and kurtosis layers after 8 to 14 months.

Figure \ref{figure: heat_up} illustrates the direction and significance\footnote{Detailed impulse response of net connectedness indicies are attached in Appendix Figure \ref{figure: ir_up} and \ref{figure: ir_down}.} of the response of net connectedness to tight and loose monetary policy shocks, respectively, where red represents positive responses and blue represents negative responses, with darker colors representing a higher level of significance. 

As shown in Figure \ref{figure: heat_up}(a), tightening of monetary policy shocks significantly amplified net outflow from the bond market benchmark indices CBI and TBI within six months. Their positive regression coefficients showcased the strongest significance, depicted by the deepest red color in Figure \ref{figure: heat_up}(a). Specifically, the effect of monetary policy shocks on CBI is most significant at 4 periods of forward forecasting and on TBI at 1 period. This could be linked to the direct impact of the increased interest rates on overall bond prices under tightening monetary policies.
Moreover, GBIG and GBIC also display a positive net connectedness response, albeit insignificant, indicated by a lighter shade of red. Simultaneously, the deeper blue shades for GBIF, KCTI, and KCEI imply significantly negative coefficients, signifying an intensified risk-absorption effect under tight monetary policy shocks. In addition, after a forecast horizon of 12 months (one year), there's a change in the direction. At this point, the net outflow effect intensifies for the equity market indices KCEI and ESGI, while the net inflow effect strengthens for the bond market indices GBIG, TBI, and CBI.

Similarly, Figure \ref{figure: heat_up}(b) illustrates the corresponding results for loosing monetary policy shocks. It reveals that within a six-month period, monetary policy shocks significantly enhance outflows from the equity market (dark blue part), and inflows in the bond market (dark red part). This aligns with the result after the Fed restarts quantitative easing in March 2020, as depicted in Figure \ref{figure: netspl_proj}. Additionally, compared to the benchmark indices (TBI and CBI for bond market; MEI for equity market), the coefficients corresponding to the green financial market indices (GBIG, GBIC, and GBIF for green bond market; KCTI, KCEI, and MEI for green equity markets) sustained significance for a longer duration. This suggests that, during the sample period, the impact of loosing monetary policy shock on the interconnectedness structure of green financial markets is noteworthy.

\section{Conclusion}
\label{sec: conclusion}

In this paper, we develop a novel multi-moment connectedness network approach to analyze the unvertainty of the green financial market, particularly focusing on how monetary policy shocks impact market interconnectedness. Our approach integrates various moments—returns, volatility, skewness, and kurtosis—to offer a comprehensive view of the market's response to policy changes. This methodology provides valuable insights for policymakers and investment managers, particularly in understanding the dynamics of green finance in the context of monetary policy fluctuations.

The multi-moment connectedness network constructed for the green financial market in this paper indicates several key findings.
Firstly, from the within-layer perspective, the connectedness effect within the return layer is the strongest, gradually diminishing as moments escalate.
Second, from the cross-layer perspective, there is an aggregation effect by market category in the network, with stronger connectedness within markets. Comparatively, the connectedness inside the green bond market is stronger, whereas the connectedness inside the green equity market is weaker and weakens further as the moment increases from the return layer to kurtosis layer. 
Additionally, the green equity indices ESGI and KCEI, together with the benchmark bonds TBI and CBI, are important nodes connecting the two markets.

We also find evidence that monetary policy affects green finance market connectedness: Based on the dynamic evolution of time-varying connectedness indices, significant changes in connectedness and its structure occurred at specific FOMC meetings, such as the emergency rate cut on March 3, 2020.
Furthermore, upon introducing U.S. monetary policy shocks, both tight and loose policies exhibit a significant positive impact on the total connectedness within the green financial market, persisting for as long as six months. Tight monetary policy shocks mainly affect the return layer, transmitting risk from the bond market benchmark indices to other nodes under shock. 
On the other hand, loose monetary policy shocks are more influential during the sample period, significantly affecting the return, volatility, and kurtosis layers, transmitting risk from the equity market to the bond market.

Our empirical evidence provides valuable information for sustainable investors.
For investors with large holdings of green equities, it is crucial to pay close attention to the uncertainties arise from tight monetary policy shock. Conversely, for those with large holdings of green bonds, quantitative easing monetary policy could bring more dramatic volatility. Hence, adjusting portfolios promptly in anticipation of corresponding monetary policy changes becomes essential to mitigate risks. In addition, although the total connectedness of higher moments appears to be low, it could be reversed under the impact of monetary policy shocks. Therefore, it is crucial to consider information on higher-order moments, such as the skewness and kurtosis of financial asset prices in risk management.




\clearpage
\vspace{1cm}
\setlength{\parskip}{2pt plus2pt minus2pt}
\footnotesize{ \setlength{\bibsep}{3.4ex}
\setstretch{0.75}
\addcontentsline{toc}{section}{References}
\bibliographystyle{apalike}
\bibliography{bibfile.bib}}

\begin{thebibliography}{}

\bibitem[Akyildirim et~al., 2022]{akyildirim2022connectedness}
Akyildirim, E., Cepni, O., Moln{\'a}r, P., and Uddin, G.~S. (2022).
\newblock Connectedness of energy markets around the world during the covid-19 pandemic.
\newblock {\em Energy economics}, 109:105900.

\bibitem[Ando et~al., 2022]{ando2022quantile}
Ando, T., Greenwood-Nimmo, M., and Shin, Y. (2022).
\newblock Quantile connectedness: modeling tail behavior in the topology of financial networks.
\newblock {\em Management Science}, 68(4):2401--2431.

\bibitem[Barigozzi and Hallin, 2017]{barigozzi2017network}
Barigozzi, M. and Hallin, M. (2017).
\newblock A network analysis of the volatility of high dimensional financial series.
\newblock {\em Journal of the Royal Statistical Society Series C: Applied Statistics}, 66(3):581--605.

\bibitem[Barun{\'\i}k et~al., 2020]{barunik2020asymmetric}
Barun{\'\i}k, J., Bevilacqua, M., and Tunaru, R. (2020).
\newblock Asymmetric network connectedness of fears.
\newblock {\em The Review of Economics and Statistics}, pages 1--41.

\bibitem[Bernardi and Catania, 2019]{bernardi2019switching}
Bernardi, M. and Catania, L. (2019).
\newblock Switching generalized autoregressive score copula models with application to systemic risk.
\newblock {\em Journal of Applied Econometrics}, 34(1):43--65.

\bibitem[Bouri et~al., 2023]{bouri2023connectedness}
Bouri, E., Lei, X., Xu, Y., and Zhang, H. (2023).
\newblock Connectedness in implied higher-order moments of precious metals and energy markets.
\newblock {\em Energy}, 263:125588.

\bibitem[Brunetti et~al., 2019]{brunetti2019interconnectedness}
Brunetti, C., Harris, J.~H., Mankad, S., and Michailidis, G. (2019).
\newblock Interconnectedness in the interbank market.
\newblock {\em Journal of Financial Economics}, 133(2):520--538.

\bibitem[Catania, 2021]{catania2021dynamic}
Catania, L. (2021).
\newblock Dynamic adaptive mixture models with an application to volatility and risk.
\newblock {\em Journal of Financial Econometrics}, 19(4):531--564.

\bibitem[Chan et~al., 2023]{chan2023optimal}
Chan, Y.~T., Ji, Q., and Zhang, D. (2023).
\newblock Optimal monetary policy responses to carbon and green bubbles: A two-sector dsge analysis.
\newblock {\em Energy Economics}, page 107281.

\bibitem[Chen et~al., 2023]{chen2023determinants}
Chen, Y.-P., Chen, Y.-L., Chiang, S.-H., and Mo, W.-S. (2023).
\newblock Determinants of connectedness in financial institutions: Evidence from taiwan.
\newblock {\em Emerging Markets Review}, 55:100951.

\bibitem[Dafermos et~al., 2018]{dafermos2018climate}
Dafermos, Y., Nikolaidi, M., and Galanis, G. (2018).
\newblock Climate change, financial stability and monetary policy.
\newblock {\em Ecological Economics}, 152:219--234.

\bibitem[Dai et~al., 2021]{dai2021multiscale}
Dai, X., Xiao, L., Wang, Q., and Dhesi, G. (2021).
\newblock Multiscale interplay of higher-order moments between the carbon and energy markets during phase iii of the eu ets.
\newblock {\em Energy Policy}, 156:112428.

\bibitem[Desalegn et~al., 2022]{desalegn2022effect}
Desalegn, G., Fekete-Farkas, M., and Tangl, A. (2022).
\newblock The effect of monetary policy and private investment on green finance: evidence from hungary.
\newblock {\em Journal of Risk and Financial Management}, 15(3):117.

\bibitem[Diebold and Yilmaz, 2014]{diebold2014network}
Diebold, F.~X. and Yilmaz, K. (2014).
\newblock On the network topology of variance decompositions: Measuring the connectedness of financial firms.
\newblock {\em Journal of Econometrics}, 182(1):119--134.

\bibitem[Diebold and Yilmaz, 2015]{diebold2015trans}
Diebold, F.~X. and Yilmaz, K. (2015).
\newblock Trans-atlantic equity volatility connectedness: Us and european financial institutions, 2004--2014.
\newblock {\em Journal of Financial Econometrics}, 14(1):81--127.

\bibitem[Dogan et~al., 2022]{dogan2022investigating}
Dogan, E., Madaleno, M., Taskin, D., and Tzeremes, P. (2022).
\newblock Investigating the spillovers and connectedness between green finance and renewable energy sources.
\newblock {\em Renewable Energy}, 197:709--722.

\bibitem[Gilchrist and Zakraj{\v{s}}ek, 2012]{gilchrist2012credit}
Gilchrist, S. and Zakraj{\v{s}}ek, E. (2012).
\newblock Credit spreads and business cycle fluctuations.
\newblock {\em American economic review}, 102(4):1692--1720.

\bibitem[Hale and Lopez, 2019]{hale2019monitoring}
Hale, G. and Lopez, J.~A. (2019).
\newblock Monitoring banking system connectedness with big data.
\newblock {\em Journal of Econometrics}, 212(1):203--220.

\bibitem[Hao and Pham, 2023]{hao2023dynamic}
Hao, W. and Pham, L. (2023).
\newblock Dynamic connectedness in the higher moments between clean energy and oil prices.
\newblock {\em Available at SSRN 4565305}.

\bibitem[Jaroci{\'n}ski and Karadi, 2020]{jarocinski2020deconstructing}
Jaroci{\'n}ski, M. and Karadi, P. (2020).
\newblock Deconstructing monetary policy surprises—the role of information shocks.
\newblock {\em American Economic Journal: Macroeconomics}, 12(2):1--43.

\bibitem[Jord{\`a}, 2005]{jorda2005estimation}
Jord{\`a}, {\`O}. (2005).
\newblock Estimation and inference of impulse responses by local projections.
\newblock {\em American economic review}, 95(1):161--182.

\bibitem[Koop and Korobilis, 2013]{koop2013large}
Koop, G. and Korobilis, D. (2013).
\newblock Large time-varying parameter vars.
\newblock {\em Journal of Econometrics}, 177(2):185--198.

\bibitem[Koop et~al., 1996]{koop1996impulse}
Koop, G., Pesaran, M.~H., and Potter, S.~M. (1996).
\newblock Impulse response analysis in nonlinear multivariate models.
\newblock {\em Journal of econometrics}, 74(1):119--147.

\bibitem[Li et~al., 2022]{li2022tracking}
Li, J., Meng, G., Li, C., and Du, K. (2022).
\newblock Tracking carbon intensity changes between china and japan: Based on the decomposition technique.
\newblock {\em Journal of Cleaner Production}, 349:131090.

\bibitem[Lin and Li, 2023]{lin2023emerging}
Lin, B. and Li, M. (2023).
\newblock Emerging industry development and information transmission in financial markets: Evidence from china's renewable energy.
\newblock {\em Energy Economics}, 128:107192.

\bibitem[Lin and Su, 2023]{lin2023uncertainties}
Lin, B. and Su, T. (2023).
\newblock Uncertainties and green bond markets: Evidence from tail dependence.
\newblock {\em International Journal of Finance \& Economics}, 28(4):4458--4475.

\bibitem[Lin and Xu, 2019]{lin2019effectively}
Lin, B. and Xu, B. (2019).
\newblock How to effectively stabilize china's commodity price fluctuations?
\newblock {\em Energy Economics}, 84:104544.

\bibitem[Lin and Zhao, 2023]{lin2023tracking}
Lin, B. and Zhao, H. (2023).
\newblock Tracking policy uncertainty under climate change.
\newblock {\em Resources Policy}, 83:103699.

\bibitem[Lundgren et~al., 2018]{lundgren2018connectedness}
Lundgren, A.~I., Milicevic, A., Uddin, G.~S., and Kang, S.~H. (2018).
\newblock Connectedness network and dependence structure mechanism in green investments.
\newblock {\em Energy Economics}, 72:145--153.

\bibitem[Maghyereh et~al., 2016]{maghyereh2016directional}
Maghyereh, A.~I., Awartani, B., and Bouri, E. (2016).
\newblock The directional volatility connectedness between crude oil and equity markets: New evidence from implied volatility indexes.
\newblock {\em Energy Economics}, 57:78--93.

\bibitem[Mensi et~al., 2022]{mensi2022spillovers}
Mensi, W., Shafiullah, M., Vo, X.~V., and Kang, S.~H. (2022).
\newblock Spillovers and connectedness between green bond and stock markets in bearish and bullish market scenarios.
\newblock {\em Finance Research Letters}, 49:103120.

\bibitem[Pham, 2021]{pham2021frequency}
Pham, L. (2021).
\newblock Frequency connectedness and cross-quantile dependence between green bond and green equity markets.
\newblock {\em Energy Economics}, 98:105257.

\bibitem[Raza et~al., 2023]{raza2023connectedness}
Raza, S.~A., Sharif, A., Kumar, S., and Ahmed, M. (2023).
\newblock Connectedness between monetary policy uncertainty and sectoral stock market returns: Evidence from asymmetric tvp-var approach.
\newblock {\em International Review of Financial Analysis}, 90:102946.

\bibitem[Reboredo and Ugolini, 2020]{reboredo2020price}
Reboredo, J.~C. and Ugolini, A. (2020).
\newblock Price connectedness between green bond and financial markets.
\newblock {\em Economic Modelling}, 88:25--38.

\bibitem[Reboredo et~al., 2020]{reboredo2020network}
Reboredo, J.~C., Ugolini, A., and Aiube, F. A.~L. (2020).
\newblock Network connectedness of green bonds and asset classes.
\newblock {\em Energy Economics}, 86:104629.

\bibitem[Su and Lin, 2022]{su2022liquidity}
Su, T. and Lin, B. (2022).
\newblock The liquidity impact of chinese green bonds spreads.
\newblock {\em International Review of Economics \& Finance}, 82:318--334.

\bibitem[Su et~al., 2022]{su2022green}
Su, T., Zhang, Z.~J., and Lin, B. (2022).
\newblock Green bonds and conventional financial markets in china: A tale of three transmission modes.
\newblock {\em Energy Economics}, 113:106200.

\bibitem[Sun et~al., 2019]{sun2019fossil}
Sun, C., Ding, D., Fang, X., Zhang, H., and Li, J. (2019).
\newblock How do fossil energy prices affect the stock prices of new energy companies? evidence from divisia energy price index in china's market.
\newblock {\em Energy}, 169:637--645.

\bibitem[Sun et~al., 2022]{sun2022analysis}
Sun, C., Khan, A., Liu, Y., and Lei, N. (2022).
\newblock An analysis of the impact of fiscal and monetary policy fluctuations on the disaggregated level renewable energy generation in the g7 countries.
\newblock {\em Renewable Energy}, 189:1154--1165.

\bibitem[Valente and Fujimoto, 2010]{valente2010bridging}
Valente, T.~W. and Fujimoto, K. (2010).
\newblock Bridging: locating critical connectors in a network.
\newblock {\em Social networks}, 32(3):212--220.

\bibitem[Wang et~al., 2023]{wang2023energy}
Wang, T., Wu, F., Zhang, D., and Ji, Q. (2023).
\newblock Energy market reforms in china and the time-varying connectedness of domestic and international markets.
\newblock {\em Energy Economics}, 117:106495.

\bibitem[Wu et~al., 2022]{wu2022complex}
Wu, F., Xiao, X., Zhou, X., Zhang, D., and Ji, Q. (2022).
\newblock Complex risk contagions among large international energy firms: A multi-layer network analysis.
\newblock {\em Energy Economics}, 114:106271.

\bibitem[Yang et~al., 2023]{caporin2023measuring}
Yang, R., Caporin, M., and Jiménez-Martin, J.-A. (2023).
\newblock Measuring climate transition risk spillovers.
\newblock {\em Review of Finance}, page Forthcoming.

\bibitem[Yang and Zhou, 2017]{yang2017quantitative}
Yang, Z. and Zhou, Y. (2017).
\newblock Quantitative easing and volatility spillovers across countries and asset classes.
\newblock {\em Management Science}, 63(2):333--354.

\bibitem[Zhang et~al., 2023]{zhang2023impact}
Zhang, W., He, X., and Hamori, S. (2023).
\newblock The impact of the covid-19 pandemic and russia-ukraine war on multiscale spillovers in green finance markets: Evidence from lower and higher order moments.
\newblock {\em International Review of Financial Analysis}, 89:102735.

\bibitem[Zheng et~al., 2023a]{zheng2023global}
Zheng, T., Gong, L., and Ye, S. (2023a).
\newblock Global energy market connectedness and inflation at risk.
\newblock {\em Energy Economics}, 126:106975.

\bibitem[Zheng and Ye, 2024]{zheng2024cholesky}
Zheng, T. and Ye, S. (2024).
\newblock Cholesky gas models for large time-varying covariance matrices.
\newblock {\em Journal of Management Science and Engineering}, 9(1):115--142.

\bibitem[Zheng et~al., 2023b]{zheng2023fast}
Zheng, T., Ye, S., and Hong, Y. (2023b).
\newblock Fast estimation of a large tvp-var model with score-driven volatilities.
\newblock {\em Journal of Economic Dynamics and Control}, 157:104762.

\bibitem[Zhou et~al., 2022]{zhou2022does}
Zhou, H., Bian, J., Qin, Q., and Yu, M. (2022).
\newblock Does us monetary policy affect the connectedness of global financial markets?
\newblock {\em Available at SSRN 4085953}.

\end{thebibliography}

\clearpage

\normalsize\setstretch{1}
\begin{appendices}

\setcounter{page}{1}
{\setstretch{1.5}
\begin{center}
\Huge\textbf{Online Appendix}\\[0.5em]
\LARGE\textbf{Monetary Roles on Green Finance Markets: Evidence from a Multi-Moment Connectedness Network
\vspace{-0.5em} 
}\\[1em]
\large\MakeUppercase{Tingguo Zheng}$^{1,2}$, \MakeUppercase{Hongyin Zhang}$^{2}$ and \MakeUppercase{Shiqi Ye$^{3}$}\\[0.5em]
\normalsize
\vspace{-0.5em} $^1$ {\textit{Department of Statistics and Data Science, School of Economics, Xiamen University}} \\
\vspace{-0.5em} $^2$\textit{Wang Yanan Institute for Studies in Economics, Xiamen University} \\
\vspace{-0.5em} $^3$ \textit{Paula and Gregory Chow Institute for Studies in Economics, Xiamen University}
\end{center}}

\numberwithin{equation}{section}
\setcounter{section}{0}
\newpage
\small

\setcounter{table}{0}
\renewcommand{\thetable}{\thesection.\arabic{table}}
\setcounter{figure}{0}
\renewcommand{\thefigure}{\thesection.\arabic{figure}}

\clearpage
\section{Additional summary statistics for the data}
\label{apdx: data}

\begin{table}[!h]
\caption{Summary statistics: volatility}
\label{tab: summary_vol}
\centering
\renewcommand\arraystretch{1.4}
\scriptsize
\resizebox{\linewidth}{!}{
\begin{tabular}{lcccccccc}
\hline
     & Mean  & Std. & Max   & Min   & Skew. & Kurt. & ADF test    & JB test       \\ \hline
GBIG & -1.93 & 0.68 & 0.54  & -3.26 & 0.55  & -0.16 & -3.06 (0.03) & 119.49 (0.00)  \\
GBIC & -2.08 & 0.66 & 1.19  & -3.29 & 0.84  & 0.82  & -3.51 (0.01) & 334.81 (0.00)  \\
GBIF & -2.10  & 0.65 & 1.11  & -3.3  & 1.17  & 1.88  & -3.83 (0.00) & 860.30 (0.00)   \\
TBI  & -2.03 & 0.69 & 0.04  & -3.37 & 0.28  & -0.57 & -3.06 (0.03) & 61.02 (0.00)   \\
CBI  & -2.26 & 0.52 & -0.72 & -3.11 & 0.86  & -0.2  & -2.94 (0.04) & 287.01 (0.00)  \\
KCTI & 1.39  & 0.75 & 3.97  & 0.36  & 0.66  & -0.31 & -2.29 (0.17) & 178.18 (0.00)  \\
KCEI & 0.25  & 0.75 & 4.08  & -1.04 & 1.19  & 2.34  & -4.64 (0.00) & 1067.61 (0.00) \\
ESGI & -0.45 & 0.73 & 3.66  & -1.68 & 1.43  & 3.88  & -5.00 (0.00)  & 2225.89 (0.00) \\
MEI  & 0.80   & 0.64 & 3.31  & -0.57 & 0.65  & 0.90   & -3.3 7(0.01) & 240.09 (0.00)  \\ \hline
\end{tabular}}
\end{table}

\begin{table}[!h]
\caption{Summary statistics: skewness}
\label{tab: summary_skew}
\centering
\renewcommand\arraystretch{1.4}
\scriptsize
\resizebox{\linewidth}{!}{
\begin{tabular}{lcccccccc}
\hline
     & Mean  & Std. & Max  & Min   & Skew.  & Kurt.  & ADF test    & JB test         \\ \hline
GBIG & -0.02 & 0.09 & 0.68 & -0.59 & -0.53  & 15.54  & -14.55 (0.00) & 23217.36 (0.00)   \\
GBIC & -0.02 & 0.08 & 0.57 & -0.61 & -1.08  & 12.02  & -14.45 (0.00) & 14275.14 (0.00)   \\
GBIF & 0.00  & 0.11 & 0.35 & -0.48 & -0.27  & 1.64   & -5.11 (0.00)  & 286.34 (0.00)     \\
TBI  & -0.01 & 0.05 & 0.17 & -0.72 & -4.78  & 57.49  & -16.38 (0.00) & 325112.12 (0.00)  \\
CBI  & -0.04 & 0.09 & 0.13 & -1.80 & -10.48 & 159.67 & -8.37 (0.00)  & 2482033.62 (0.00) \\
KCTI & -0.14 & 0.18 & 0.31 & -0.31 & 0.62   & -1.11  & -2.43 (0.13) & 266.55 (0.00)     \\
KCEI & -0.11 & 0.11 & 0.09 & -0.66 & -0.54  & 0.13   & -6.43 (0.00)  & 112.75 (0.00)     \\
ESGI & -0.33 & 0.39 & 0.36 & -2.51 & -0.95  & 1.67   & -7.48 (0.00)  & 611.82 (0.00)     \\
MEI  & -1.3  & 0.93 & 0.61 & -4.67 & 0.08   & -0.62  & -2.4 (0.14)  & 39.52 (0.00)     \\ \hline
\end{tabular}}
\end{table}

\begin{table}[!h]
\caption{Summary statistics: Kurtosis}
\label{tab: summary_kurt}
\centering
\renewcommand\arraystretch{1.4}
\scriptsize
\resizebox{\linewidth}{!}{
\begin{tabular}{lcccccccc}
\hline
     & Mean  & Std.  & Max   & Min   & Skew. & Kurt. & ADF test    & JB test      \\ \hline
GBIG & 4.06  & 1.72  & 12.71 & -4.34 & 1.01  & 3.38  & -5.26 (0.00)  & 1488.88 (0.00) \\
GBIC & 4.27  & 1.74  & 12.21 & -3.27 & 0.82  & 3.13  & -4.86 (0.00)  & 1191.78 (0.00) \\
GBIF & 4.07  & 2.34  & 11.84 & -4.26 & -0.12 & 1.08  & -4.58 (0.00)  & 117.54 (0.00)  \\
TBI  & 2.88  & 1.07  & 8.81  & -3.46 & -0.48 & 6.51  & -10.93 (0.00) & 4141.78 (0.00) \\
CBI  & 10.92 & 6.19  & 33.25 & 1.16  & 0.70   & 0.25  & -2.87 (0.05) & 194.70 (0.00)   \\
KCTI & 3.45  & 0.46  & 5.34  & 2.86  & 2.08  & 5.01  & -3.85 (0.00)  & 4048.54 (0.00) \\
KCEI & 4.81  & 1.99  & 14.47 & 3.13  & 1.78  & 3.39  & -4.05 (0.00)  & 2316.50 (0.00)  \\
ESGI & 5.06  & 2.45  & 16.57 & -0.22 & 1.37  & 1.68  & -3.66 (0.00)  & 987.99 (0.00)  \\
MEI  & 15.23 & 10.68 & 74.70  & 2.67  & 1.70   & 4.42  & -4.62 (0.00)  & 2977.63 (0.00) \\ \hline
\end{tabular}}
\end{table}

\clearpage
\section{Additional connectedness table}
\label{apdx: conntable}

\begin{table}[!h]
\caption{Static local connectedness table of the projection layer: inside bond market}
\label{table: local1}
\centering
\renewcommand\arraystretch{1.5}
\scriptsize
\begin{tabular}{lcccccc}
\hline
          & GBIG  & GBIC  & GBIF  & TBI   & CBI   & From others \\ \hline
GBIG      & 24.66 & 22.17 & 18.46 & 14.29 & 12.42 & 67.34       \\
GBIC      & 21.69 & 25.14 & 18.79 & 13.07 & 13.2  & 66.75       \\
GBIF      & 19.8  & 20.46 & 27.22 & 12.37 & 10.98 & 63.61       \\
TBI       & 16.6  & 16.07 & 12.94 & 32.92 & 12.42 & 58.03       \\
CBI       & 14.03 & 15.67 & 11.37 & 13.01 & 36.52 & 54.08       \\ \hline
To others & 72.12 & 74.37 & 61.56 & 52.74 & 49.02 & 61.962      \\ \hline
Net       & 4.78  & 7.62  & -2.05 & -5.29 & -5.06 &             \\ \hline
\end{tabular}
\end{table}

\begin{table}[!h]
\caption{Static local connectedness table of the projection layer: inside equity market}
\label{table: local2}
\centering
\renewcommand\arraystretch{1.5}
\scriptsize
\begin{tabular}{lccccc}
\hline
          & KCTI  & KCEI  & ESGI  & MEI   & From others \\ \hline
KCTI      & 55.36 & 12.17 & 11.73 & 7.24  & 31.14       \\
KCEI      & 12.02 & 53.93 & 11.19 & 7.56  & 30.77       \\
ESGI      & 9.72  & 9.99  & 49.97 & 12.78 & 32.49       \\
MEI       & 7.52  & 8.75  & 14.81 & 54.02 & 31.08       \\ \hline
To others & 29.26 & 30.91 & 37.73 & 27.58 & 31.37       \\ \hline
Net       & -1.88 & 0.14  & 5.24  & -3.5  &             \\ \hline
\end{tabular}
\end{table}

\begin{table}[!h]
\caption{Static local connectedness table of the projection layer: from bond to equity market}
\label{table: local3}
\centering
\renewcommand\arraystretch{1.5}
\scriptsize
\begin{tabular}{lcccccc}
\hline
          & GBIG  & GBIC           & GBIF  & TBI            & CBI            & From others    \\ \hline
KCTI      & 2.47  & 2.74           & 2.65  & 2.9            & 2.75           & 13.51          \\
KCEI      & 2.88  & 3.03           & 2.7   & 3.11           & 3.6            & \textbf{15.32} \\
ESGI      & 3.3   & 3.78           & 3.7   & 3.4            & 3.37           & \textbf{17.55} \\
MEI       & 2.8   & 3.21           & 2.8   & 3.16           & 2.94           & 14.91          \\ \hline
To others & 11.45 & \textbf{12.76} & 11.85 & \textbf{12.57} & \textbf{12.66} &                \\ \hline
\end{tabular}
\end{table}

\begin{table}[!h]
\caption{Static local connectedness table of the projection layer: from equity to bond market}
\label{table: local4}
\centering
\renewcommand\arraystretch{1.5}
\scriptsize
\begin{tabular}{lccccc}
\hline
          & KCTI  & KCEI           & ESGI           & MEI  & From others   \\ \hline
GBIG      & 1.97  & 2.32           & 1.93           & 1.79 & 8.01          \\
GBIC      & 1.9   & 2.3            & 2.17           & 1.73 & 8.1           \\
GBIF      & 2.23  & 2.52           & 2.61           & 1.81 & \textbf{9.17} \\
TBI       & 2.3   & 2.44           & 2.58           & 1.73 & \textbf{9.05} \\
CBI       & 2.08  & 2.64           & 2.5            & 2.17 & \textbf{9.39} \\ \hline
To others & 10.48 & \textbf{12.22} & \textbf{11.79} & 9.23 &               \\ \hline
\end{tabular}
\end{table}

\begin{table}[!h]
\caption{Static connectedness table of the return layer}
\label{table: rn}
\centering
\renewcommand\arraystretch{1.4}
\scriptsize
\resizebox{\linewidth}{!}{
\begin{tabular}{l|ccccc|cccc|c}
\hline
          & GBIG  & GBIC  & GBIF  & TBI   & CBI   & KCTI  & KCEI  & ESGI  & MEI   & From others \\ \hline
GBIG      & 21.50 & 20.31 & 19.01 & 16.78 & 16.41 & 1.49  & 1.19  & 1.98  & 1.33  & 78.50       \\
GBIC      & 19.42 & 21.38 & 18.99 & 15.56 & 17.51 & 1.74  & 1.39  & 2.35  & 1.66  & 78.62       \\
GBIF      & 20.20 & 21.06 & 21.03 & 14.79 & 15.40 & 1.76  & 1.42  & 2.59  & 1.75  & 78.97       \\
TBI       & 19.02 & 18.51 & 15.61 & 23.81 & 17.84 & 1.36  & 0.89  & 1.59  & 1.36  & 76.19       \\
CBI       & 17.02 & 19.13 & 14.83 & 16.34 & 23.92 & 2.21  & 1.75  & 2.67  & 2.12  & 76.08       \\ \hline
KCTI      & 1.86  & 2.13  & 1.35  & 1.79  & 2.35  & 42.12 & 19.93 & 17.17 & 11.31 & 57.88       \\
KCEI      & 2.09  & 2.27  & 1.65  & 1.57  & 2.49  & 19.81 & 41.61 & 17.23 & 11.28 & 58.39       \\
ESGI      & 3.21  & 3.53  & 2.85  & 2.53  & 3.28  & 15.62 & 15.29 & 36.51 & 17.19 & 63.49       \\
MEI       & 1.56  & 1.78  & 1.58  & 1.88  & 1.80  & 12.23 & 12.24 & 20.63 & 46.30 & 53.70       \\ \hline
To others & 84.38 & 88.73 & 75.87 & 71.22 & 77.09 & 56.22 & 54.10 & 66.21 & 47.99 & 69.09       \\ \hline
Net       & 5.89  & 10.11 & -3.10 & -4.97 & 1.01  & -1.65 & -4.29 & 2.72  & -5.71 &             \\ \hline
\end{tabular}}
\end{table}

\begin{table}[!h]
\caption{Static connectedness table of the volatility layer}
\label{table: vol}
\centering
\renewcommand\arraystretch{1.4}
\scriptsize
\resizebox{\linewidth}{!}{
\begin{tabular}{l|ccccc|cccc|c}
\hline
          & GBIG  & GBIC  & GBIF  & TBI   & CBI   & KCTI  & KCEI  & ESGI  & MEI    & From others \\ \hline
GBIG      & 22.65 & 20.31 & 19.06 & 15.36 & 12.03 & 3.07  & 2.87  & 2.97  & 1.67   & 77.35       \\
GBIC      & 19.57 & 22.50 & 19.36 & 14.07 & 13.22 & 3.48  & 2.99  & 3.33  & 1.49   & 77.50       \\
GBIF      & 19.51 & 20.94 & 23.38 & 13.30 & 11.65 & 3.36  & 2.86  & 3.46  & 1.55   & 76.62       \\
TBI       & 16.11 & 15.11 & 13.75 & 28.02 & 12.71 & 4.17  & 3.46  & 4.40  & 2.26   & 71.98       \\
CBI       & 14.48 & 16.33 & 14.03 & 15.83 & 27.89 & 3.56  & 3.09  & 3.32  & 1.47   & 72.11       \\ \hline
KCTI      & 2.98  & 3.77  & 3.76  & 2.40  & 3.47  & 44.74 & 14.31 & 15.47 & 9.11   & 55.26       \\
KCEI      & 3.10  & 3.49  & 3.24  & 2.54  & 4.31  & 14.13 & 46.08 & 13.64 & 9.47   & 53.92       \\
ESGI      & 3.36  & 4.20  & 4.44  & 3.02  & 4.16  & 12.12 & 10.69 & 46.06 & 11.95  & 53.94       \\
MEI       & 4.26  & 4.93  & 4.37  & 4.08  & 3.29  & 9.55  & 9.64  & 14.94 & 44.94  & 55.06       \\ \hline
To others & 83.37 & 89.08 & 82.01 & 70.59 & 64.84 & 53.44 & 49.91 & 61.52 & 38.96  & 65.97       \\ \hline
Net       & 6.03  & 11.58 & 5.39  & -1.39 & -7.26 & -1.82 & -4.00 & 7.58  & -16.10 &             \\ \hline
\end{tabular}}
\end{table}

\begin{table}[!h]
\caption{Static connectedness table of the skewness layer}
\label{table: skew}
\centering
\renewcommand\arraystretch{1.4}
\scriptsize
\resizebox{\linewidth}{!}{
\begin{tabular}{l|ccccc|cccc|c}
\hline
          & GBIG  & GBIC  & GBIF  & TBI   & CBI   & KCTI   & KCEI  & ESGI   & MEI    & From others \\ \hline
GBIG      & 25.97 & 24.03 & 17.53 & 12.39 & 13.09 & 0.72   & 2.41  & 1.13   & 2.73   & 74.03       \\
GBIC      & 23.01 & 27.46 & 17.44 & 11.83 & 13.39 & 0.60   & 2.41  & 1.40   & 2.45   & 72.54       \\
GBIF      & 19.15 & 19.35 & 31.58 & 12.60 & 9.13  & 0.90   & 3.05  & 1.93   & 2.30   & 68.42       \\
TBI       & 14.70 & 15.53 & 12.24 & 36.86 & 11.67 & 1.09   & 3.54  & 2.76   & 1.62   & 63.14       \\
CBI       & 15.48 & 16.85 & 9.13  & 12.11 & 38.49 & 0.42   & 2.78  & 1.54   & 3.21   & 61.51       \\ \hline
KCTI      & 2.65  & 2.36  & 2.53  & 5.96  & 2.83  & 68.60  & 5.65  & 6.02   & 3.41   & 31.40       \\
KCEI      & 4.18  & 4.22  & 3.23  & 5.94  & 5.09  & 4.96   & 65.40 & 4.31   & 2.66   & 34.60       \\
ESGI      & 4.03  & 4.47  & 4.78  & 4.09  & 3.21  & 4.95   & 6.65  & 59.56  & 8.25   & 40.44       \\
MEI       & 3.59  & 4.39  & 2.96  & 3.65  & 4.83  & 3.23   & 5.78  & 8.68   & 62.89  & 37.11       \\ \hline
To others & 86.79 & 91.20 & 69.84 & 68.58 & 63.25 & 16.88  & 32.26 & 27.77  & 26.63  & 53.69       \\ \hline
Net       & 12.76 & 18.66 & 1.41  & 5.44  & 1.74  & -14.53 & -2.34 & -12.67 & -10.49 &             \\ \hline
\end{tabular}}
\end{table}

\begin{table}[!h]
\caption{Static connectedness table of the kurtosis layer}
\label{table: kurt}
\centering
\renewcommand\arraystretch{1.4}
\scriptsize
\resizebox{\linewidth}{!}{
\begin{tabular}{l|ccccc|cccc|c}
\hline
          & GBIG  & GBIC  & GBIF  & TBI   & CBI   & KCTI  & KCEI  & ESGI  & MEI   & From others \\ \hline
GBIG      & 30.88 & 25.54 & 17.83 & 11.15 & 6.07  & 2.56  & 3.12  & 1.29  & 1.56  & 69.12       \\
GBIC      & 26.70 & 32.04 & 19.25 & 9.27  & 6.32  & 1.42  & 2.52  & 1.14  & 1.35  & 67.96       \\
GBIF      & 20.39 & 20.13 & 37.15 & 7.04  & 5.39  & 2.89  & 3.07  & 2.22  & 1.72  & 62.85       \\
TBI       & 15.89 & 14.38 & 8.47  & 49.42 & 4.54  & 2.45  & 2.02  & 1.22  & 1.62  & 50.58       \\
CBI       & 7.05  & 7.96  & 4.83  & 4.85  & 66.19 & 1.68  & 3.20  & 2.16  & 2.07  & 33.81       \\ \hline
KCTI      & 2.44  & 2.62  & 3.14  & 1.68  & 2.24  & 75.46 & 4.89  & 4.71  & 2.83  & 24.54       \\
KCEI      & 2.23  & 2.09  & 2.89  & 2.91  & 2.46  & 5.35  & 70.70 & 6.49  & 4.88  & 29.30       \\
ESGI      & 2.47  & 2.70  & 2.63  & 4.49  & 2.54  & 2.81  & 4.81  & 64.92 & 12.63 & 35.08       \\
MEI       & 1.59  & 1.47  & 2.19  & 3.17  & 1.90  & 2.40  & 5.63  & 13.02 & 68.63 & 31.37       \\ \hline
To others & 78.76 & 76.88 & 61.23 & 44.56 & 31.46 & 21.57 & 29.26 & 32.26 & 28.65 & 44.96       \\ \hline
Net       & 9.63  & 8.92  & -1.63 & -6.03 & -2.35 & -2.98 & -0.04 & -2.82 & -2.72 &        \\  \hline    
\end{tabular}}
\end{table}

\clearpage 
\section{Additional impluse response}
\label{apdx: ir}

\begin{figure}[!h]
    \centering
    \begin{subfigure}{0.325\textwidth}
        \includegraphics[width=\linewidth]{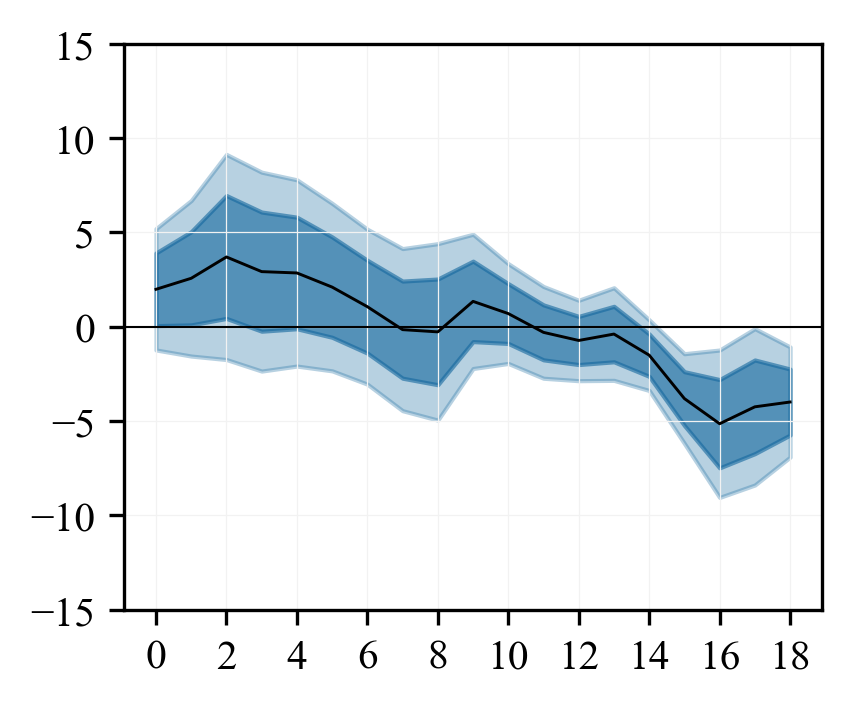}
        \caption{\footnotesize{GBIG}}
    \end{subfigure}
    \vspace{0.25cm}
    \begin{subfigure}{0.325\textwidth}
        \includegraphics[width=\linewidth]{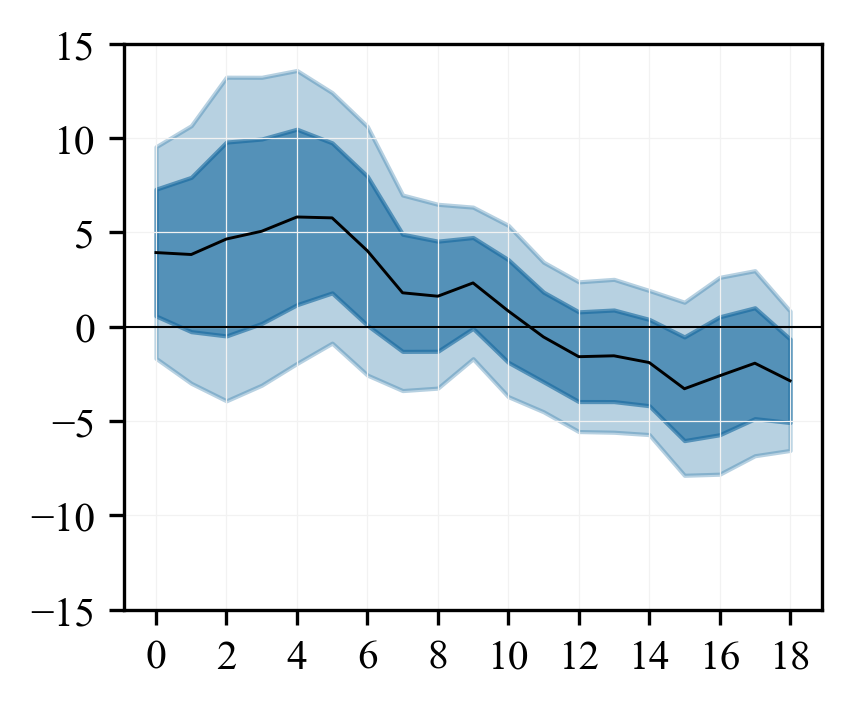}
        \caption{\footnotesize{GBIC}}
    \end{subfigure}
    \begin{subfigure}{0.325\textwidth}
        \includegraphics[width=\linewidth]{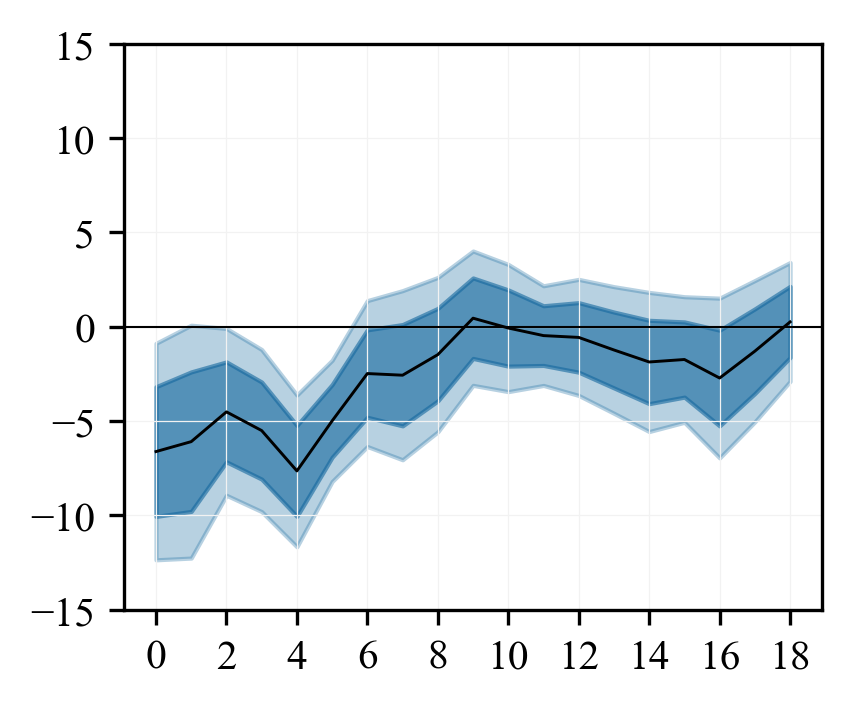}
        \caption{\footnotesize{GBIF}}
    \end{subfigure}
    
    \begin{subfigure}{0.325\textwidth}
        \includegraphics[width=\linewidth]{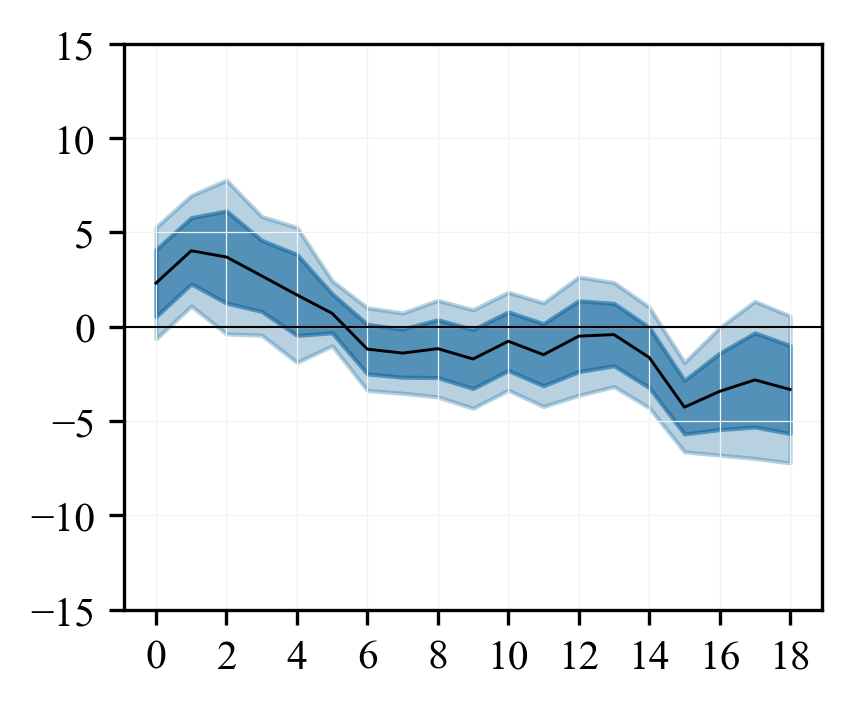}
        \caption{\footnotesize{TBI}}
    \end{subfigure}
    \vspace{0.25cm}
    \begin{subfigure}{0.325\textwidth}
        \includegraphics[width=\linewidth]{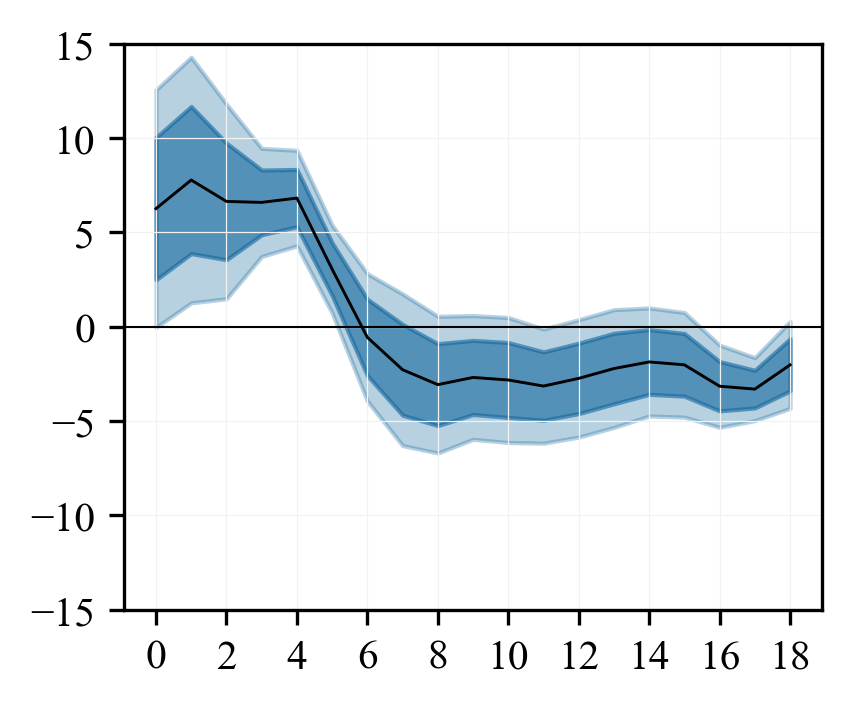}
        \caption{\footnotesize{CBI}}
    \end{subfigure}
    \begin{subfigure}{0.325\textwidth}
        \includegraphics[width=\linewidth]{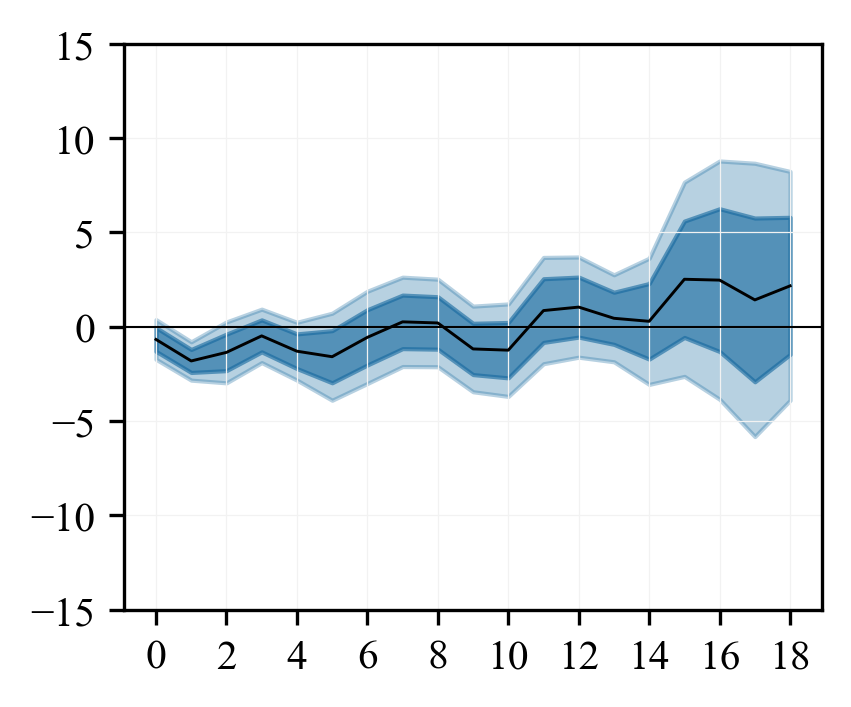}
        \caption{\footnotesize{KCTI}}
    \end{subfigure}

    \begin{subfigure}{0.325\textwidth}
        \includegraphics[width=\linewidth]{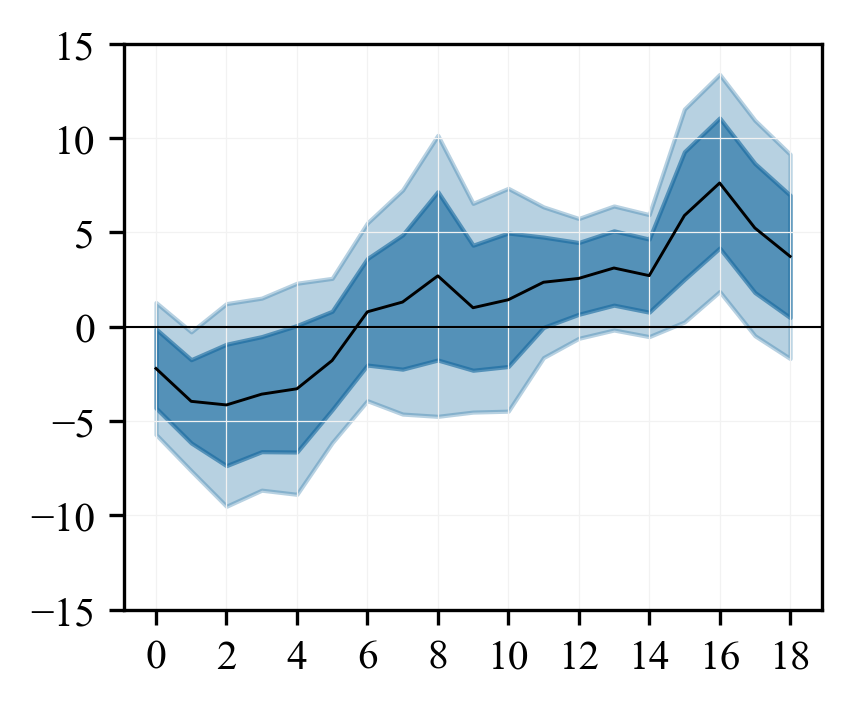}
        \caption{\footnotesize{KCEI}}
    \end{subfigure}
    \begin{subfigure}{0.325\textwidth}
        \includegraphics[width=\linewidth]{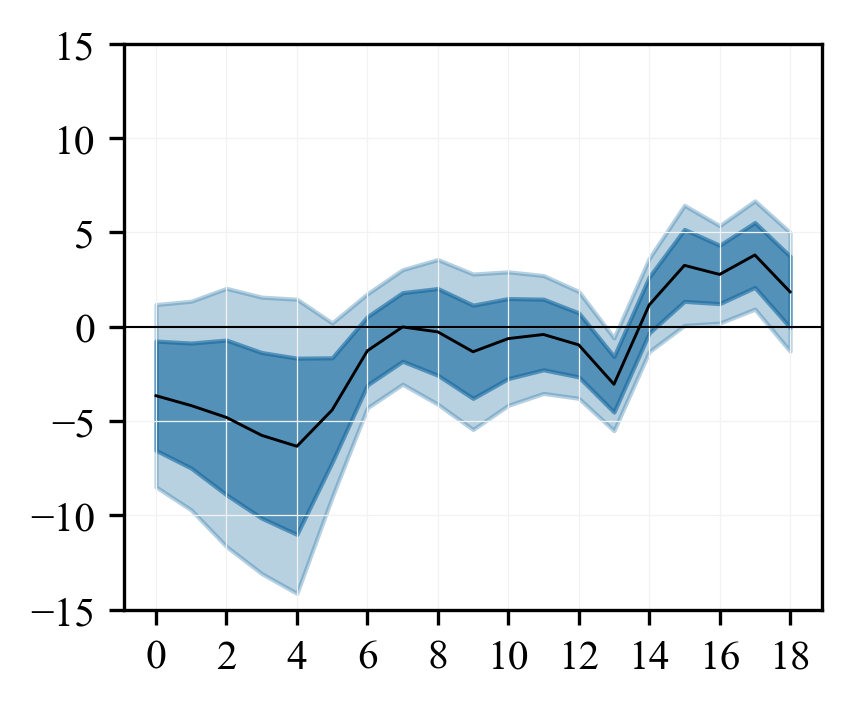}
        \caption{\footnotesize{ESGI}}
    \end{subfigure}
    \begin{subfigure}{0.325\textwidth}
        \includegraphics[width=\linewidth]{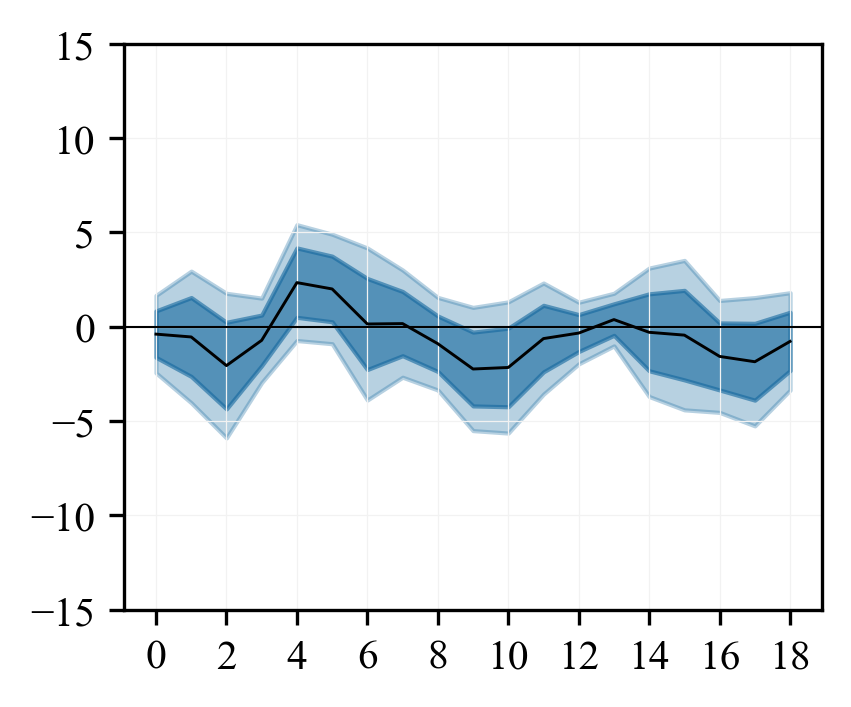}
        \caption{\footnotesize{MEI}}
    \end{subfigure}

    \caption{Impluse response of nspl: Fed rate hike}
    \label{figure: ir_up}
    \raggedright{\justifying{\footnotesize{\textit{Notes}: The black line is the impulse response of net connectedness index in a projection layer to monetary policy shocks during Fed rate hike. The dark blue band represents the 68\% confidence interval, while the light blue band represents the 90\% confidence interval.}}}
\end{figure}

\begin{figure}[!h]
    \centering
    \begin{subfigure}{0.325\textwidth}
        \includegraphics[width=\linewidth]{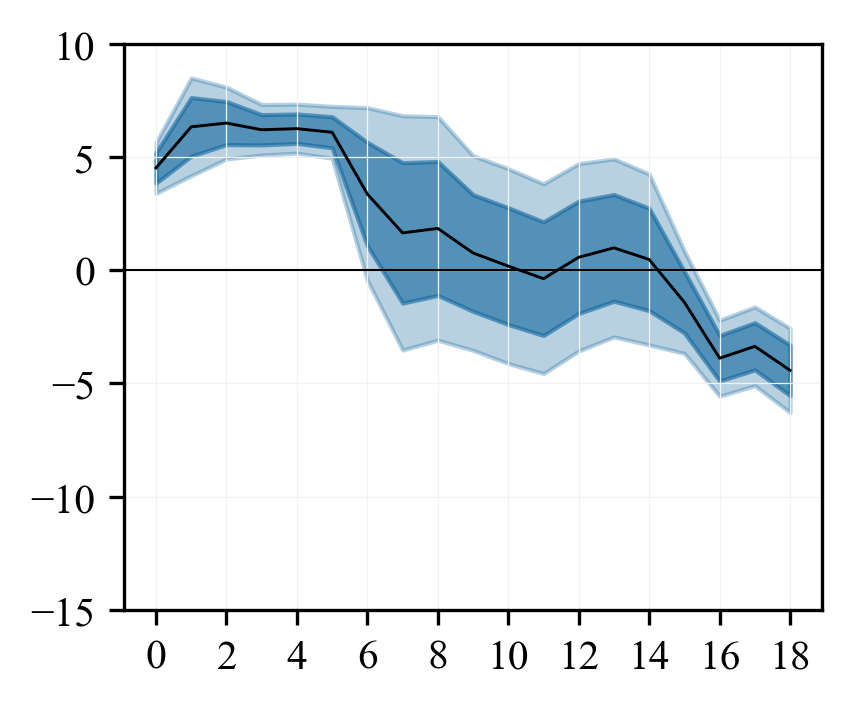}
        \caption{\footnotesize{GBIG}}
    \end{subfigure}
    \vspace{0.25cm}
    \begin{subfigure}{0.325\textwidth}
        \includegraphics[width=\linewidth]{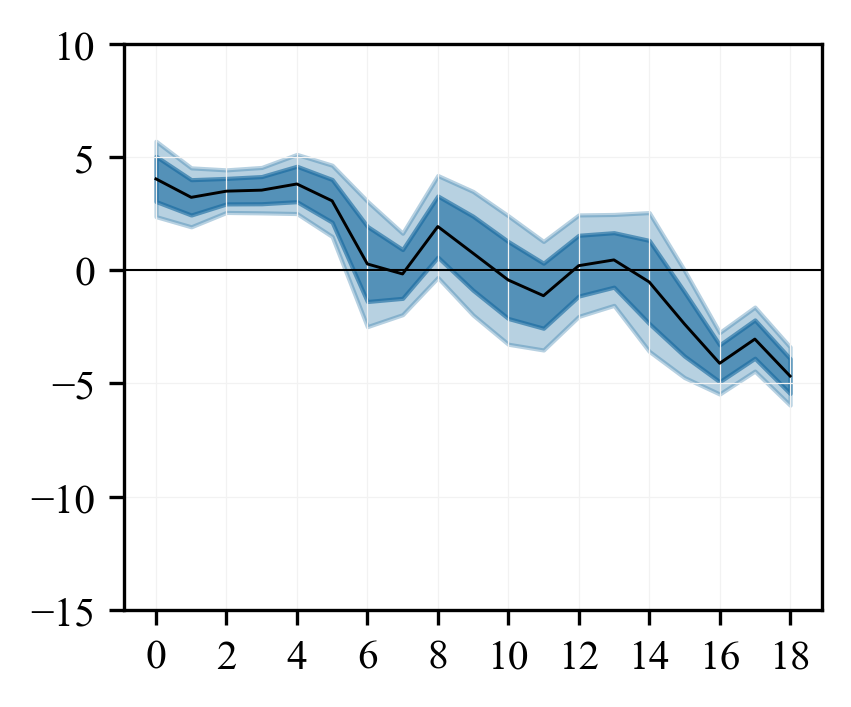}
        \caption{\footnotesize{GBIC}}
    \end{subfigure}
    \begin{subfigure}{0.325\textwidth}
        \includegraphics[width=\linewidth]{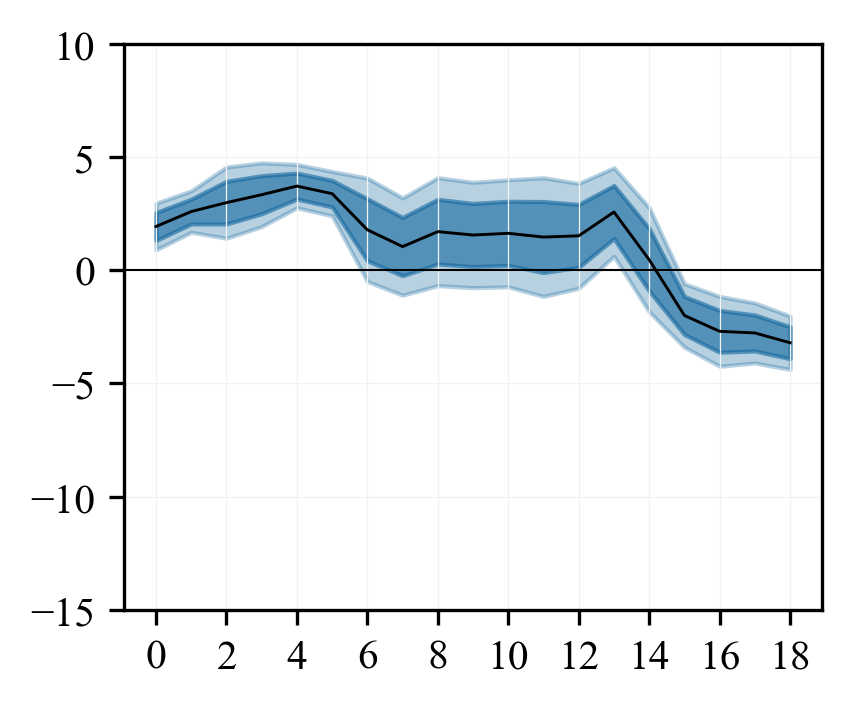}
        \caption{\footnotesize{GBIF}}
    \end{subfigure}
    
    \begin{subfigure}{0.325\textwidth}
        \includegraphics[width=\linewidth]{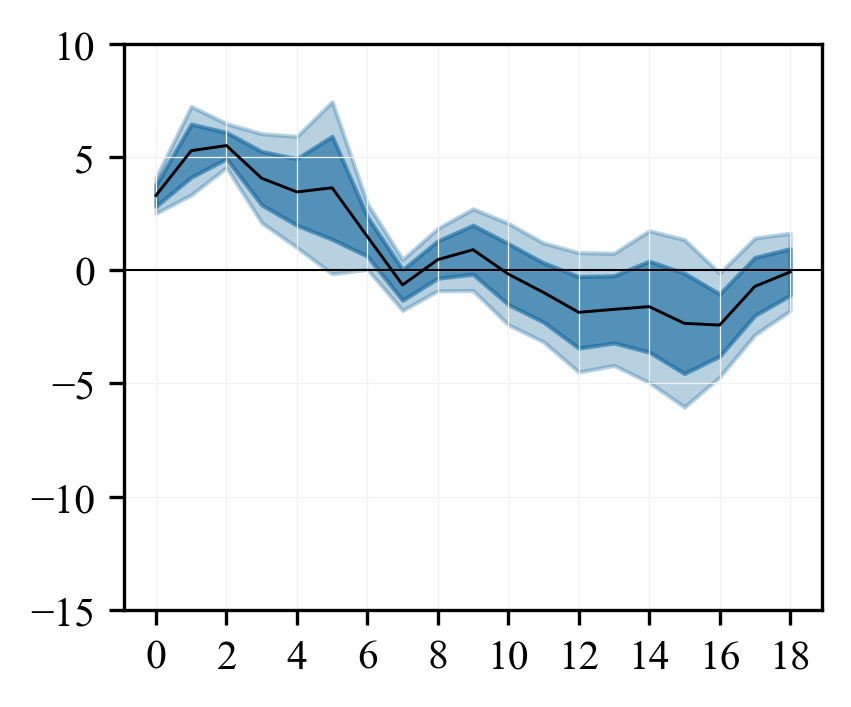}
        \caption{\footnotesize{TBI}}
    \end{subfigure}
    \vspace{0.25cm}
    \begin{subfigure}{0.325\textwidth}
        \includegraphics[width=\linewidth]{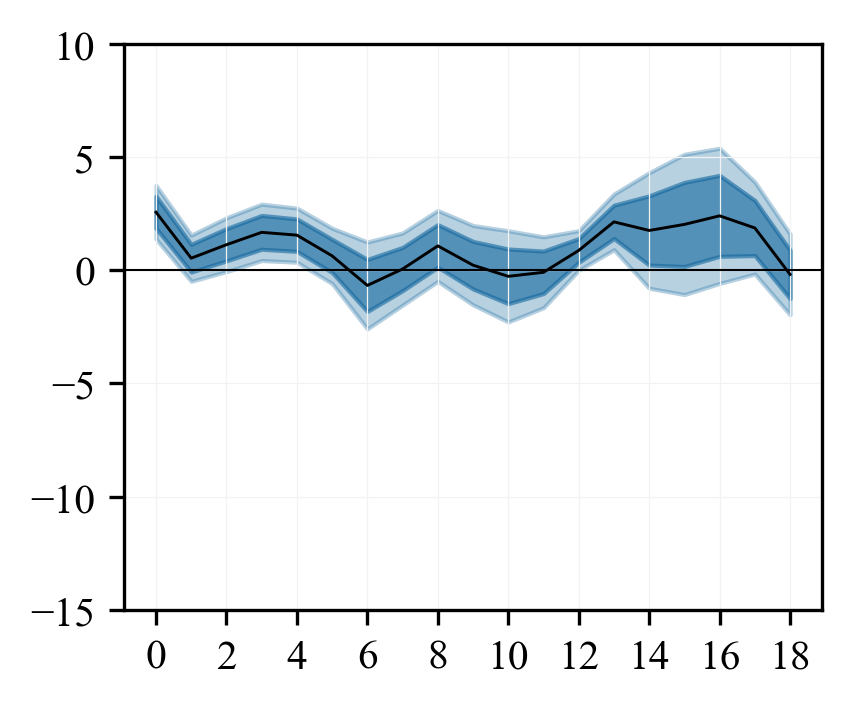}
        \caption{\footnotesize{CBI}}
    \end{subfigure}
    \begin{subfigure}{0.325\textwidth}
        \includegraphics[width=\linewidth]{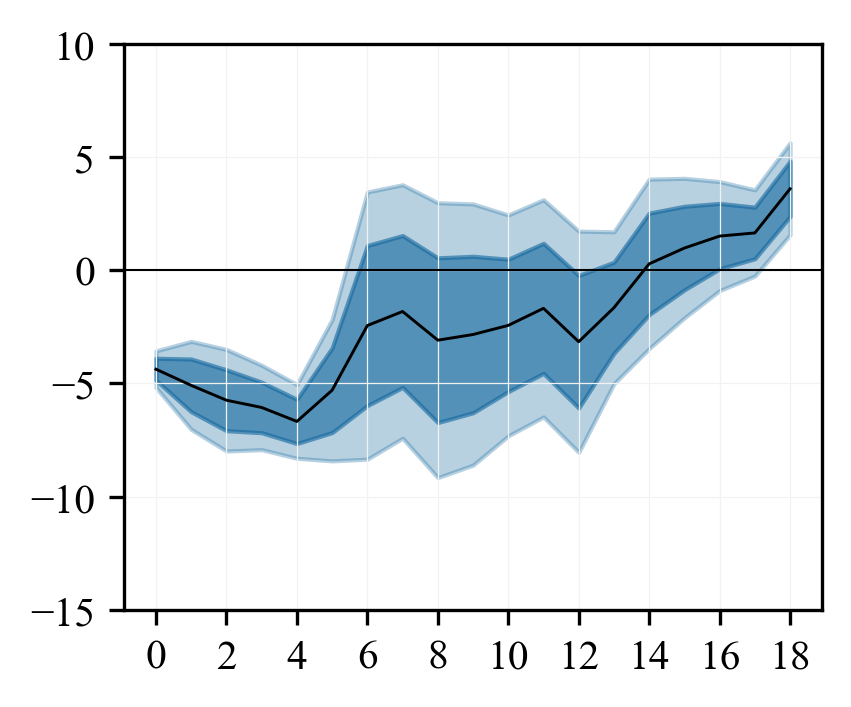}
        \caption{\footnotesize{KCTI}}
    \end{subfigure}

    \begin{subfigure}{0.325\textwidth}
        \includegraphics[width=\linewidth]{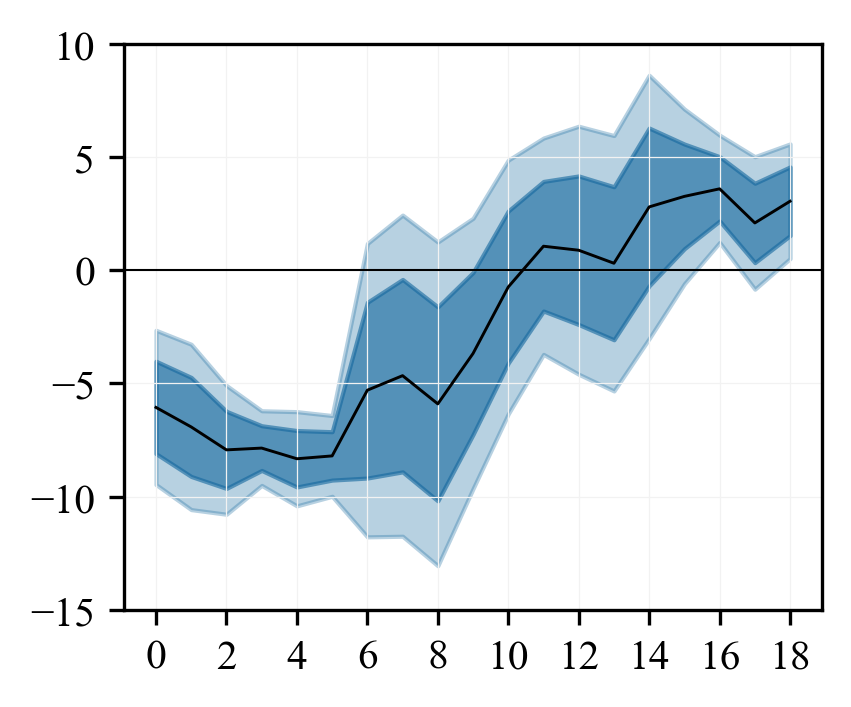}
        \caption{\footnotesize{KCEI}}
    \end{subfigure}
    \begin{subfigure}{0.325\textwidth}
        \includegraphics[width=\linewidth]{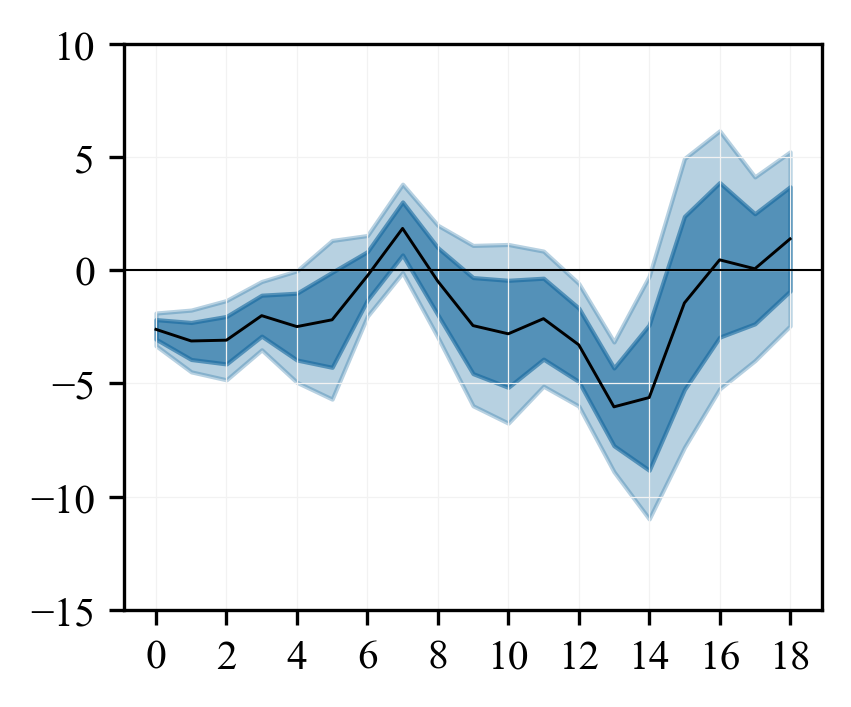}
        \caption{\footnotesize{ESGI}}
    \end{subfigure}
    \begin{subfigure}{0.325\textwidth}
        \includegraphics[width=\linewidth]{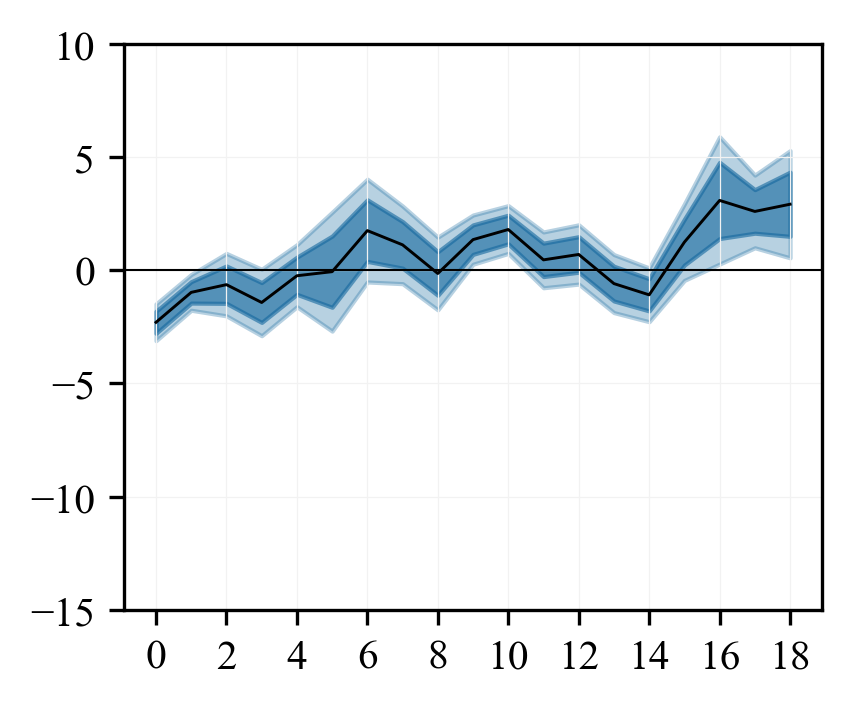}
        \caption{\footnotesize{MEI}}
    \end{subfigure}

    \caption{Impluse response of nspl: Fed rate cut}
    \label{figure: ir_down}
    \raggedright{\justifying{\footnotesize{\textit{Notes}: The black line is the impulse response of net connectedness index in a projection layer to monetary policy shocks during Fed rate cut. The dark blue band represents the 68\% confidence interval, while the light blue band represents the 90\% confidence interval.}}}
\end{figure}
\clearpage 

\section{Complete Estimation Procedures for TVP-VAR}
\label{apdx: TVP-VAR formula}

Let $N\times1$ vector of $\hy_t = (y_{1t},\ldots, y_{Nt})'$ for $t=1,\ldots, T$ denotes the collection of log-returns, log-volatilities, skewness, or kurtosis of the $N$ indicators, we consider the following reduced-form TVP-VAR($p$) model:
\begin{equation} \label{equ: TVP-VAR(p)_app}
    \h y_t = \h c_t+\h B_{1t}\h y_{t-1}+\ldots+\h B_{pt}\h y_{t-p}+\h \varepsilon_t,\qquad
	\h \varepsilon_t\sim \h{\mathcal{N}}(\h 0,\h \Sigma_t),
\end{equation}
where $\h c_t$ is an $N\times 1$ vector of time-varying intercepts, $\h B_{it}$ are $N\times N$ time-varying coefficient matrices for $i=1,\ldots, p$, $\h \varepsilon_t$ denotes a $N\times 1$ vector of random disturbance that follows multivariate Gaussian distribution with mean $\h 0$ and time-varying covariance matrix $\h \Sigma_t$.  

Given the model in Eq.\eqref{equ: TVP-VAR(p)_app}, by some linear transformations, we then obtain the following contemporaneous-form state space representation:
\begin{align*}
    \label{equ: TVP-VAR-obs_app}
	\h y_t &= \h Z_t\h \beta_t + \h\varepsilon_t,\qquad \h\varepsilon_t\sim N(0,\h \Sigma_t),\\
	\h \beta_t &= \h \beta_{t-1} + \h v_t,\qquad \h v_t\sim N(0,\h Q_t),
\end{align*}
where $\h Z_t = \hI_N\otimes [\h 1', \hy_{t-1}',\ldots, \hy_{t-p}']'$ and $\h \beta = \mathrm{vec}([\h c_t, \h B_{1t},\ldots, \h B_{pt}]')$. Note that here we follow \cite{koop2013large} and assume that the time-varying coefficients $\h \beta_t$ follow a random walk process, where the $N\times 1$ random error $\h v_t$ follows a zero-mean multivariate Gaussian distribution with time-varying covariance matrix $\h Q_t$.

The linear state space representation allows the immediate implementation of the Kalman filter to estimate the conditional mean and variance of the time-varying parameters:
\begin{align*} 
        \widetilde{\h y}_t &= \h y_t-\h Z_t\h \beta_{t\vert t-1},\qquad\qquad\quad \h F_t = \h Z_t\h P_{t}\h Z_t^{\prime}+\h \Sigma_t,\\
    	\h K_t &= \h P_{t}\h Z_t^{\prime}\h F_t^{-1},\qquad\qquad\qquad\quad \h L_t = \h I-\h K_t\h Z_t,\\
    	\h \beta_{t\vert t} &= \h \beta_{t\vert t-1}+\h K_t\widetilde{\h y}_t,\qquad\qquad\quad \h P_{t|t}= \h P_t\h L_t^{\prime},
\end{align*}
for $t = 1,\ldots, T$, where $\h \beta_{t\vert t-1} = \E(\h \beta\vert \mathcal{F}^{t-1})$, $\h P_{t\vert t-1} = \Var(\h \beta \vert \mathcal{F}^{t-1})$ and $\mathcal{F}^{t-1} = \sigma(\h y_{t-1}, \h y_{t-2},\ldots, \h y_{1-p})$ denotes the $\sigma-$field generated by past information. 

Three issues need to be addressed to ensure the implementation of the Kalman filter, the estimation of time-varying covariance matrices $\h \Sigma_t$, $\h Q_t$, and the prior setting of $\h \beta$. Here we follow the standard solutions provided in \cite{koop2013large}. For $\h \Sigma_t$, we assume it follows an exponential weighted moving average (EWMA) process:
$
    \h \Sigma_t = \kappa \Sigma_{t-1} + (1-\kappa) \widetilde{\h y}_t \widetilde{\h y}_t'.
$
For $\h Q_t$, \cite{koop2013large} introduce forgetting factor by letting $\h Q_t = (\lambda^{-1}-1)\h P_{t-1\vert t-1}$. For the prior distribution of $\h \beta_0$, we use a Minnesota-type prior with prior mean $\E(\h\beta_0) = 0$ and let the variance $\Var(\h\beta_0) = \h V_0$ be a diagonal matrix with its $i$th diagonal element be
\begin{align*}
	(\h V_{0})_i = \left\{ \begin{array}{ll}
	\frac{\gamma}{l^2},
	&\mbox{for coefficients on lag $l$ for $l = 1,\ldots p$};
	\\
	100,& \mbox{for the intercept},
	\end{array}\right.
\end{align*}
We follow \cite{akyildirim2022connectedness} and set $\kappa = 0.99$, $\lambda = 0.99$, $\gamma = 0.01$ to aviod numerical instability. 

Given the estimates of $\h c_t$ and $\h B_{lt} $ for $l = 1,\ldots, p$, we first transform the TVP-VAR($p$) model into the TVP-VMA($\infty$) one $\h y_t = \sum_{i=0}^{\infty} \h\Psi_{i, t}\hvarepsilon_t$. Then the generalized forecast error variance decomposition (GFEVD) is calculated using the standard approach in \cite{koop1996impulse}.

\end{appendices}
\end{document}